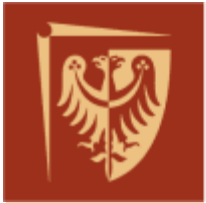

## Wrocław University of Technology

**Faculty of Computer Science and Management**
**Institute of Informatics**

# A Method for Group Extraction and Analysis in Multi-layered Social Networks

Piotr Bródka, MSc.

Supervisor:
Przemysław Kazienko, PhD, DSc, Prof. of WrUT

Thesis submitted for the Degree of Doctor of Philosophy at the
Wrocław University of Technology, 2012

***This dissertation is dedicated to my parents, my brother and the rest of my family.***

*For their endless love, encouragement
and the fact that they always support my decisions
no matter how crazy they seem to be.*

*Thank you!!!!*

# Acknowledgment

I would like to start with thanks to my supervisor *Przemysław Kazienko*. I am grateful for his directness, his ideas, for development opportunities, for a well-organized inspiring research environment and showing me that I should always reach further.

I am thankful to *Tomasz Kajdanowicz* and *Tomasz Filipowski* for endless discussion, help and support in moments of "PhD crisis", and finally for dragging me from the office to try something else than research.

I am grateful *to Katarzyna Musiał* who despite the fact that she is always very busy and lives thirteen hundred kilometres from me always has managed to find the time for scientific discussion or simple chat.

Special thanks to my former students and now colleagues: *Stanisław Saganowski* for his help in *GED* method development, *Tomasz Grecki* for his help in m*LFR* Benchmark development, *Krzysztof Skibicki* and *Paweł Stawiak* for their help in development of measures for multi-layered social networks analysis.

I would like also to thank my friend *Agata Zielińska* for dragging me to the "behind the firewall" world, where we could have a lot of fun, but also for being a person whom I could always talk and confide.

Last, but not least, I would like to thank to rest of my friends and colleagues who supported me and always at least pretended that they are interested in my research ☺

# Content

# **Abstract**

The main subject studied in this dissertation is a multi-layered social network (*MSN*) and its analysis. MSN is a structure in which two nodes can be connected by more than one relationship (edge). Despite the fact that multi-layered profile of networks is an absolutely natural concept, for many years, researchers have refrained from analysing more than one layer at once. However, some relations are so intertwined that it is impossible to analyse them separately. Moreover, if considered together, they provide additional vital information about the network. Especially nowadays, when it is so important to analyse information diffusion and social dependencies between people in the global network, very often the data from only one communication channel (one type of relationship) are insufficient.

One of the crucial problems in multi-layered social network analysis is community extraction. In order to cope with this problem the *CLECC* measure (Cross Layered Edge Clustering Coefficient) was proposed in the thesis. It is an edge measure which expresses how much the neighbours of two given users are similar each other. Based on this measure the *CLECC* algorithm for community extraction in the multi-layered social networks was designed. The algorithm was tested on the real single-layered social networks (*SSN*) and multi-layered social networks (*MSN*), as well as on benchmark networks from *GN* Benchmark (*SSN*), *LFR* Benchmark (*SSN*) and m*LFR* Benchmark (*MSN*) – a special extension of *LFR* Benchmark, designed as a part of this thesis, which is able to produce multi-layered benchmark networks.

The second research problem considered in the thesis was group evolution discovery. Studies on this problem have led to development of the inclusion measure and the Group Evolution Discovery (*GED*) method, which are designed to identify events between two groups in successive time frames in the social network. The method was tested on real social network and compared with two well-known algorithms in terms of accuracy, execution time, flexibility and ease of implementation.

Finally, a new approach to prediction of group evolution in the social network was developed. The new approach, involves usage of the outputs of the *GED* method. It is shown, that using even a simple sequence, which consists of several preceding groups' sizes and events, as an input for the classifier, the learnt model is able to produce very good results also for simple classifiers.

# Streszczenie

Sieci społeczne (social networks) *i* ich analiza (SNA) od dawna budzą zainteresowanie badaczy z całego świata. Jednakże w ciągu ostatnich kilku lat nastąpił znaczny wzrost zainteresowania wielowarstwowymi sieciami społecznymi (multi-layered social network). Związane jest to z szeroko rozumianym pojęciem dyfuzji informacji w sieci. Poprzez informację możemy rozumieć zarówno opinię na temat produktu krążącą w internecie, wirusa komputerowego rozprzestrzeniającego się w sieci, jak *i* wirusa grypy rozprzestrzeniającego się w mieście. Gdy rozważano ten problem, okazało się, iż analizując jedną warstwę sieci bardzo często dochodzi do sytuacji gdy nie możemy zrozumieć jak informacja trafiła od użytkownika A do użytkownika B, gdyż nie są oni na danej warstwie w żaden sposób połączeni. Natomiast analizując wszystkie dostępne warstwy jednocześnie możemy uzyskać dodatkową informację, np. że istnieje użytkownik C który jest połączony z A na warstwie pierwszej (np. Facebook) *i* z B na drugiej (np. LinkedIn) *i* to on przekazał informację. Dzisiejszą wizję wielowarstwowych sieci społecznych wykreowali amerykanie (Air Force Institute of Technology), którzy bardzo mocno rozwinęli zagadnienie w celu analizy siatek terrorystycznych *i* różnych kanałów komunikacji między terrorystami po 11.09.2001.

Jednym z najbardziej aktualnych problemów w dziedzinie SNA jest wyszukiwanie grup (groups, communities) w sieciach wielowarstwowych (pierwsze teoretyczne prace zaczęły się pojawiać dwa lata temu). Obecnie oprócz opracowanego *i* zaprezentowanego w tej pracy algorytmu *CLECC* istnieje jeszcze jeden będący w dalszym ciągu w fazie testowania. Dodatkowym problemem związanym z grupowaniem jest brak możliwości testowania poprawności działania zaproponowanych metod. Dla sieci jednowarstwowych istnieją sieci referencyjne, przebadane przez socjologów, którzy określili w nich grupy, oraz istnieją generatory sztucznych sieci referencyjnych (*GN* Benchmark, *LFR* Benchmark), przy użyciu których można testować nowe metody. Natomiast dla sieci wielowarstwowych w ramach pracy niezbędne było opracowane rozszerzenie do *LFR* Benchmarku, które pozwala na generowanie wielowarstwowych sieci referencyjnych. Dodatkowo wykonano testy przy użyciu danych z wirtualnego świata polskiej gry, Timik.pl.

Drugim problemem podejmowanym w pracy jest analiza grupy a konkretnie określanie zmian jakie przeszła grupa w swojej historii. Zostało zaproponowane nowe podejście nazwane *GED* (Group Evolution Discovery). Przy pomocy opracowanej w ramach pracy metryki zwanej inkluzją, metoda *GED* pozwala na określenie jaka zmiana zaszła dla dwóch grup pomiędzy kolejnymi oknami dynamicznej sieci społecznej. Metodę przetestowano przy wykorzystaniu rzeczywistych sieci społecznych *i* porównano do wiodących metod określania zmian jakie przeszła grupa w swojej historii.

Powyższe zagadnienie jest o tyle ważne, że w momencie kiedy określona zostanie historia zmian grupy, można na jej podstawie próbować przewidywać przyszłe zmiany. W pracy podjęto próbę wykorzystania wyników metody *GED* w celu predykcji przyszłej zmiany dla danej grupy. Przy pomocy sekwencji składającej się z wielkości grupy w kolejnych oknach czasowych, zmian pomiędzy tymi oknami oraz wykorzystując proste klasyfikatory udało się uzyskać bardzo dobre wyniki.

# List of the Most Important Abbreviations and Symbols

| | |
|---|---|
| *CDC* | Cross layered Degree Centrality |
| *CLCC* | Cross Layered Clustering Coefficient |
| *CLECC* | Cross Layered Edge Clustering Coefficient |
| *DSN* | Dynamic Social Network |
| *GED* | Group Evolution Discovery |
| *ME* | Multi-layered Edge |
| *MDC* | Multi-layered Degree Centrality |
| *MN* | Multi-layered Neighbourhood |
| *MSN* | Multi-layered Social Network |
| *NMI* | Normalized Mutual Information measure |
| *SN* | Social Network |
| *SNA* | Social Network Analysis |
| *SSN* | Single-layered Social Network |
| *E* | The connections in the social network *SSN* or *MSN* |
| *G* | The group (community) in the social network |
| $k_i$ | The degree of the user *i* ($k_i$=*DC*(*i*) for *SSN* and $k_i$=*CDC*(*i*) for *MSN*) |
| $k_i^{(in)}$ | The number of connection to the nodes from the same community as node *i* |
| $k_i^{(out)}$ | The number of connection to the nodes not belonging to node *i* community |
| *L* | The layers in the social network *MSN* |
| *n* | The number of nodes in the social network |
| *m* | The number of connections between two nodes in the social network |
| $\mu$ | The mixing parameter |
| *V* | The vertices in the social network *SSN* or *MSN* |

# List of Figures

# List of Tables

# 1. Introduction

## 1.1 Research Domain

It is obvious that in the real world, between two actors – social entities (humans, or groups of people) more than one kind of relationships (e.g. family, friendship and work ties) can exist. These ties can be so intertwined that it is impossible to analyse them separately [Fienberg 85], [Minor 83], [Szell 10]. A network where more than one type of relation exists are not new in the world of science [Wasserman 94], but they have been analysed mainly at the small scale, e.g. in [McPherson 01], [Padgett 93], and [Entwisle 07]. Just like in the case of regular single-layered social network presented in section 2.1 there is no widely accepted definition or even common name. At the beginning, such networks have been called multiplex network [Haythornthwaite 99], [Monge 03]. The term was derived from communications theory, which defines multiplex as combining multiple signals into one in such a way that it is possible to separate them [Hamill 06]. Recently, the area of large-scale multi-layered social networks has started attracting more and more attention of researchers from different fields [Kazienko 11], [Szell 10], [Rodriguez 07], [Rodriguez 09], and the meaning of multiplex network has expanded and covers not only social relationships but any kind of connections, e.g. based on geography, occupation, kinship, hobbies, etc. [Abraham 12]. Nowadays, social networks with more than one kind of relationship have many different names. The most common name is m*ulti-layered (or just layered) social networks* [**Bródka** 11b], [Geffre 09], [Hamill 06], [Kennedy 09], [Magnani 11], [Schneider 11] but also *multi-relational social networks* [Szell 10], *multi-dimensional social networks* [Kazienko 11], *multidimensional dynamic social network* [Kazienko and **Bródka** 11a], [Kazienko and **Bródka** 11a] or *multivariate social networks* [Szell 10] are in use.

Despite the fact that multi-layered nature of networks is an absolutely natural concept, researchers have refrained from analysing more than a single layer at once for many years. For example, Wasserman and Fraust [Wasserman 94] recommended that common centrality and prestige measures should be calculated for each relation (layer) separately and suggested not to perform any aggregation of the relations. Unfortunately, they do not explain why they advised such approach. The possible arguments for that are: (i) potential loss of information, which may occur during aggregation process and (ii) not sufficient computational power in 1994.

However, as mentioned at the beginning, some relation are so intertwined that it is impossible to analyse them separately and if considered together, they may reveal additional vital information about the network. Especially nowadays, when it is so important to analyse information diffusion in the global network. For example, let's consider the following case:

User *A* publishes information on YouTube[1] (in this case the movie). Next, user *B,* who subscribes user's *A* channel on YouTube, forwards this information and puts out the movie on the Facebook[2] board. After that, user *C* who is user's *B* Facebook friend and does not know user *A*, watches the movie and sends it via Twitter[3] to user *D* who knows neither *A* nor *B*.

If the data from each above systems (YouTube, Facebook, Twitter) is analysed as separate social networks, then it would be hard to find how the information (the movie) has circulated from user *A* to user *D.* However, if the data from all three systems could be merged into one multi-layered social network, then the path from user *A* to user *D* can be quite easily traced. Hence, the possibility of extracting the new information made multi-layered social networks so popular these days.

## 1.2 Problem Description

One of the most important problems in analysis of multi-layered social network (*MSN*) is community detection. Only few researchers have tried to address this issue. The first method to find the community structure was presented in [Mucha 10]. The authors have proposed a general framework to detect community in time-dependent, multiscale, and multiplex networks using generalized Louvain method (see Section 3.3.3). Unfortunately, the method description in the article is too general to be implemented independently and the 1.0 version in MATLAB source code[4] was released on January 5, 2012 but it is available only to the limited number of people since it is still under development. Moreover, the dataset on which authors were testing their solution, *Tastes, Ties, and Time*, is not currently available due to privacy concerns.

Other attempts to community detection [Carchiolo 11], [Barigozzi 11], [Berlingerio 11] are mostly theoretical by now. It is worth noticing that only one of the presented papers is more than one year old, so the problem of community extraction in the multi-layered social network is quite new.

---

[1] http://www.youtube.com/
[2] http://www.facebook.com/
[3] https://twitter.com/
[4] http://netwiki.amath.unc.edu/GenLouvain/GenLouvain

After community extraction the most interesting research topic is the dynamics of social groups, it refers to analysis of group evolution over time. In recent years, several methods for tracking changes in social groups have been proposed. Sun et al. have introduced GraphScope [Sun 07], Chakrabarti et al. have presented another original approach in [Chakrabarti 06], Lin et al. have provided the framework called FacetNet [Lin 08] using evolutionary clustering, Kim and Han in [Kim 09] have introduced the concept of nono-communities, Hopcroft et al. have also investigated group evolution, but no method which can be implemented have been provided [Hopcroft 04].However two methods stand out from the rest Asur et al. a[Asur 07] and Palla et al. [Palla 07] but even those two have not been good enough so far.

Creating efficient and flexible method for determining group history is very useful because having this knowledge, one may attempt to predict the future of the group, and then manage it properly in order to discover or even change this predicted future according to specific needs. Such ability would be a powerful tool in the hands of human resource managers, personnel recruitment, marketing, etc.

## 1.3 Thesis Objectives and Contribution

**The main goal of this thesis is to develop a set of tools: measures, algorithms and methods which facilitate the group extraction and analysis of its evolution in multi-layered social networks.** In order to achieve the defined goal the list of objectives were established and the realization of them is the main contribution to the development of the research area called complex networked systems. The objectives are:

1. to develop the new measures and algorithms which allow to analyse the multi-layered social networks and groups within them,

2. to develop a new method and an algorithm which are able to extract groups from multi-layered social networks,

3. to develop a new benchmark which enable to test the new algorithm for group extraction in multi-layered social networks,

4. to test the new algorithm of group extraction against available reference networks, existing benchmarks, real world multi-layered social networks and the new benchmark,

5. to develop a method for group evolution extraction together with a new measure which allow to determine the inclusion of one group in another,

6. to test the new method for group history extraction on real world social networks in comparison to other existing methods for determining group history,

7. to propose and evaluate new method for predicting future group changes using new group evolution extraction method.

## 1.4 Outline of the Thesis

This dissertation on the group extraction, and analysis of its evolution in multi-layered social networks (*MSN*) consists of five major parts.

The first part (Chapter 2) presents the general state-of-the art of social network research with special focus on the multi-layered social network area. Additionally, in this part the definition of multi-layered social network is presented together with the number of measures designed for multi-layered social network analysis. The most important measure is *CLECC* – cross layered edge clustering coefficient, which is utilized to create the new group extraction algorithm in the following part.

The state-of-the art of the group extraction problem in social network and multi-layered social network research are described in the second part (Chapter 3). The group (community) definition is proposed together with a new algorithm for group extraction in the multi-layered social network called a *CLECC* algorithm. The large part of this chapter is devoted to tests of the new method. Experiments starts form special but most common case of the multi-layered social network, i.e. single-layered social network (*SSN*). The *CLECC* algorithm is tasted on reference networks (karate club, football and dolphins network), *GN* Benchmark and *LFR* Benchmark. The second part of experiments involves the real word multi-layered social networks, extracted from the virtual world of Polish game Timik.pl, and multi-layered social networks generated by means of m*LFR* Benchmark – an extension to *LFR* Benchmark. This extension was developed as a part of this thesis, and enables to generate multi-layered social networks. The results of all these experiments are also described in this chapter.

The third part (Chapter 4) is devoted to the development of a new method for group evolution analysis called *GED* (group evolution discovery). At the beginning, a short introduction to problem of group evolution extraction is provided. Next, the concepts of dynamic social network (*DSN*) and a new measure called inclusion measure are described, followed by the presentation of the *GED* method. Finally, the experiments involving real world social network and two leading methods of group evolution extraction are presented together with the test results.

The next, fourth part (Chapter 5) is dedicated to presentation and evaluation of a new method for predicting future group changes. The new approach, involves usage of the results produced by the *GED* method. It is shown that using a simple sequence, which consists of several preceding groups' sizes and events, as an input for the classifier, enables the learnt model to produce very good results, even for simple classifiers.

In the last, fifth part (Chapter 6) the conclusions that were drawn during the performed research and the possible future work are presented.

## 2. Multi-layered Social Network

### 2.1 Introduction to Social Networks

#### 2.1.1 General Concept of Social Network

For the first time the term "social network" was used by Barnes [Barnes 54]. According to his definition a social network is a group of people drawn together by family, work or hobby where the size of the group is about 100-150 people. Nowadays, researchers define a *social network SN (also called a single-layered social network SSN in opposite to a more complex multi-layered social network MSN, see Section 2.*3) in a many different ways:

- Wasserman and Faust in [Wasserman 94] define a social network as a finite set or sets of actors and one or more relations defined on them. An actor is a discrete individual, corporate or collective social unit and a relation is a linkage between a pair of actors.

- Newman in [Newman 03] describes a social network as a graph, $G = (N,A)$, where $N$ is a set of $n$ nodes representing the individuals, and A is the set of arcs representing ties, relationships, bonds, or some other contextually dependent connections between two individuals.

- Hatala in [Hatala 06] claims that it is a set of actors with some patterns of interaction or "ties" between them, represented by graphs or diagrams illustrating the dynamics of the various connections and relationships within the group. Actors are people or groups of people.

- Garton, Haythorntwaite, and Wellman in [Garton 97] propose the following definition of social network – a set of social entities (people, organizations etc.) connected by a set of social relationships (friendship, co-working, information exchange etc.).

- Liben-Nowell and Kleinberg define a social network as a structures whose nodes represent entities embedded in the social context, and whose edges represent interaction, collaboration, or influence between entities [Liben-Nowell 03].

- Hanneman and Riddle in [Hanneman 05] describe social network as a set of points (nodes or agents) that may have relationships with one another

- Yang, Dia, Cheng, and Lin in [Yang 06] claim that social network is an undirected, unweighted graph, where a node represents a customer and an edge denotes the connectedness between two nodes.

As we can see, there is no commonly approved definition of a social network. However, from the definitions presented above, we can draw a conclusion that a social network represents social entities and relations between them. For that reason in this dissertation, the following definition will be further used:

*Definition 2.1*:

*A social network (SN)[5] is defined as a tuple <V,E>, where:*

*V – is a not-empty set of nodes (vertices, actors representing social entities: humans, organizations, departments etc. called also vertices or members);*

*E – is a set edges (relations between actors called also arcs or connections) where single edge is represented by a tuple <x,y>, x,y∈V, x≠y and for two edges <x,y> and <x',y'> if x=x' then y≠y'.*

Since social networks usually represent one kind of relationships they are also called *Single-layered social network SSN* [Magnani 11].

Based on literature and own observations, several examples of social networks can be enumerated: a family [Bott 03], a friendship network of students [Amaral 03], a community of scientists or other professionals in the given discipline, who collaborate with each other [Newman 01] or prepare common scientific papers, a corporate partnership network [Lazega 01], a set of business leaders who cooperate with each other [Liben-Nowell 03], a company director network [Robins 04], a group of acquaintances who share similar interests, etc.

---

[5] In this thesis *a social network* (*SN*) is also called *a single-layered social network* (*SSN*) to distinguish it from *a multi-layered social network* (*MSN*), see Sec.2.3

### 2.1.2 Notation and Representation of Social Network

Three main types of notations can be distinguished: graph, sociometric, and algebraic approach.

The most common representation is a graph. Graph theory has been widely studied by many researches e.g. [Biggs 86], [Chartrand 85], [Harary 69] and the social network analysis (SNA) has commonly adopted this method of representation because of its usefulness for calculation the centrality and prestige within the network, identification of cohesive subgroups, etc. [Scott 00], [Wasserman 94]. Flament in [Flament 63] and Harary in [Harary 65] were among the first scientists who analysed the usage of graphs for social networks. The basic definition of a graph, and in consequence also a social network $SN=(V,E)$, is as follows: it is a finite set of nodes (network members) $V$ and the set of arcs (relationships) $E$ that connects them [Degenne 99], see Figure 2.1. Such graph $SN$ depending on the character of the connections can be either undirected or directed. The former consists of nodes and arcs that fulfil the condition: for each arc $(m_i,m_j) \in E$: $(m_i,m_j)=(m_j,m_i)$ or in other words, an arc is a set $\{m_i,m_j\}$, not a tuple. Hence, in the case of undirected graph, if there is a connection from $m_i$ to $m_j$ then simultaneously exists an arc from $m_j$ to $m_i$ [Wasserman 94]. In the directed graph, we have: $(m_i,m_j) \neq (m_j,m_i)$. It means that the existence of the connection from $m_i$ to $m_j$ does not entail the existence of the opposite relation $(m_j,m_i)$. [Wasserman 94], [Degenne 99]. Graphs can also be either weighted (also called valued) or unweighted. In social network analysis, the relations within an unweighted graph are called binary ones, and they indicate only the fact of the existence of the symmetric relation between two nodes. In the weighted graph, the weights denote the strength or importance of the connections (relations) between two nodes (members).

For better understanding of this dissertation, a few basic terms need to be introduced:

**Walk** – a consistent sequence of following actors and connections which starts and ends with an actor. Closed walk is a walk which starts and ends with the same actor [Rupnik 06].

**Trail** – is a walk between two actors which contains a given connection only once (however one actor can be a part of a trail many times). **Length** of the trail is a number of connections it contains [Rupnik 06].

**Path** – is a walk in which the single actor and single connection can be used only once. The exception is a closed path which starts and ends with the same actor. **Length** of the path is the number of connections it contains. Two paths are independent if their actor sets are disjunctive (they share no actors), only the first and the last actor can be the same [Rupnik 06].

**Neighbourhood of actor A** – is a set of all actors which are directly connected with actor A (the path length between them and actor A is 1).

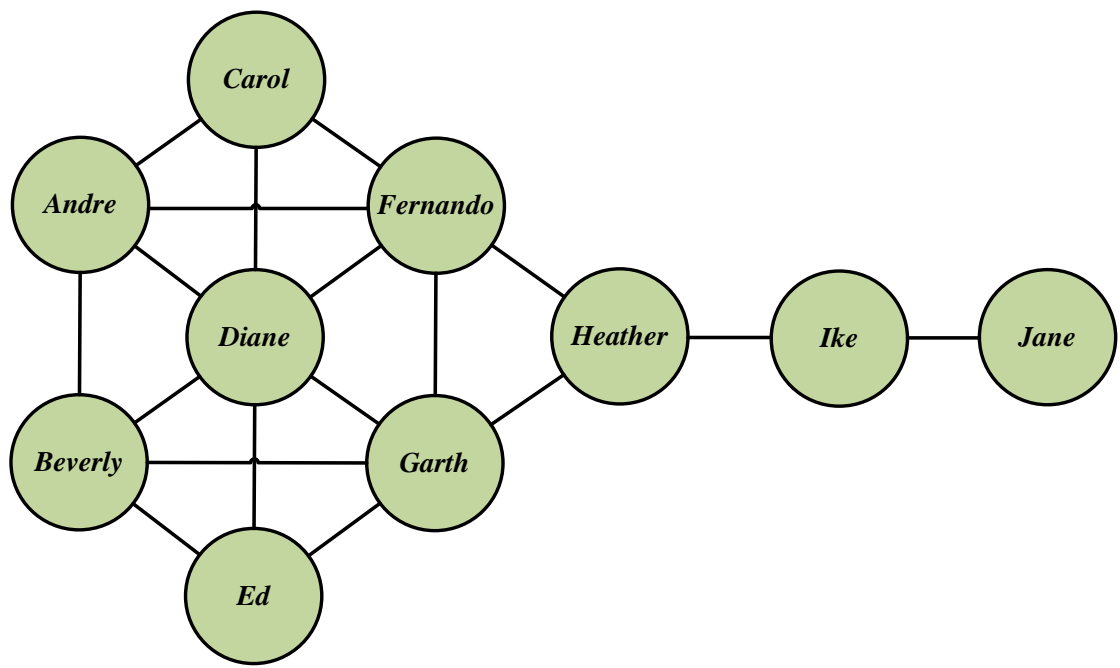

Figure 2.1 Example of a simple social network. Nodes are people and the edges represent data exchange or information flow [Krebs 00].

In the sociometric notation, a social network is represented by a sociomatrix, which is an adjacency matrix for the graph [Wasserman 94], [Degenne 99]. Sociometric notation, introduced by Moreno [Moreno 34], is used to study the structural equivalence and blockmodels [Wasserman 94]. In sociomatrix, each row and column corresponds to a node from graph *SN*. The nodes are taken in the same order for both rows and columns. An element of the matrix denotes the fact of the existence of the connection between two nodes and it contains the strength of the relation in the case of valued networks. For example, the unweighted and directed graph can be represented by the matrix, which elements can have two values: 1 if there is a connection from $m_i$ to $m_j$ and 0 when such a relation does not exist. The matrix can be either symmetrical, if it represents an undirected graph or asymmetrical when it describes the directed graph. Moreover, it will contain only 1 and 0 values, if the social network is unweighted one. The sociometric notation facilitates algebraic computations and transformations on matrixes.

An algebraic approach is most appropriate for role and positional analyses, relational algebras, and is used to study multiple relations [Wasserman 94]. This notation is designed for one-mode networks [Wasserman 94] and was first utilized in [White 63] and [Boyd 69] .

## 2.2  Social Network Analysis

Social network analysis stems from traditional social analysis used by sociologists and anthropologists in the first half of the 20th century. After introducing mathematical interpretation of social networks, scientists have started developing another, more specific domain – social network analysis (SNA).

One of the most popular definition of social network analysis was proposed in [Krebs 00]: "*Social network analysis (SNA) is the mapping and measuring of relationships and flows between people, groups, organizations, computers, web sites, and other information/knowledge processing entities. The nodes in the network are the people and groups while the links show relationships or flows between the nodes. SNA provides both a visual and a mathematical analysis of human relationships*".

The regular social data (Table 2.1) is quite different than social network data (Table 2.2). Traditional social data describes actors whereas social network data can contain social data but mainly describes connections between actors rather than actors themselves [Hanneman 05].

| Name | Gender | Age | Marital status |
|---|---|---|---|
| **Carol** | Female | 32 | Married |
| **Jane** | Female | 26 | Single |
| **Richard** | Male | 30 | Single |
| **Andre** | Male | 45 | Married |

Table 2.1 Example of simple social data.

| Who likes whom? | | | | |
|---|---|---|---|---|
| **Name A\B** | Carol | Jane | Richard | Andre |
| **Carol** | - | 0 | 1 | 0 |
| **Jane** | 1 | - | 0 | 1 |
| **Richard** | 1 | 1 | - | 0 |
| **Andre** | 1 | 0 | 1 | - |

Table 2.2 Example of social network data. 0 – person A does not like person B, 1 – person A likes person B.

Because of the fact that social network analysis focuses on investigation of connections it does not mean that this analysis is not interested in actors. After drawing conclusions social analysis may study actors to retrieve additional information and to better understand this network.

In social network analysis four main steps can be distinguished [Garton 97]: (i) selecting a sample, (ii) collecting data, (iii) choosing and applying the method of social network analysis, (iv) drawing conclusions.

In order to identify and investigate the patterns that occur within the network, first the selection of groups of people should be done. The possibility of analysing every node of the network (especially for huge and heterogeneous networks) is sometimes limited by the available computational resources and because of that only the representative group of actors ought to be chosen for further analysis. This group of actors is called population [Hanneman 05] or sample [Garton 97]. After that, the appropriate data is collected. Many methods of gathering data such as questionnaires, interviews, observation, and artefacts exist [Garton 97]. However, most of researches agree that the best method is the hybrid one that copes with the shortcomings of the enumerated methods and combines them all [Rogers 87]. The researches distinguish the types of data that should be investigated. The data for analyses also called units of analysis are as follow: relations (ties) and actors [Garton 97].

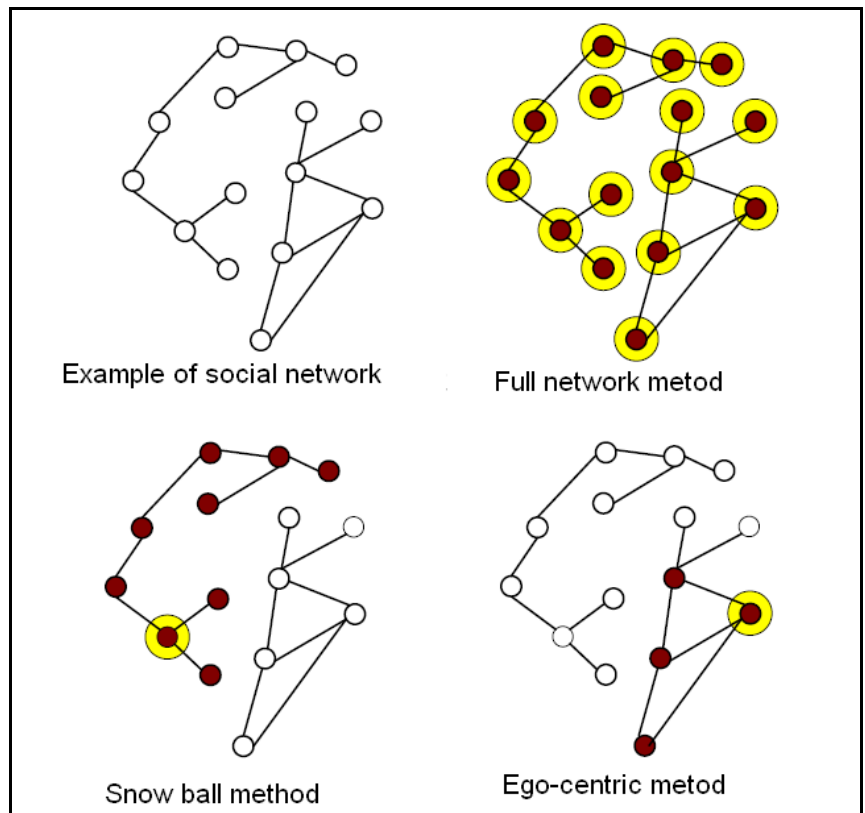

Figure 2.2 Visualisation of social network analysis methods [Hanneman 05].

The next step in social network analysis is to choose the most suitable method of analysis. Three main approaches for analysing the process may be identified in social network analysis (Figure 2.2):

**Full network methods –** those methods collect and investigate data about the entire network (each actor and each connection). This approach gives the best results but is the most expensive, very time-consuming and sometimes it is impossible to collect the full data. However, full network methods are necessary to calculate some measures (e.g. betweenness) [Hanneman 05].

**Snowball methods –** methods start with one local actor or small set of actors. For each actor some or all their connections to other actors have to be identified. Actors picked up in the second step have to do the same like for first actors. The whole process ends when no new connections are revealed or after the predefined number of iteration. This method is very useful in finding a strongly connected group in the big network but it has few weaknesses. Firstly, if a person is isolated or very loosely connected, then it might be never found by this method. Secondly, if the first actor will not be chosen properly, the method can result with nothing. Because of that the snowball method is usually used after pre-study, which locates the good starting point (e.g. president/governor for country or CEO of company) [Hanneman 05].

**Ego-centric method** – this method investigates only one actor (ego) and its neighbourhood (also connections between the neighbours). This method can provide quite good information about the local network and how this network affects this actor. If an ego was chosen randomly, it gives the incomplete view of the whole network. Additionally, the method is efficient both in time and resources consumed [Hanneman 05].

The last step that enables to identify patterns existing within the particular social network is to draw the conclusion from the investigation. The issue that has to be emphasized is that collecting the network data and picking the right method of analysis is a challenging task.

Nevertheless, due to its potential, the social network analysis is becoming one of the main techniques in modern sociology, anthropology, sociolinguistics, geography, economics, social psychology, communication studies, information science, organizational studies, and biology.

### 2.2.1 Measures in Social Network Analysis

Measures (also called metrics[6]) are used in social network analysis to describe the actors' or ties' features, characteristic within the social network as well as to indicate personal importance of individuals in the social network. In further sections, descriptions of the most popular and useful measures are provided.

**Degree Centrality**

A centrality degree is the simplest and the most intuitive measure among all others. It is the number of links that directly connect one node with others. In an undirected graph, it is the number of edges, which are connected with the single node. In a directed graph, in turn, degree is divided into *indegree* for edges which are incoming to the given node and *outdegree* for edges which are outgoing from the given node. For the example from Figure 2.1, Diane has the biggest centrality degree because she has 6 direct ties. A centrality degree is determined using:

---

[6] In fact, formally a metric (equivalent to distance function) needs to fulfill four strict conditions. However, less formally, this term is used instead of the measure even though such measure may not satisfy some of these conditions

$$DC(x) = k_x = d(x) \quad (2.1)$$

where $d(x)$ is the number of nodes, which are directly connected to node $x$. A centrality degree may be normalized using the following formula

$$DC^N(x) = \frac{d(x)}{n-1}, \quad (2.2)$$

where: $n$ – the total number of members in the social network, i.e. n=|V|.

Indegree centrality $IDC(x)$ of node $x$, in turn, takes into account only edges incoming to node $x$, in the following way:

$$IDC(x) = \frac{i(x)}{n-1}, \quad (2.3)$$

where: $i(x)$ – the number of the first level neighbours that are directly connected to x.

Another measure is outdegree centrality $ODC(x)$ that respects only edges outgoing from node $x$:

$$ODC(x) = \frac{o(x)}{n-1}, \quad (2.4)$$

where: $o(x)$ – the number of the first level neighbours $y$ of node $x$, for which exist edges from $x$ to $y$.

Note that in the case of weighted one-layer social network it is possible to use sum of edges weights between $x$ and its neighbours instead of number of the first level neighbours. [Carrington 05], [Degenne 99], [Scott 00].

The centrality degree (*DC*) values for the social network from Figure 2.1 is presented in Table 2.3.

| [Name\Measure | $DC^N$ | $CC^N$ | $BC^N$ |
|---|---|---|---|
| **Diane** | 0.666 | 0.600 | 0.102 |
| **Fernando** | 0.556 | 0.643 | 0.231 |
| **Garth** | 0.556 | 0.643 | 0.231 |
| **Andre** | 0.444 | 0.529 | 0.023 |
| **Beverly** | 0.444 | 0.529 | 0.023 |
| **Carol** | 0.333 | 0.500 | 0.000 |

| [Name\Measure | $DC^N$ | $CC^N$ | $BC^N$ |
|---|---|---|---|
| **Ed** | 0.333 | 0.500 | 0.000 |
| **Heather** | 0.333 | 0.600 | 0.389 |
| **Ike** | 0.222 | 0.429 | 0.222 |
| **Jane** | 0.111 | 0.310 | 0.000 |

Table 2.3 Normalized centrality measures values for the social network from Figure 2.1.

### Closeness Centrality

A centrality closeness describes how close a node is to all other nodes in the network and tells how quick this node can be reach from all other nodes (for example to spread some information over entire network). This measure emphasizes quality (position in the network) rather than quantity (the number of links, like in a centrality degree measure). On the example from Figure 2.1, Fernando and Garth have the best closeness despite having fewer direct ties than Diane. Centrality closeness is determined using:

$$CC(x) = \frac{1}{\sum_{\substack{y \neq x \\ x, y \in V}} c(x, y)} \quad (2.5)$$

where $c(x,y)$ is a function describing the length of the path between nodes $x$ and $y$. Usually it is the length of the shortest path. Closeness is normalized using

$$CC^N(x) = \frac{n-1}{\sum_{\substack{y \neq x \\ x, y \in V}} c(x, y)} \quad (2.6)$$

where $n$ is the number of nodes in the network [Carrington 05], [Degenne 99], [Krebs 00].

Table 2.3 presents the centrality closeness (*CC*) values for the social network from Figure 2.1.

### Betweenness Centrality

A betweenness centrality denotes how often a node is between two other nodes and how many shortest paths go through this node. Actors with high centrality betweenness are very important in the network because many other actors can connect with each other only through them. For example Ike and Jane would be cut off the rest of the network without Heather, see Figure 2.1. Betweenness for node *n* is calculated by

$$BC(x) = \sum_{\substack{i \neq x \neq j \\ i,j \in V}} \frac{b_{ij}(x)}{b_{ij}} \quad (2.7)$$

where $b_{ij}(x)$ is number of shortest paths from $i$ to j that pass through $x$, and $b_{ij}$ is the number of shortest paths from $i$ to $j$. Centrality betweenness is normalized using:

$$BC^N(x) = \frac{\sum_{\substack{i \neq x \neq j \\ i,j \in V}} \frac{b_{ij}(x)}{b_{ij}}}{n-1} \quad (2.8)$$

where *n* is the number of nodes in a network [Carrington 05], [Degenne 99], [Scott 00], [Krebs 00].

Table 2.3 presents the centrality betweenness (*BC*) values for social network from Figure 2.1.

### Degree Prestige

A degree prestige measures how popular is an individual by counting how many direct connections are directed to this individual, so degree prestige has the same meaning as indegree measure [Wasserman 94].

### Rank Prestige

A rank prestige (also called a status prestige) of an actor A is a function of the prestige that ranks others actors from the social network. If many individuals with a high rank value are in contact with one actor, then this actor has higher prestige than actors connected to individuals with lower rank value. "It's not what you know, but whom you know" [Wasserman 94].

### Social Position

Social position $SP(x)$ of individual $x$ in the social network can be used to evaluate importance of $x$ in the community. It respects the values of the node positions of $x$'s direct acquaintances as well as their activities towards $x$ [**Bródka** 09a], [Musiał and **Bródka** 09a]. The social position for the network $SN(V,E)$ is calculated in the iterative way, as follows:

$$SP_{n+1}(x) = (1-\varepsilon) + \varepsilon \cdot \sum_{y \in V} SP_n(y) \cdot C(y \to x) \qquad (2.9)$$

where $SP_{n+1}(x)$ and $SP_n(x)$ is the social position of member $x$ after the $n+1^{st}$ and $n^{th}$ iteration, respectively, and $SP_0(x)=1$ for each $x \in V$; $\varepsilon$ is the fixed coefficient from the range $(0;1)$; $C(y \to x)$ is the commitment function, which expresses the strength of the relation from $y$ to $x$ – the weight of edge $<y,x>$.

## 2.3 Multi-layered Social Network

It is obvious that in the real world, more than one kind of relationship can exist between two actors (e.g. family, friendship and work ties) and that those ties can be so intertwined that it is impossible to analyse them separately [Fienberg 85], [Minor 83], [Szell 10]. A network where more than one type of relation exists are not new in the world of science [Wasserman 94] but they were analysed mainly at the small scale e.g. in [McPherson 01], [Padgett 93], and [Entwisle 07]. Just like in the case of regular single-layered social network presented earlier (see section 2.1) there is no widely accepted definition or even common name. At the beginning such networks have been called multiplex network [Haythornthwaite 99], [Monge 03]. The term is derived from communications theory which defines multiplex as combining multiple signals into one in such way that it is possible to separate them if needed [Hamill 06]. Recently the area of multi-layered social network has started attracting more and more attention in researchers from different fields [Kazienko 11], [Szell 10], [Rodriguez 07], [Rodriguez 09], and the meaning of multiplex network has expanded and covers not only social relationships but any kind of connection, e.g. based on geography, occupation, kinship, hobbies, etc. [Abraham 12].

As mentioned before, nowadays social networks with more than one kind of relationship have many different names. The most common name is *Multi-Layered* (or just *Layered*) *social networks* [**Bródka** 11b], [Geffre 09], [Hamill 06], [Kennedy 09], [Magnani 11], [Schneider 11] but also *Multi-relational social networks* [Szell 10], *Multi-dimensional social networks* [Kazienko 11], *Multidimensional Dynamic social network* [Kazienko and **Bródka** 11a], [Kazienko and **Bródka** 11a] or *Multivariate social networks* [Szell 10] are in use.

Additionally, the researchers in the field of multi-layered networks also try to develop new models of networks that capture not only the multi-layered characteristics of social data. Authors in [Cantador 06] proposed a multi-layered semantic social network model that enables to investigate human interests in more details than when they are analysed all together. Jung at al. [Jung 07] describe Multiplex Social Networks as multi-labeled semantic social networks with semantic relations between actors personal ontologies. Wong at al. in [Wong-Jiru 07] propose multi-layered model, where only one layer describes relation between people, the other layers are processes, applications, systems and a physical network.

Despite the fact that many researchers investigate multi-layered social networks there is surprisingly few definitions or descriptions going beyond the statement "It is the network where two nodes are connected by more than one connection, relation, tie". Newman [Newman 10] defines such a network as a multigraph with multiedges between nodes. A network is then represented by adjacency matrix where mutiedge is denoted by setting corresponding matrix with element $A_{ij}$ equal to the number of edges between the node $i$ and the node $j$. The second possible representation is an adjacency list where multiedge is represented by multiple identical entries in the list of neighbours. The main problem with Newman's definition is that he has never mentioned any labels on those edges so in this model the information about which edge represents what network is lost.

A different approach is presented by Magnani and Rossi in [Magnani 11]. They describes two concepts: (1) *Pillar Multi-Network* and (2) *Multi Layer Network* (*ML-Model*). The first concept defines social network with multiple relation as a set of single-layered networks $\{<V_1,E_1>, <V_2,E_2>, …, <V_k,E_k>\}$ and some mapping relation by means which a user in one network is mapped to another user in the second network. It means that one user from one single-layered network $SN_1$ can correspond to only one user in another single-layered social network $SN_2$. This case is typical for most web-based services, e.g. a Facebook account may correspond to the Twitter account. The second concept is almost the same but the mapping function is slightly different, i.e. many users from one single-layered social network $SN_1$ can correspond to a single user in another single-layered social network $SN_2$. For example, if one single-layered social network $SN_1$ represents relations between co-workers from one company and another network $SN_2$ represents relations between departments from the same company, then many employees working in a given department $D$ in $SN_1$ are mapped to this department $D$ in the second network $SN_2$.

Kazienko at al. in [Kazienko and **Bródka** 11a] and [Kazienko and **Bródka** 11a] present yet another model of multidimensional temporal social network, which considers three distinct dimensions of social networks: layer, time and group dimension. All the dimensions share the same set of nodes that corresponds to social entities: single humans or groups of people. A layer dimension describes all kinds of relationships between users of the system; a time dimension presents the dynamics of the social network and a group dimension focuses on interactions within separated social communities (groups). At the intersection of all this dimensions is a small social network, which contains only one kind of interactions (one layer) for a particular group in a given time. This concept allows to analyse systems, where people are linked by many different relationship types (layers in the layer dimension) like in complex social networking sites (e.g. Facebook). It means, people may be connected as friends, via common groups, "like it", etc. It can also refer complex relationships within regular companies: department colleagues, best friends, colleagues from the company trip, etc.. Multidimensionality provides an opportunity to analyse each layer separately and at the same time investigate different aggregations over instances of the layer dimension. For example, let's consider a network consisting of six layers, three from the real word: family ties, work colleagues and gym friends and three from the virtual world, i.e. friends from Facebook and fiends from the MMORPG game and friends from some forum. Now, one has many different possibilities for studies on such a network, for example: (i) to analyse each layer separately, (ii) to aggregate layers from the real world and compare them to the virtual world layers aggregation, and finally, (iii) to aggregate all layers together. The time dimension provides possibility to investigate the network evolution and its dynamics. For example, the analysis (i) how users neighbourhoods change when one of the neighbours leaves the network and how it affects the network in longer period, (ii) how roles of group leaders (e.g. project team leaders) change over time (are overtaken by different people), or (iii) how changes on one layer affect the other layers. Finally, the group dimension allows studying groups existing within the social network. Using multidimensionality, not only the usual social groups can be analysed (friend family, school, work, etc.) but also groups created upon various member features like gender, age, location etc. Moreover, the model allows to compare the results of different community extraction methods, e.g. by means of social community extraction or typical data mining clustering. To conclude, the multidimensional social network enables to analyse all three dimensions at the same time, e.g. how interaction on different five layers of two social groups changes over three selected periods.

Rodriguez in [Rodriguez 07] define a multi-relational social network as a tuple $G = (N, E, W)$, where $N$ is the set of nodes in the network, $E$ is a set of directed edges, $W$ is the set of weights associated with each edge of the network $|W|=|E|$. However, two years later in [Rodriguez 09] he define the same network as $M = (V, E)$, where $V$ is the set of vertices in the network, $E = \{E_1, E_2, ..., E_m\}$ is a family of edge sets in the network, and any $E_k \subseteq (V \times V): 1 \leq k \leq m$. Each edge set in $E$ has a different semantic interpretation.

Therefore, like in the case of a single-layered social network, it is crucial to define the concept of the multi-layered social network, which will be used in this thesis. The idea is very similar to *Pillar Multi-Network* presented in [Magnani 11], but was introduced one year earlier in [Kazienko and **Bródka** 10a]. The main difference is that instead of mapping function, in this model the set of nodes is unified, i.e. it is common for all layers.

*Definition 2.2*:

*A multi-layered social network MSN is defined as a tuple <V, E, L> where:*

*V – is a not-empty set of nodes (social entities);*
*E – is a set of tuples <x,y,l>, x,y$\epsilon$V, l$\in$L, x≠y and for any two tuples <x,y,l>, <x',y',l'> $\in$E if x=x' and y=y' then l≠l';*
*L – is a fixed set of distinct layers.*

Each layer corresponds to one type of relationships between users. Different relationships can result from the character of connections, types of communication channel, or types of collaborative activities that users can perform within a given system. The examples of different relationships can be: friendship, family or work. Different communication channels that result in different types of connections are: email exchange, VoIP calls, instant messenger chats, etc. The separate relationship types can be also defined based on users' common activities within photo publishing services, such as publishing photos, commenting photos, adding photos to favourites where photo is so called meeting object, etc. The last to enumerate types of relations possess a semantic meaning as for example publishing photos is a much more proactive action than just adding photos to favourites. Another example where information about users' activities has a clear semantic meaning can be an internet forum where people, who are very active and post a lot of queries can be perceived as new to a field. On the other hand, people who comment a lot but do not post any queries can be seen as experts in a field.

Nodes *V* and edges $E_l \subseteq E$ from only one layer $l \in L$ correspond to a simple, single-layered social network *SSN*= <*V, $E_l$, {l}*>. As mentioned above a multi-layered social network *MSN*=<*V,E,L*> may be represented by a multigraph, where multiple relations are represented by multiedge [Newman 10]. Hence, all the below proposed structural measures can also be applied to other kinds of complex networks that are described by means of multi-graphs.

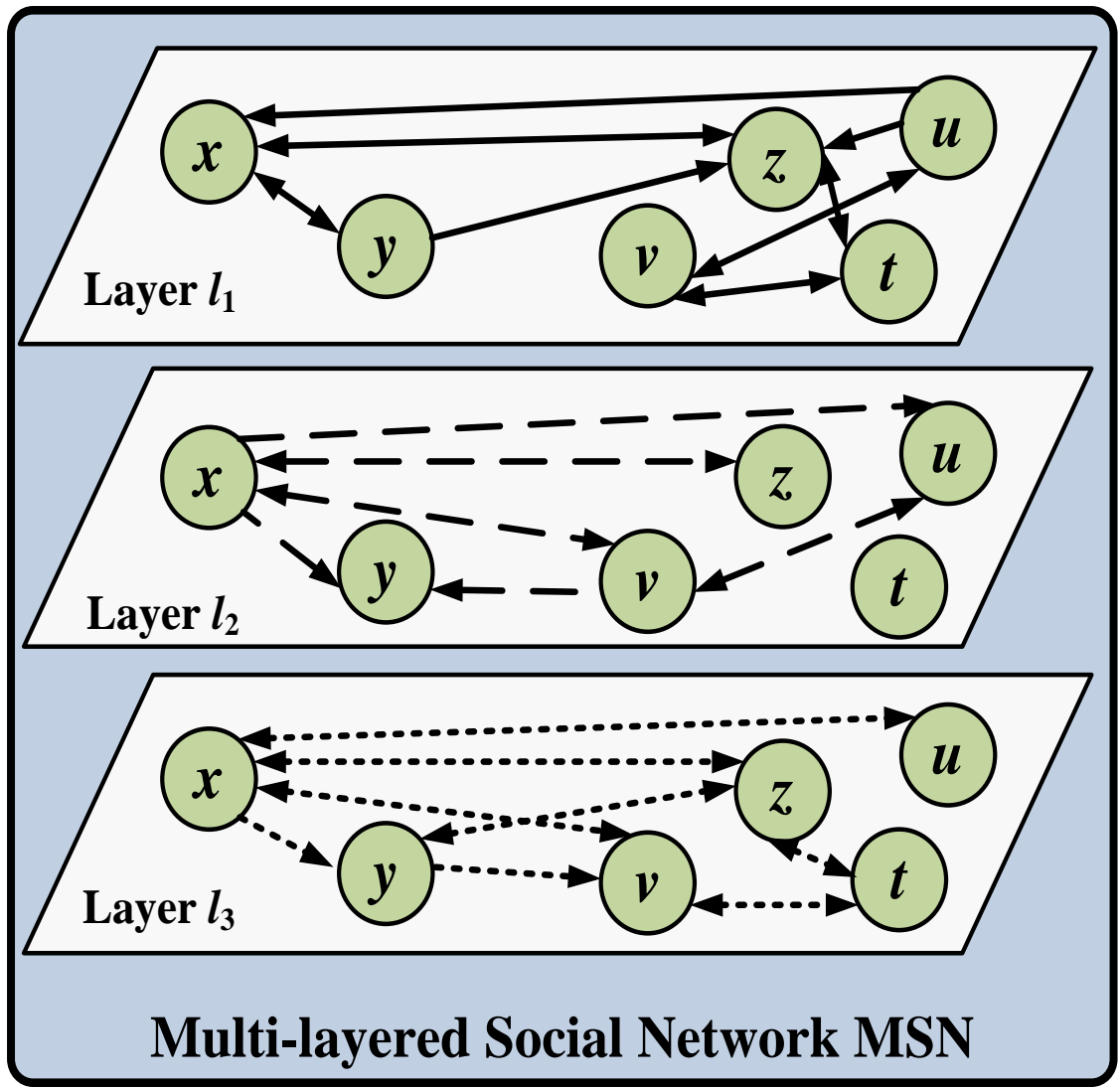

Figure 2.3 An example of the multi-layered social network *MSN*.

In Figure 2.3, the example of three-layered social network is presented. The set of nodes consists of {*t, u, v, x, y, z*} so there are five users in the network that can be connected with each other's on three layers: $l_1$, $l_2$ and $l_3$. On the layer $l_1$, eight relationships (tuples) between users: <*x,y,$l_1$*>, <*y,x,$l_1$*>, <*x,z,$l_1$*>, <*z,x,$l_1$*>, <*y,z,$l_1$*>, <*u,z,$l_1$*>, <*u,v,$l_1$*>, <*v,u,$l_1$*> can be distinguished, 6 edges on layer $l_2$ and 7 on layer $l_3$.

Social networks emerging from different types of social media or social networking sites are good examples of multi-relational networks. One reason for such a big interest in this area is the fact that these systems offer large datasets including information about peoples' profiles and activities that can be analysed. Due to the fact that this data reflect users' behaviours in the virtual world, the networks extracted from this data are called online social networks [Garton 97], web-based social networks [Golbeck 06], or computer-supported social networks [Wellman 96].

Bibliographic data [Girvan 02], blogs [Agarwal 10], photos sharing systems like Flickr [Kazienko 11], e-mail systems [**Bródka** 09b], telecommunication data [Blondel 08], social services like Twitter [Huberman 09] or Facebook [Ellison 07], video sharing systems like YouTube [Cheng 08], Wikipedia [Capocci 06] and many more are the examples of data sources which are used by many researches to analyse the underlying social networks. However, this vast amount of data and especially its multi-relational character are the source of new research challenges related to processing problems of this data [Domingos 03]. Although most of the existing methods work properly for single-layered networks, there is a lack of well-established tools for multi-layered network analysis. Development of new measures is very important from the perspective of further advances in the web science as the multi-relational networks can be found almost everywhere. They are more expressive in terms of the semantic information and give opportunity to analyse different types of human relationships [Rodriguez 09].

## 2.4 Measures in Multi-layered Social Network

Despite this fact that multi-layered nature of networks is an absolutely natural concept, researchers have refrained from analysing more than a single layer at once for many years For example, Wasserman and Fraust [Wasserman 94] recommended that common centrality and prestige measures should be calculated for each relation (layer) separately and suggested not to perform any aggregation of the relations. Unfortunately, they do not explain why they advised such approach. The possible arguments for that are: (i) potential loss of information, which may occur during aggregation process and (ii) not sufficient computational power in 1994.

However, as mentioned at the beginning, some relation are so intertwined that it is impossible to analyse them separately and **if considered together, they reveal additional** **vital information about the network**. Especially nowadays, when it is so important to analyse information diffusion in global network. For example, let's consider the following case:

> User *A* publishes information (in this case the movie) on the YouTube[7]. Next user *B*, who subscribes user's *A* channel on the YouTube, forwards this information and puts out the movie on the Facebook[8] board. After that user *C* who is user's *B* Facebook friend and does not know user *A*, watches the movie and sends it via Twitter[9] to user *D* who knows neither *A* nor *B*.

If the data from each above systems (YouTube, Facebook, Twitter) will be analyse as the separate social networks, then it will be hard to find how the information (the movie) has circulated from user *A* to user *D*. However, if the data from all three system could be merged into one multi-layered social network, then the path from user *A* to user *D* can be quite easily traced.

The second scenario is from the real world: viral infection:

> *A* infects its spouse *B*, then *B* goes to work and infects *B*'s colleague *C,* and finally *C* infects its spouse *D* at home.

And once again, if only family ties or work relations are analysed, it is hard to understand why the disease has spread from *A* to *D*. But when both networks are analysed simultaneously, it is easy to spot the connection between *A* and *D*.

The data about actors activities and interactions, collected in different systems, enable to extract complex social networks, in which different types of relations exist. All these relations can be analysed separately or parallel as the knowledge is not only hidden in individual layers but also in cross layered relationships. The information about what happens on one layer can influence the actions on other ones. So it is completely natural that recently (especially for the last 6 years) scientists work very hard to develop a variety of measures and methodologies, which will allow to analyse multi-layered social networks.

Therefore, researchers try to cope with multi-layered networks by analysing layers separately by means of existing methods for single-layered networks and then comparing the results using some correlation measure, e.g. Jaccard coefficient or cosine measure. In [Szell 10], authors distinguished 6 different relation types between users of the massive multiplayer online games. First, they analysed the characteristics of each layer separately and after that they studied correlations and overlap between the extracted types of relations. One of the interesting findings is that users tend to play different roles in different networks. From the structural perspective, authors found that different types of interactions are characterised by different patterns of connectivity, e.g. according to their study power-law degree distributions indicate aggressive actions. Another example is the analysis on Flickr [Kazienko 11], where authors distinguished eleven types of relationships between users. First, the authors investigated layers separately and then used the correlation measures to compare these networks. Their main finding was that relations may be either semantically or socially driven. Wong-Jiru at al. in [Wong-Jiru 07] computed the number of metrics (geodesic distance, no. of geodesic paths, maximum flow, point connectivity, in/out-degree centrality, closeness centrality, flow betweenness centrality, reachability, density, node betweenness centrality, and edge betweenness) for each layer separately, next asking the weight (called: composite network score) to each layer and based on that they aggregated the network. It is also possible to create a single-layered network from multi-layered network and then to apply the existing methods to such a structure. In [Rodriguez 09], authors presented the path algebra which purpose is to transform a multi-relational network to many single-relational networks that are "semantically-rich".

---

[7] http://www.youtube.com/
[8] http://www.facebook.com/
[9] https://twitter.com/

Another way to deal with multi-layered networks is to develop new methods for their analysis. Such work is usually, to some extent, based on the existing methods for single-layered networks. The investigated topics in that research field are among others: community mining [Cai 05], [Mucha 10], ranking network's nodes [Zhuge 03], paths [Aleman-Meza 05], shortest paths [**Bródka** 11c] and unique paths finding [Lin 04]. Magnani and Rossi in [Magnani 11] introduced the in-degree centrality and distance measures for their ML-Model. Geffre et al. in [Geffre 09] defined three measures: (i) terrorist social connectedness across multi-layered affiliations, (ii) their involvement in operation and (iii) their emergence during periods and at locations of interest, all in order to determine, which terrorist is a critical one to cancel, trigger or impact the operations. Kennedy in [Kennedy 09] was trying to resolve a similar problem. He developed methods, which should be able to identify potential targets within a multi-layered social network in order to cut the resource from them and maximize either the protection or disruption of the organization.

In this thesis, the set of the most popular measures, which are utilized to analyse single-layered were adapted to the multi-layered social network needs. The measures described below were presented, analysed and tested in [**Bródka** 10a], [**Bródka** 11a], [**Bródka** 11b], [**Bródka** 11c], [**Bródka** 12a].

### 2.4.1 Multi-layered Neighbourhood

Neighbourhood $N(x,l)$ of a given node $x$ on a given layer $l$ for multi-layered social network $MSN=<V,E,L>$ is defined as:

$$N(x,l) = \{y :< y,x,l > \in E \vee < x,y,l > \in E\} \tag{2.10}$$

Set $N(x,l)$ is equivalent to the simple neighbourhood for regular single-layered SN

| Node\Layer | $l_1$ | $l_2$ | $l_3$ |
|---|---|---|---|
| **x** | {u,y,z} | {u,v,y,z} | {u,v,y,z} |
| **y** | {x,z} | {v,x} | {v,x,z} |
| **z** | {t,u,x,y} | {x} | {t,x,y} |
| **u** | {v,x,z} | {v,x} | {x} |
| **t** | {v,z} | { } | {v,z} |
| **v** | {t,u} | {u,x,y} | {t,x,y} |

Table 2.4 Node neighbourhoods for each layer for *MSN* from Figure 2.3.

A multi-layered neighbourhood of a given node $x$ with a minimum number of layers required – $\alpha$, $1 \leq \alpha \leq |L|$, is a set of nodes, which are neighbours of node $x$ on at least $\alpha$ layers in the *MSN*. Five different versions of multi-layered neighbourhood may be distinguished.

The first one is the multi-layered neighbourhood $MN^{In}(x,\alpha)$ derived from the edges incoming to node $x$, in the following way:

$$MN^{In}(x,\alpha) = \{y : |\{< y,x,l > \in E\}| \geq \alpha\} \tag{2.11}$$

The value of $MN^{In}(x,\alpha)$ denotes the set of neighbours that are connected to node $x$ with at least $\alpha$ edges, i.e. on at least $\alpha$ layers of *MSN*. For $\alpha=1$, we need an edge on only one layer, while for $\alpha=|L|$, if node $y \in MN^{In}(x,\alpha)$, then user $y$ must have edges to a given node $x$ on all existing layers. For the example *MSN* from Figure 2.3, $MN^{In}(x,1)=\{u,v,y,z\}$, $MN^{In}(x,2)=\{u,v,z\}$, $MN^{In}(x,3)=\{z\}$.

Another multi-layered neighbourhood $MN^{Out}(x,\alpha)$ respects only edges outgoing from node $x$:

$$MN^{Out}(x,\alpha) = \{y : |\{<x,y,l> \in E\}| \geq \alpha\} \quad (2.12)$$

For the *MSN* from Figure 2.3, we have $MN^{Out}(x,1)=\{u,v,y,z\}$, $MN^{Out}(x,2)=\{u,v,y,z\}$, $MN^{Out}(x,3)=\{y,z\}$.

If we consider incoming or outgoing edges on any layers, then we obtain $MN^{InOutAny}(x,\alpha)$:

$$MN^{InOutAny}(x,\alpha) = \{y : |\{<x,y,l> \in E\}| \geq \alpha \wedge |\{<y,x,l> \in E\}| \geq \alpha\} \quad (2.13)$$

Neighbourhood $MN^{InOutAny}(x,\alpha)$ includes nodes that have at least $\alpha$ incoming and $\alpha$ outgoing edges to and from node $x$, respectively, but these edges may occur on different layers. For the network from Figure 2.3, there are following sets: $MN^{InOutAny}(x,1)=\{u,v,y,z\}$, $MN^{InOutAny}(x,2)=\{u,v,z\}$ $MN^{InOutAny}(x,3)=\{z\}$.

The next type of multi-layered neighbourhood $MN^{InOut}(x,\alpha)$ is quite similar to $MN^{InOutAny}(x,\alpha)$ but it is more restrictive. Each neighbour $y \in MN^{InOut}(x,\alpha)$ must have bidirectional connections on at least $\alpha$ layers in *MSN*, i.e. both the incoming and outgoing edge have to occur on the same layer to satisfy the condition, as follows:

$$MN^{InOut}(x,\alpha) = \{y : |\{l : <x,y,l> \in E \wedge <y,x,l> \in E\}| \geq \alpha\} \quad (2.14)$$

In the example *MSN*, Figure 2.3, we have $MN^{InOut}(x,1)=\{u,v,y,z\}$, $MN^{InOut}(x,2)=\{v,z\}$ $MN^{InOut}(x,3)=\{z\}$.

The final, fifth neighbourhood $MN(x,\alpha)$ is the least restrictive. It takes into consideration, any incoming or outgoing edges on any layer but the total number of these layers should be at least $\alpha$:

$$MN(x,\alpha) = \{y : |\{l : <x,y,l> \in E \vee <y,x,l> \in E\}| \geq \alpha\} \quad (2.15)$$

In the example social network, the neighbourhoods are as follows: $MN(x,1)=\{u,v,y,z\}$, $MN(x,2)=\{u,v,y,z\}$ $MN(x,3)=\{u,y,z\}$, see Figure 2.3, the rest of multi-layered neighbourhoods $MN(x,\alpha)$ can be found in Table 2.5.

| Node\Metric | MN(x, 1) | MN(x, 2) | MN(x, 3) |
|---|---|---|---|
| x | {u,v,y,z} | {u,v,y,z} | {u,y,z} |
| u | {x,v,z} | {x,v} | {x} |
| z | {t,u,x,y} | {t,x,y} | {x} |
| u | {z,v,x} | {x,v} | {x} |
| t | {v,z} | {v,z} | { } |
| v | {t,u,x,y} | {t,u,x,y} | { } |

Table 2.5 Multi-layered neighbourhoods for all nodes from Figure 2.3.

$MN(x,\alpha)$ is utilized for studies described in further sections of this thesis. However, the other neighbourhoods may also be used and it would require only small modifications of the measures proposed below.

Note that $MN^{InOutAny}(x,\alpha) = MN^{In}(x,\alpha) \cap MN^{Out}(x,\alpha)$ and $MN^{InOut}(x,\alpha) \subseteq MN^{InOutAny}(x,\alpha) \subseteq MN(x,\alpha)$, $MN^{InOut}(x,\alpha) \subseteq MN^{In}(x,\alpha) \subseteq MN(x,\alpha)$ and $MN^{InOut}(x,\alpha) \subseteq MN^{Out}(x,\alpha) \subseteq MN(x,\alpha)$. The smallest neighbourhood is $MN^{InOut}(x,\alpha)$ while the largest is $MN(x,\alpha)$. It means that $MN^{InOut}(x,\alpha)$ is the most restrictive whereas $MN(x,\alpha)$ is the least.

Although multi-layered neighbourhood is a structural measure, it also has a semantic meaning. For example, a large $MN(x,\alpha)$ for a big value of $\alpha$ will be an indicator that person $x$ is a communication hub. Neighbourhood $MN^{In}(x,\alpha)$ is more restrictive so it provides some more detailed semantic meaning. Its large value when $\alpha$ is high, means that there is a lot of incoming interaction towards $x$. For example, in the context of the company it will mean that $x$ is a line manager or another person, for who a group of people is reporting using different communication channels. These channels can be treated as separate layers in the network. On the other hand, a high value of $MN^{Out}(x,\alpha)$ for greater $\alpha$, means that person $x$ is probably responsible for propagating information in the network, e.g. a person responsible for bulletin distribution in the organisation. It also helps to investigate which communication channels are neglected. Such examples can be multiplied. Please note that it can serve to investigate both positive and negative human behaviours. For example, a person within a company with high $MN^{Out}(x,\alpha)$ who is not responsible for propagating information can be seen as a "chatterbox" who spend more time on communication than on doing the job.

Also at the level of the whole network, the multi-layered neighbourhood can be analysed, e.g. power-law distribution of $MN(x,\alpha)$ depending on $\alpha$, means that not all types of relations are fully used and people tend to focus on only few relation types and neglect others. Complete evaluation of this measure can be found in [**Bródka** 12a].

### 2.4.2 Cross Layered Clustering Coefficient

A cross layered clustering coefficient $CLCC(x,\alpha)$ was introduced and investigated in [**Bródka** 10a] and [**Bródka** 12a]. The idea was to allow calculation of clustering coefficient [Watts 98], [Fronczak 09] for the multi-layered social network *MSN*. For a given node *x*, and *x*'s non-empty neighbourhood $MN(x,\alpha)$, cross layered clustering coefficient $CLCC(x,\alpha)$ is computed in the following way:

$$CLCC(x,\alpha) = \frac{\sum_{l \in L} \sum_{y \in MN(x,\alpha)} \left( in(y, MN(x,\alpha), l) + out(y, MN(x,\alpha), l) \right)}{2 \cdot |MN(x,\alpha)| \cdot |L|} \quad (2.16)$$

where: $in(y, MN(x,\alpha), l)$ – the weighted indegree of node *y* in the multi-layered neighbourhood $MN(x,\alpha)$ of node *x* within the simple single-layered network $<V, E, \{l\}>$, i.e. within only one layer *l*; $out(y, MN(x,\alpha), l)$ – the weighted outdegree of node *y* in the multi-layered neighbourhood $MN(x,\alpha)$ of node *x* in the network containing only one layer *l*. If neighbourhood $MN(x,\alpha)=\{\}$, then $CLCC(x,\alpha)=0$.

The weighted indegree $in(y, MN(x,\alpha), l)$ for a given node *y* in the network $<V, E, \{l\}>$ containing one layer *l* is the sum of all weights $w(z,y,l)$ of edges $<z,y,l>$ incoming to node *y* from other nodes *z* that are from layer *l* and belong to multi-layered neighbourhood $MN(x,\alpha)$:

$$in(y, MN(x,\alpha), l) = \sum_{z \in MN(x,\alpha)} w(z, y, l) \quad (2.17)$$

Likewise, the weighted outdegree $out(y, MN(x,\alpha), l)$ for a given node *y* and neighbourhood $MN(x,\alpha)$ is the sum of all weights $w(y,z,l)$ of the outgoing edges $<y,z,l>$ that come from *y* to *x*'s neighbours *z* on layer *l*:

$$out(y, MN(x,\alpha), l) = \sum_{z \in MN(x,\alpha)} w(y, z, l) \quad (2.18)$$

Note that if the sum of weights of outgoing edges for a given node is 1 (this is the usual practice in social network analysis, see [**Bródka** 09a], [Kazienko 11]), then $CLCC(x,\alpha)$ is always from the range [0;1]. The value $CLCC(x,\alpha)=1$, if each neighbour $y \in MN(x,\alpha)$ has outgoing relationships towards all other nodes $z \in MN(x,\alpha)$ and only to them. The value of $CLCC(x,\alpha)$ equals 0 occurs when $x$ has only one neighbour, i.e. $|MN(x,\alpha)|=1$, or when none of the $x$'s neighbours $y \in MN(x,\alpha)$ has any relationship with any neighbour $z \in MN(x,\alpha)$.

For node $t$ from the example social network, Figure 2.3, $MN(t,1)=MN(t,2)=\{v,z\}$ but $CLCC(t,1)=CLCC(t,2)=0$ because there are no edges between $v$ and $z$. Due to $MN(t,3)=\{\}$ we have $CLCC(t,3)=0$. For node $z$: $MN(z,1)=\{t,u,x,y\}$, $MN(z,2)=\{t,x,y\}$, $MN(z,3)=\{x\}$. If weights of all edges equal 1, then $CLCC(z,1)=^{16}/_{24}=^{2}/_{3}$ and $CLCC(z,2)=^{8}/_{18}=^{4}/_{9}$. Since there is only one neighbour $x$ in $MN(z,3)$, the value $CLCC(z,3)=0$.

The formula for another measure – multi-layered clustering coefficient (*MCC*) as well as the two special cases of cross layered clustering coefficient $CLCC(x,\alpha)$ were described in [**Bródka** 10a]. One of them is multi-layered clustering coefficient in extended neighbourhood (*MCCEN*). It is, in fact, the cross layered clustering coefficient for only one layer($\alpha=1$), i.e. $MCCEN(x)=CLCC(x,1)$. Multi-layered clustering coefficient in reduced neighbourhood (*MCCRN*) presented in [**Bródka** 10a] is equivalent to the cross layered clustering coefficient for all layers, $\alpha=|L|$ i.e. $MCCRN(x)=CLCC(x,|L|)$.

Cross Layered Clustering Coefficient can also be interpreted in the semantic context. For example people who are in the professional relationships (e.g. co-workers) prefer to communicate via e-mail as it enables to keep track of their interactions. It means that their clustering coefficient will be higher at one layer (assuming that their neighbours communicate with each other) than at the rest of the layers. Different situation occurs in social situations where people tend to use different communication channels and it can result in big value of *CLCC* although the clustering coefficient for a single layer may not be particularly high. It is a consequence of the fact that our neighbours can interact with each other using different communication layers (phone call, e-mail, text messages, etc.).

### 2.4.3 Multi-layered Degree Centrality

Apart from clustering coefficient, there exists another measure commonly used in social network analysis – degree centrality see Section 2.2.1. This measure for multi-layered social network, were introduced and examined in [**Bródka** 11b] and [**Bródka** 12a].

*Cross layered Degree Centrality*

The first multi-layered degree centrality is called cross layered degree centrality (*CDC*). It is defined as a sum of edge weights both incoming to and outgoing from node $x$ towards multi-layered neighbourhood $MN(x,\alpha)$ divided by the number of layers and total network members:

$$CDC(x,\alpha) = \frac{\sum_{y \in MN(x,\alpha)} w(x,y,l) + \sum_{y \in MN(x,\alpha)} w(y,x,l)}{(m-1)|L|}, \quad (2.19)$$

where: $w(x,y,l)$ – the weight of edge $\langle x,y,l \rangle$.

Similarly to different versions of degree centrality $DC(x)$ - $IDC(x)$ and $ODC(x)$ (se Equation 2.1, 2.3, 2.4), we can define cross layered indegree centrality $CDC^{In}(x,\alpha)$ in the multi-layered social network *MSN*:

$$CDC^{In}(x,\alpha) = \frac{\sum_{y \in MN(x,\alpha)} w(y,x,l)}{(m-1)|L|}, \quad (2.20)$$

and cross layered outdegree centrality $CDC^{Out}(x,\alpha)$:

$$CDC^{Out}(x,\alpha) = \frac{\sum_{y \in MN(x,\alpha)} w(x,y,l)}{(m-1)|L|}. \quad (2.21)$$

As in the case of the cross layered clustering coefficient $CLCC(x,\alpha)$, the value of $CDC(x,\alpha)$ directly depends on the parameter $\alpha$, which determines the multi-layered neighbourhood of a given social network member $x$.

### *Multi-layered Degree Centrality Version 1*

The other three multi-layered degree centralities are not calculated based on $MN(x,\alpha)$ but using the local neighbourhood in particular layer $N(x,l)$. The first of them $MDC^{(1)}(x)$ is defined as a sum of *x's* local weighted degree centralities in each layer *l* divided by the number of layers:

$$MDC^{(1)}(x) = \frac{\sum_{l \in L} \left( \sum_{y \in N(x,l)} w(x,y,l) + \sum_{y \in N(x,l)} w(y,x,l) \right)}{(m-1)|L|}. \quad (2.22)$$

The first multi-layered indegree centrality $MDC^{(1)In}(x)$ is defined as follows:

$$MDC^{(1)In}(x) = \frac{\sum_{l \in L} \sum_{y \in N(x,l)} w(y,x,l)}{(m-1)|L|}, \quad (2.23)$$

and the first multi-layered outdegree centrality $MDC^{(1)Out}(x)$:

$$MDC^{(1)Out}(x) = \frac{\sum_{l \in L} \sum_{y \in N(x,l)} w(x,y,l)}{(m-1)|L|}. \quad (2.24)$$

### *Multi-layered Degree Centrality Version 2*

The next multi-layered degree centrality $MDC^{(2)}(x)$ is a sum of *x's* local weighted degree centralities in each layer but in opposite to $MDC^{(1)}(x)$, it is divided by the quantity of the union of *x*'s neighbourhoods from all layers.

$$MDC^{(2)}(x) = \frac{\sum_{l \in L} \left( \sum_{y \in N(x,l)} w(x,y,l) + \sum_{y \in N(x,l)} w(y,x,l) \right)}{(m-1)|MN(x,1)|}. \quad (2.25)$$

Note that the union of neighbourhood sets from all layers for a given member *x* is the same as the multi-layered neighbourhood $MN(x,\alpha)$ for $\alpha=1$, i.e. $\bigcup_{l \in L} N(x,l) = MN(x,1)$.

The multi-layered indegree centrality $MDC^{(2)In}(x)$ is:

$$MDC^{(2)In}(x) = \frac{\sum_{l \in L} \sum_{y \in N(x,l)} w(y,x,l)}{(m-1)|MN(x,1)|}, \quad (2.26)$$

and the first multi-layered outdegree centrality $MDC^{(2)Out}(x)$:

$$MDC^{(2)Out}(x) = \frac{\sum_{l \in L} \sum_{y \in N(x,l)} w(x,y,l)}{(m-1)|MN(x,1)|}. \quad (2.27)$$

### *Multi-layered Degree Centrality Version 3*

The last multi-layered degree centrality $MDC^{(3)}(x)$ is quite similar to $MDC^{(2)}(x)$ but instead of the neighbourhood sets from all layers, the sum of *x's* local weighted degree centralities in each layer is divided by the sum of neighbourhood quantities on each layer:

$$MDC^{(3)}(x) = \frac{\sum_{l \in L}\left(\sum_{y \in N(x,l)} w(x,y,l) + \sum_{y \in N(x,l)} w(y,x,l)\right)}{(m-1)\sum_{l \in L}|N(x,l)|} \quad (2.28)$$

The third version of first multi-layered indegree centrality $MDC^{(3)In}(x)$:

$$MDC^{(3)In}(x) = \frac{\sum_{l \in L}\sum_{y \in N(x,l)} w(y,x,l)}{(m-1)\sum_{l \in L}|N(x,l)|}, \quad (2.29)$$

and third version of first multi-layered outdegree centrality $MDC^{(3)Out}(x)$:

$$MDC^{(3)Out}(x) = \frac{\sum_{l \in L}\sum_{y \in N(x,l)} w(x,y,l)}{(m-1)\sum_{l \in L}|N(x,l)|}. \quad (2.30)$$

Both Cross Layered and Multi-Layered Degree Centralities are helpful in the interpretation of the social network semantics. They provide more information than $MN(x,\alpha)$ as they take into consideration not only the number of relations but also their quality, i.e. connection strengths.

### 2.4.4 Cross Layered Edge Clustering Coefficient

The cross layered edge clustering coefficient (*CLECC*) is an edge measure which was developed based on the idea of edge clustering coefficient measure introduced by Radicchi et. al [Radicchi 04].

Edge clustering coefficient for an edge $<x,y>$ expresses how much the neighbours of the user $x$, and neighbours of the user $y$ are connected to both $x$ and $y$. The edge clustering coefficient is defined as:

$$ECC(x,y) = \frac{z_{x,y} + 1}{s_{x,y}}, \quad (2.31)$$

where $x$ and $y$ are the users connected by the edge $<x,y>$, $z_{x,y}$ is the number of triangles built upon the edge $<x,y>$ and all edges between x, $y$ and their neighbours, $s_{x,y}$ is the possible number of triangles that one could build based on edge<x,y> and all possible edges (even those that do not exist) between *x, y* and their neighbours.

*CLECC* measure [**Bródka** 11a] was created on the same idea as edge clustering coefficient and expresses the similar neighbours interconnectivity but for the multi-layered social network, using multi-layered neighbourhood.

$$CLECC(x, y, \alpha) = \frac{|MN(x,\alpha) \cap MN(y,\alpha)|}{|(MN(x,\alpha) \cup MN(y,\alpha))/\{x, y\}|} \quad (2.32)$$

Thus, it can be described as a proportion between the common multi-layered neighbours and all multi-layered neighbours of *x* and *y*.

*CLECC,* by utilizing multi-layered neighbourhood, considers all layers at the same time. The *α* parameter allows to adjust the measure strictness depending on differences in the density of each layer. This latter feature is particularly important when there are very large differences in the density of each layer.

For instance, imagine that we have four layers and 1000 users. Two of them are very dense (50,000 edges), while the two other quite sparse (5,000 edges). Now, thanks to the *α* parameter the measure can be adjusted. It can be either very restrictive and require the connection to exist on all layers (*α*=4), or it can be more gentle and require connections only in few of them (*α*=2). We can also choose a middle ground and assume that the connection exists on two existed dense layers and one of the sparse ones (*α*=3).

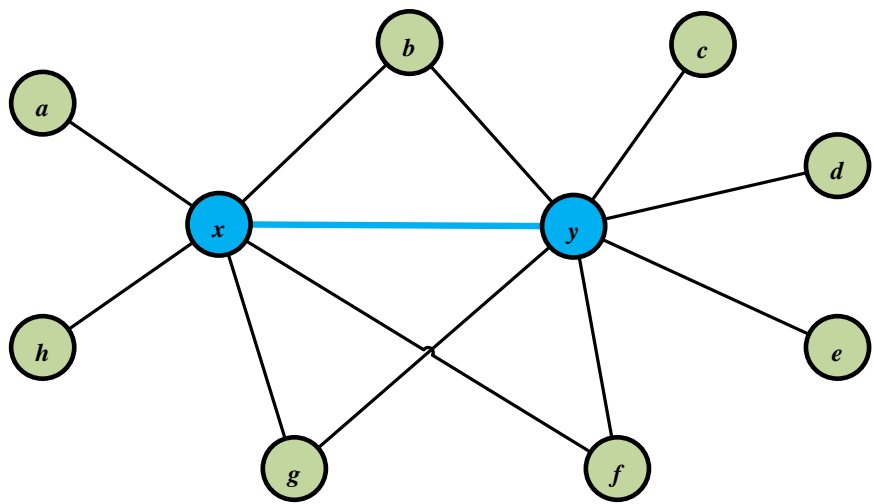

Figure 2.4 The example of single-layered social network [Fortunato 10].

For the network presented in Figure 2.4, edge clustering coefficient introduced by Radicchi et al. for the blue edge <*x,y*> is 3/5 [Fortunato 10]. The two blue vertices *x* and *y* have five and six neighbours (except each other) and because *x* has degree five there cannot be more than five triangles built on the edge connecting the blue vertices and out of all possible five three are actually there (via *b*, *f*, and *g*).

Meanwhile, *CLECC* for the same edge is equal to 3/8,

$$CLECC(x,y,1) = \frac{|MN(x,1) \cap MN(y,1)|}{|(MN(x,1) \cup MN(y,1))/\{x,y\}|} = \frac{|\{a,b,y,g,h\} \cap \{b,c,d,e,f,g,x\}|}{|\{a,b,y,g,h\} \cup \{b,c,d,e,f,g,x\})/\{x,y\}|} =$$

$$= \frac{|\{b,f,g\}|}{|\{a,b,c,d,e,f,g,h,x,y\})/\{x,y\}|} = \frac{3}{|\{a,b,c,d,e,f,g,h\}|} = \frac{3}{8}.$$

Both measures reflect how much neighbours of user *x* and *y* are connected to both *x* and *y* but *CLECC* was designed to be as simple as possible.

## 2.4.5 Algorithms in Multi-layered Social Network, Example of Shortest Path Discovery

The last two measures important for social network analysis are closeness centrality and betweenness centrality (see Equation 2.5 and 2.7). Unfortunately they are quite complicated because to calculate them the shortest paths within *SSN* or in this case *MSN* are needed. That is why it became necessary to develop an approach which will allow to calculate shortest paths in *MSN*.

This approach was introduced in [**Bródka** 11c]. To extract the shortest paths from *MSN*, some existing algorithms can be utilized. However, since there is no single edge between pairs of members in *MSN*, it is required to calculate the distance between any two neighbours in *MSN*. Basically, in most of the algorithms the shortest path is extracted based on the cost of transition from one node to another (negative connection), i.e. the greater cost the longer path value. On the other hand, in most social networks, the edges (relations) are considered to be positive, so, they express how close the two nodes are to each other, the greater weight the shorter path. That is why each typical 'positive weight', $w(x,y,l) \in [0,1]$ assigned to edges in *MSN*, has to be converted to 'negative' distance values by subtracting *w* from one, Figure 2.5. If there is no relationship between user *x* and user *y* then $w(x,y,l) = 0$.

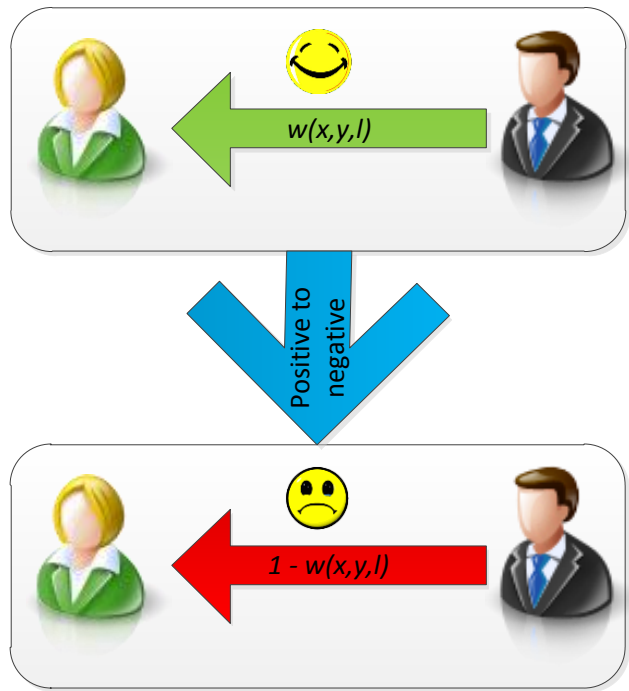

Figure 2.5 Transformation of 'positive' closeness of the social relationship into 'negative' distance (strangeness).

The distance (strangeness) $d(x,y)$ between $x$ and $y$ in *MSN* aggregated over all layers is calculated using the weights $w(x,y,l) \in [0,1]$ of all existing edges $\langle x,y,l \rangle$ from $x$ to $y$, i.e. from all layers. The sum of these weights is subtracted from the number of layers - $|L|$ and normalized by the total number of layers in *MSN*, as follows:

$$d(x,y) = \frac{\sum_{l \in L}(1-w(x,y,l))}{|L|} = \frac{|L| - \sum_{l \in L} w(x,y,l)}{|L|} = 1 - \frac{\sum_{l \in L} w(x,y,l)}{|L|}. \qquad (2.33)$$

Of course, if *MSN* has been built based on negative relations like: apathy, hatred, hostility, etc. the weight does not have to be subtracted from 1 to obtain the distance because it already reflects the distance.

### *Multi-layered Edge*

Once, we have distance characterized, the multi-layered edge *ME* between two users can be defined. Overall, three different versions of multi-layered edge *ME* may be distinguished:

1. A multi-layered edge $ME^l$ based on the number of layers is computed in following way:

$$ME^l(x, y, \alpha) = <x, y>: |\{<x,y,l>: l \in L\}| \geq \alpha \quad . \tag{2.34}$$

A multi-layered edge $ME^l$ from $x$ to $y$ exists if there are at least $\alpha$ edges $<x,y,l>$ from $x$ to $y$ in *MSN*, i.e. $|\{<x,y,l>: l \in L\}| \geq \alpha$. The weight of such a multi-layered edge $ME^l$ is equal to the distance from $x$ to $y$, $d(x,y)$, see Equation 2.33.

2. A multi-layered edge $ME^d$ based on the distance between users is computed as follows:

$$ME^d(x, y, \beta) = <x, y>: d(x, y) \leq \beta \quad . \tag{2.35}$$

A multi-layered edge $ME^d$ from $x$ to $y$ exists, if the distance from $x$ to $y$ is lower than or equal $\beta$: $d(x,y) \leq \beta$. The weight of such an edge $ME^d$ is equal to the distance from $x$ to $y$, $d(x,y)$, see Equation 2.33.

3. A multi-layered edge *ME* calculated based on the number of layers and on the distance between users is calculated in the following way:

$$ME(x, y, \alpha, \beta) = <x, y>: |\{<x,y,l>: l \in L\}| \geq \alpha \wedge d(x, y) \leq \beta \quad . \tag{2.36}$$

A multi-layered edge *ME* from $x$ to $y$ exists, if there exist at least $\alpha$ edges, $<x,y,l>$, from $x$ to $y$, $|\{<x,y,l>: l \in L\}| \geq \alpha$ in the *MSN* and simultaneously the distance from $x$ to $y$ is not greater than $\beta$, i.e. $d(x,y) \leq \beta$. The weight of such a multi-layered edge *ME* is equal to the distance from $x$ to $y$, $d(x,y)$, see Equation 2.33.

### *Multi-layered Path*

The multi-layered path is a list of nodes and from each of these nodes there is a multi-layered edge to the next node in the list. The length of such path from node $x$ to $v$ is a sum of distances of all multi-layered edges from the path. The shortest multi-layered path is the path with the smallest length $SSP(x,v)$ among all paths existing from node $x$ to $v$ in multi-layered social network.

### *Shortest Path Calculation*

Now, the concept presented above can be applied to calculate shortest paths in two different approaches.

The first approach is to calculate multi-layered edges and their length at the beginning of calculation. Next, any commonly used shortest path calculation algorithm can be utilized in the same way like for "normal" social network. For example, the "regular" Dijkstra algorithm provides all shortest paths with their lengths $SSP(x,v)$ outgoing from a single node $x$. If applied to multi-layered edges, it is as follows:

---

**DAP – Dijkstra Algorithm with Preprocessing**

**Input**: $MSN=<V,E,L>$, $x \in V$ – source node
**Output**: the list of shortest paths outgoing from $x$ to all $v \in V$ and their lengths $SSP(x,v)$

Calculate distances $d(y,z)$ for each $y,z \in V$ using Equation 2.33
Calculate all multi-layered edges $ME$ using Equation 2.36

$P=\{\}$ (empty set)          /* $P$ – set of nodes already processed */
$T=V$                         /* $T$ – set of nodes to process */
$SSP(x,v)=\infty$ for all $v \in V$
$SSP(x,x)=0$
$pred(x,x)=0$           /* $pred(x,v)$ – predecessor of selected */
                        /* node $v$ on the path from $x$ to $v$     */
**while** $P \neq V$ **do**
  **begin**
        $v=\arg\min\{SSP(x,v) \mid v \in T\}$
        $P:=P \cup v$, $T:=T \setminus v$
        **if** $SSP(x,v)= \infty$ **then** end while
        **for** $w \in N^{out}(v)$ **do**
  **if** $SSP(x,w) > SSP(x,v)+ d(v,w)$ **then**
                **begin**
                        $SSP(x,w):= SSP(x,v)+ d(v,w)$
                        $pred(x,w)=v$
                **end**
  **end**

---

The second approach is to modify the selected algorithm and process edges "on the fly". This approach is slightly better because additional multi-layered measures can be added, for example multi-layered neighbourhood. Example on Dijkstra algorithm modified into multi-layered Dijkstra algorithm.

**MDA - Multi-layered Dijkstra Algorithm**

**Input**: $MSN=<V,E,L>$, $x \in V$ – source node
**Output**: the list of shortest paths outgoing from $x$ to all $v \in V$ and their lengths $SSP(x,v)$

$P=\{\}$ (empty set)
$T=V$
$SSP(x,v)=\infty$ for all $v \in V$
$SSP(x,x)=0$
$pred(x,x)=0$
**while** $P \neq V$ **do**
  **begin**
        $v=\mathrm{argmin}\{SSP(x,v) \mid v \in T\}$
        $P:=P \cup v$,
        $T:=T \setminus v$
        **if** $SSP(x,v)= \infty$ **then** end while
        **for** $w \in MN^{out}(v,\alpha)$ **do**
   **if** $SSP(x,w) > SSP(x,v)+ d(v,w)$ **then**
                **begin**
                        $SSP(x,w):= SSP(x,v)+ d(v,w)$
                        $pred(x,w)=v$
                **end**
  **end**

Now when it is possible to calculate shortest paths for *MSN* we can calculate closeness centrality and betweenness centrality by utilizing regular equations developed for *SSN* (see Equation 2.5 and 2.7). The complete evaluation of described above solutions can be found in [**Bródka** 11c].

## 3. Group Extraction in Multi-layered Social Network

### 3.1 Introduction to Group Extraction in Single-layered Social Network

The existence of groups (communities) in social networks is intuitively obvious [Porter 09] and have been studied by many researchers for a long time. There is no universally acceptable definition of groups in the social network [Coleman 64], [Fortunato 10]. Nevertheless, there are several of them, which are used depending on the authors' needs [Coleman 64], [Freeman 04], [Fortunato 10], [Kottak 04]. In addition, some of them cannot be even called definitions because they introduce only some criteria for the group existence. In the biological terminology, a group, often also called a community is a collection of cooperating organisms, sharing a common environment. In sociology, in turn, it is traditionally defined as a group of people living and cooperating in a single location. However, due to the fast growing and spreading Internet, the concept of social community has lost its geographical limitations. In computer science, groups are the product of unsupervised machine learning algorithms – group extraction methods.

Overall, a general idea of the social community (group in a given population) is a set of social entities, who more frequently collaborate with each other rather than with other members of the population (the entire social network). The concept of the group (community) can be easily transposed to the graph theory, in which the social network is a graph and a group is a subset of vertices with high density of edges inside the group, and lower edge density between nodes from two separate groups [Evans 09], [Fortunato, 10], [Porter 09]. Others try to define a group as a set of closely interrelated links rather than a set of nodes[Evans 09], [Ahn 10]. However, another problem arises in the quantitative definition of a community, there is a number of conditions for group existence which try to achieve that (see Section 3.2). Anyway, most definitions of groups are built based on the general idea presented above.. Additionally, groups can also be algorithmically determined, as the outcome of the specific clustering algorithm, i.e. without a precise a priori definition [Moody 03]. Thus, there is a difficulty to find in literature an unequivocal definition of a group, acceptable to everybody [Wasserman 94], [Agarwal 09], [Tang 10a], so the term has been widely used without formal definition.

In this dissertation, the following definition will be used:

*Definition 3.1*

   *A group G in social network SSN(V,E$_l$) or MSN(V,E,L) is a subset of vertices from V (G⊆V), which fulfil weak community requirements i.e.* $\sum_{i \in G} k_i^{(in)} > \sum_{i \in G} k_i^{(out)}$ .

Weak community is relaxed condition of strong community (see Section 3.2.3) defined as, the internal degree of the subgraph exceeds its external degree, as follows:

$$\sum_{i \in G} k_i^{(in)} > \sum_{i \in G} k_i^{(out)} ,$$

where $k_i^{(in)}$ is the number of connections with nodes from the same group *G* and $k_i^{(out)}$ is the number of connections between *G*'s nodes and nodes outside group *G*.

## 3.2 Condition for Group Existence

As mentioned before, in the literature exists a number of conditions for group existence, which try to quantify the group. Almost all of them are described in [Fortunato 10], below only most important are presented, i.e. complete mutuality, reachability, adjacency of vertices, comparison of internal and external cohesion, spatial measures and random walks.

### 3.2.1 Complete Mutuality

Complete mutuality in social community terms would correspond to the case of the group, whose members are all friends to each other. In the graph theory domain, it is called a clique, i.e. a subset of vertices, which are all adjacent to each other. The simplest clique example is a triangle and it is very frequent in real networks, where larger cliques are much less frequent. Due to that fact, complete mutuality is very strict condition. As a matter of fact, a subgraph with all possible edges except one, would be extremely cohesive, but it would not be considered as a community according to this measure. Moreover, vertices of communities found under this recipe are all the same (the same degree, closeness etc.), where it is expected and desirable that within community exist vertices with different roles, e.g. core vertices coexisting with peripheral ones. However, it is possible to relax the notion of clique. An n-clique has been proposed by Alba [Alba 73] and Luce [Luce 50]. It is based on **reachability**, i.e. existence and length of paths between vertices. It means that an n-clique is a maximal subgraph in which the distance of each pair of its vertices is not larger than *n*. For *n* = 1 one recovers the definition of clique. Of course the geodesic path does not need to run only on vertices which are a part of the subgraph under study. To avoid this problem Mokken [Mokken 79] has suggested an alternative to n-clique, the n-clan. An n-clan restricts n-clique by insisting that a path distance between any two members of an n-clique does not exceed *n*.

### 3.2.2 Adjacency of Vertices

The idea of adjacency of vertices criterion is quite self-explanatory. In order to be a part of the community, a vertex has to be adjacent to some minimum number of other vertices in the subgraph. There are two complementary ways of expressing that. A k-plex, introduced by Seidman and Foster [Seidman 78] is a maximal subgraph, in which each vertex is adjacent to all other vertices of the subgraph except at most *k* of them. Similarly, a k-core, proposed by Seidman in 1983 [Seidman 83], is a maximal subgraph, in which each vertex is adjacent to at least *k* other vertices of the subgraph. Definitions impose conditions on the minimal number of absent or present edges, respectively.

### 3.2.3 Comparison of Internal and External Cohesion

Comparison of internal and external cohesion can be considered as a next criterion. A highly cohesive subgraph does not necessary form a community, while there is strong cohesion between the subgraph and the rest of graph. Therefore, it is important to compare the internal and external cohesion of the subgraph. In fact, this is what is usually done in the most recent definitions of community. We can distinguish two most important concepts here: a strong community and weak community introduced by Radicchi et al. [Radicchi 04]. The strong condition denotes a community such that the internal degree of each vertex is greater than its external degree and it is calculated in the following way:

$$k_i^{(in)} > k_i^{(out)}, \quad \forall i \in G, \tag{3.1}$$

where $k_i^{(in)}$ is the number of connections with nodes from the same group *G* and $k_i^{(out)}$ is the number of connections between *G*'s nodes and nodes outside group *G*.

Weak community is relaxed condition of strong community defined as, the internal degree of the subgraph exceeds its external degree, as follows:

$$\sum_{i \in G} k_i^{(in)} > \sum_{i \in G} k_i^{(out)}, \quad (3.2)$$

where $k_i^{(in)}$ is the number of connections with nodes from the same group $G$ and $k_i^{(out)}$ is the number of connections between $G$'s nodes and nodes outside group $G$.

### 3.2.4  Spatial Measures

For graphs that are embedded in a n-dimensional Euclidean space, i.e graph vertices have assigned a position in space distance between pairs of vertices can be used as a measure of their similarity, or more precisely dissimilarity (similar vertices are expected to be closer to each other). Given the two data points $A = (a_1, a_2, ..., a_n)$ and $B = (b_1, b_2, ..., b_n)$ any norm $L_m$ can be used to compute the distance.

- The Manhattan distance ($L_1$ norm), i.e.: $d_{AB}^M = \sum_{i=1}^{n} |a_i - b_i|$.

- the Euclidean distance ($L_2$ norm), defined as: $d_{AB}^E = \sum_{i=1}^{n} \sqrt{(a_i - b_i)^2}$ and

- the $L_\infty$ norm, i.e. $d_{AB}^\infty = \max_{i \in [1,n]} |a_i - b_i|$.

The biggest problem with this methods is the fact that the network need to be transformed into Euclidian space, and despite the fact that there is a number of such methods [Bronstein 06], [Harel 04], [Shavitt 04], [Shaw 07] they are very time consuming and not deterministic i.e. for the same network they could produce different results. Therefore spatial measures can be applied only to small networks.

### 3.2.5  Random Walks

The next class of measures of vertex similarity is based on properties of random walks on graphs. A random walker on the move from a vertex follows each adjacent edge with equal probability. In the literature a number of different criteria based on random walks can be found. Commute-time is the average number of steps needed for a random walker, starting at either vertex, to reach the other vertex for the first time and to come back to the starting vertex. Commute-time is a dissimilarity measure – the larger the time, the farther the vertices. Commute-time issue and its variants have been extensively studied in work by Saerens and his co-workers [Fouss 07], [Saerens 04].

## 3.3 The Most Frequently Used Methods for Community Extraction in Single-layered Social Networks

In the literature, a growing interest in research related to identification and understanding groups and communities in social networks has been observed [Agarwal 09], [Tang 10b], [Newman 10], [Wasserman 94]. A major breakthrough was done in 2002 as a result of the paper by Girvan and Newman with a proposal of the graph partitioning algorithm [Girvan 02] which became very attractive for broad group of researchers, especially physicists and mathematicians. Additionally the discussions have been on-going whether the groups are disjoint or overlapping [Palla 05], and if such partitions are at one level or form some kind of hierarchical structure (each partition could be divided recursively) [Fortunato 10], [Girvan 02], [Porter 09]. The last approach better reflects the hierarchical nature of many real networks [Ahn 10], [Lancichinetti 09c], [Evans, 09]. In [Lancichinetti 09c], a method finding simultaneously both hierarchical and overlapping groups was proposed. That method finds local maxima of a fitness function by local, iterative searching and the group is recognised as a peak in a fitness histogram.

Many methods of finding coherent groups have been proposed, most of them are proposed for specific applications. An interesting approach to systematize these methods into four categories: node-centric, group-centric, network-centric and hierarchy-centric has been proposed in [Tang 10a], [Tang 10b]. Methods based on node-centric criteria require each node in a group to satisfy certain properties (such as complete mutuality or reachability see Section 3.2). CFinder [Palla, 05] is a good example of such a method. In turn, methods based on group centric criteria consider connections inside a group as a whole. It is acceptable, for example, that some nodes in a group are loosely connected as far as a whole group satisfies certain properties (e.g. week community see Section 3.2.3). Group identification using network-centric criteria takes into account global network topology as a whole. Nodes of the network are divided into some number of disjoint sets. Methods based on graph partitioning can be a good example. The last category – hierarchy-centric – consists of methods, which build a hierarchical structure of groups based on the network structure. An example of this group can be the popular edge betweenness algorithm [Newman 04b], [Newman 10]. Recently, in [Yang 11], authors tried to compare and evaluate several community detection algorithms on different small data sets. They came to the conclusion that different algorithms have different performance on different social networks and the quality of communities detected by algorithms is hard to evaluate. A number of similar surveys has been recently published which deeply describe existing community extraction methods [Agarwal, 09], [Fortunato 10] and struggle to compare them. That is why in this dissertation only brief descriptions of most commonly used methods are presented.

### 3.3.1 Girvan – Newman Method

In opposition to attempts for construction of a measure which determines the edges most central to communities, Girvan and Newman [Girvan 02] focused on the edges which are at least central. The authors have generalized betweenness centrality and proposed the edge betweenness. The edge betweenness is the number of shortest paths between pairs of vertices that run along particular edge. In a case of more than one shortest path between a pair of vertices existing, authors propose assign to each path equal weight such the total weight of all paths is unity.

In a network containing communities or groups that are loosely connected it is clear that all shortest paths between vertices in different communities have to go through the few intercommunity edges, which therefore have higher edge betweenness value, i.e. high edge betweenness implicates that a given edge connects communities. By removing such edges, Girvan and Newman algorithm separates groups from one another and, by doing this, reveals the underlying community structure of a graph.

Girvan and Newman (GN) algorithm for identifying communities is stated as follows:

**Girvan and Newman Algorithm**

1. Calculate betweenness scores for all edges in the network
2. Remove the edge with the highest score. In case of two or more edges with the same edge betweenness choose one of them at random and remove that.
3. Recalculate betweennesses for all edges affected by the removal.
4. Repeat from step 2 until no edges remain.

Since the edge betweenness is a global measure GN algorithm can be considered computationally costly. For calculation edge betweenness for all *m* edges in a graph of *n* vertex a method uses the fast algorithm of Newman with computational complexity $O(mn)$. The iteration of the procedure and recalculating measure once for every edge removed leads in the worst case to a total scaling of the computational time $O(nm^2)$. However, since recalculating edge betweenness is done only for edges affected by the removal the running time of GN algorithm may be better than in the worst case for the networks with the strong community structure (ones which rapidly break up into separate components after the first few iterations of the algorithm). Though, causing computational complexity, recalculating of betweenness for all edges is preferable. For two communities connected by more than one edge, there is no guarantee that all of those edges will have high betweenness. By recalculating betweenness after removal of each edge, the GN algorithm ensures that at least one of the remaining edges between two communities will always have a high betweenness.

### 3.3.2 Radicchi et al. Method

Another approach to detect communities relies on presence of cycles, i.e. closed path whose vertices and edges are all distinct. Communities are characterized with high density of edges so intuition prompts that cycles should be formed among them, where edges lying between communities will hardly be part of cycles. Radicchi in [Radicchi 04] proposed new measure, the edge clustering coefficient, that is based on that idea. Low values of this measure are likely to correspond to intercommunity edges. The measure was presented in Section 2.4.4. Additionally, edge clustering is closely related to edge betweenness. Even though, correlation cannot be considered perfect, edges with low edge clustering coefficient usually have high betweenness and vice versa.

The method works on the same bases as the algorithm by Girvan and Newman. At each iteration, the edge with the smallest edge clustering coefficient is removed. Then, the measure is recalculated again, and the next iteration begins. If the removal of an edge leads to the split of the subgraph, the method enforces the community condition on both clusters, i.e. the split is accepted only if both clusters are strong or weak communities (Section 3.2.3). This condition is verified on the full adjacency matrix of the initial graph. The algorithm stops when all clusters produced by the edge removals are communities in a strong or weak sense, and further splits would violate this condition.

**Radicchi et al. Method**

1. Calculate full adjacency matrix of the initial graph
2. Calculate edge clustering coefficient score for all edges in the network
3. Remove the edge with the lowest score. In case of two or more edges with the same edge clustering coefficient betweenness choose one of them at random and remove that. If the

> removal of an edge leads to split of a subgraph, accept removal only if both clusters are weak or strong communities.
> 4. If the edge has been removed recalculate edge clustering coefficient.
> 5. Repeat from step 2 until all clusters produced by the edge removals are communities in the strong or weak sense.

The running time of the algorithm is $O\left(\frac{m^2}{n^2}\right)$, or $O(n^2)$ for a sparse graph [Fortunato 10]. The method requires recalculating edge clustering coefficient at every iteration. Since this measure is local, involving at most extended neighbourhood of the edge, the process is quick and the method running time is much shorter than running time of the GN method.

### 3.3.3 Fast Modularity Optimization

The growing number of large networks has created a need for a very fast group extraction algorithm. Responding to this demand, Blondel, Guillaume, Lambiotte and Lefebvre [Blondel 08]. have created the *Fast Modularity Optimization method* also called *Blondel et al.* or *Louvain method*. Computational complexity of the method is $O(m)$, where $m$ the number of edges in the networks, so it is very fast and a greater problem for it is the disk write speed performance rather than the calculations speed.

The method originates from the modularity of the network that is a measure describing whether the network is well grouped. The modularity Q is defined as follows:

$$Q = \frac{1}{\sum_{x,y \in V} w(x,y)} \cdot \sum_{x,y \in V} \left[ \left( w(x,y) + w(y,x) - \frac{DC(x)DC(y)}{\sum_{i,j \in V} w(x,y)} \right) \delta(G_x, G_y) \right],$$

where: $V$ – the set of network nodes; $w(x,y)$ – the weight of the edge from $x$ to $y$; $DC(x)$ – degree centrality of node $x$ and similarity measure $\delta(G_1,G_2)$ for two groups $G_1$ and $G_2$ is:

$$\delta(G_1, G_2) = \begin{cases} 1 \text{ when } G_1 = G_2 \\ 0 \text{ when } G_1 \neq G_2 \end{cases}.$$

Since the optimization of this measure is NP-complete [Brandes 06], the approximating algorithms are used for large networks.

Fast optimization algorithm is as follows:

**Blondel et al. Algorithm**

1. Place each node in a separate group
2. For each vertex $x$ remove it from its group, put it in a group $G_y$ of its neighbour $y$ separately for each neighbour $y$ and calculate their modularity increase $\Delta Q(G_y,x)$. Leave neighbour $x$ in the group for which the modularity increase is the highest. If modularity increase $\Delta Q(G_y,x)$ is not positive for all neighbours $y$ ($\Delta Q(G_y,x) \leq 0$) than node $x$ stays in its original group.
3. Repeat step 2 until the modularity can no longer grow, i.e. for all nodes $x$ in the network and all their neighbours $y$ their $\Delta Q(G_y,x) \leq 0$.
4. Build a new network by replacing the separate groups with the super-nodes. The super-nodes are connected if at least one vertex in the two super-nodes are connected. However, the edge weight is the sum of weights of all edges between nodes located in super-nodes.
5. Repeat steps 1-4 until there are no more changes and a maximum of modularity is achieved.

The modularity increase $\Delta Q(G,x)$ is calculated as follows (see [Newman 04a] for derivation of this formula):

$$\Delta Q(G_y,x) = \left[\frac{D^{in}(G_y) + d^{in}(x)}{2m} - \left(\frac{D(G_y) + DC(x)}{2m}\right)^2\right] - \left[\frac{D^{in}(G_y)}{2m} - \left(\frac{D(G_y)}{2m}\right)^2 - \left(\frac{DC(x)}{2m}\right)^2\right]$$

where: $m = \sum_{x,y \in V} w(x,y)$; $D^{in}(G_y)$ – group internal degree; $D(G_y)$ – group degree; $d^{in}(x)$ – node internal degree in the group $G_y$; $DC(x)$ – node degree centrality in the entire network.

The only downside of this algorithm is the fact that it is dependent on the order of the processed nodes. However, this dependency is not yet fully known.

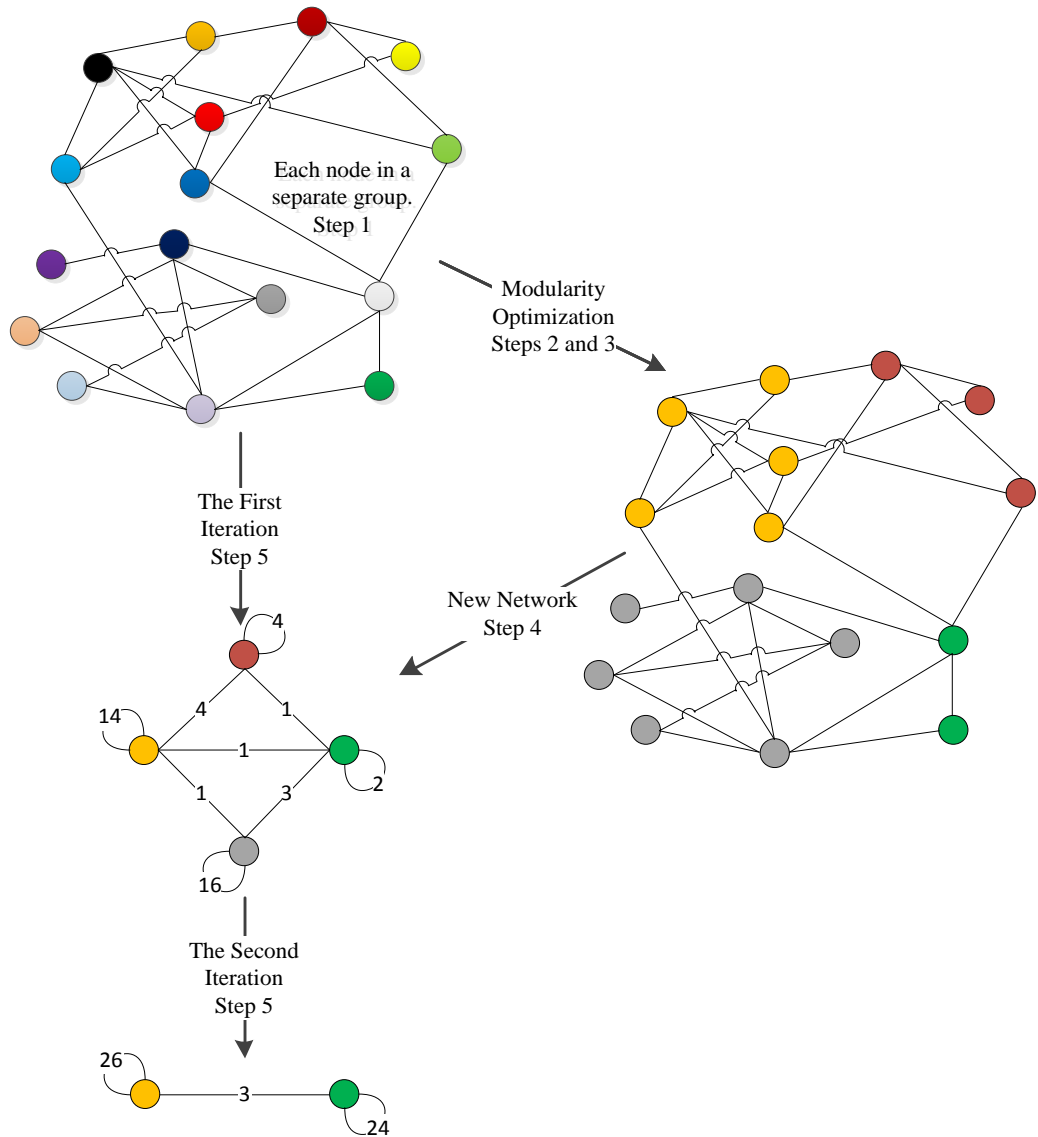

Figure 3.1. The example of Fast Modularity Optimization [Blondel 08].

### 3.3.4  Infomap

In the Infomap method introduced by Rosvall and Bergstrom, [Rosvall 08] the information theoretic approach is used to the clustering problem. It focuses on information diffusion across the graph and compression of the information flow description obtained from a random walker, which is chosen as a mean of information diffusion. Infomap changes the problem of finding the best cluster structure into finding partition with the minimum description length of an infinite random walk. It follows the intuitive idea that if the community structure is present, the random walker will spend more time inside a community because of the higher edges density. It means that the transition to another cluster will be less probable.

Infomap uses a two-level description. The first level assigns unique names to the clusters whereas the second one to the vertices within the same cluster. The names chosen for the vertex can be reused in the different clusters. This can be easily compared to the number of town maps. Each town with unique name would correspond to the particular structure, where street names to the vertices inside this structure. It is not unlikely for the towns to have the same street names, but it can be hardly imagined that two street with the same name can exist in the same town. Thus, town name and street name together yields full identification of the element. To clarify, in this comparison town names are the first level description and street names correspond to the second level.

Codes, i.e. unique names, are assigned to the vertices using Huffman code [Huffman 52]. It is characterized by assigning short codes to the frequently occurring events and longer to the rare ones. In case of the Infomap algorithm shorter codes are assigned to the most frequently visited vertices, within each structure. Moreover, the exit code is chosen for each structure, every time, the random walker, exits the structure and enters another. When transitioning a random walker has to use another special code word, describing which cluster it enters. Both names are unique among each cluster.

### 3.3.5 Ronhovde and Nussinov

An approach presented by Ronhovde and Nussinov [Ronhovde 09] is based on Reichardt and Bornholdt Potts Model (RBPM) [Reichardt 04]. Authors presented an Absolute Potts Model (APM). The first difference if compared to RBPM is the absence of null model term. Instead of defining community by comparison with a random graph, the method uses edge density as a determinant of the community. Due to that APM is a local measure of the community structure, though, the model computes a global energy sum. What is more, APM is free from resolution limit problem.

However the general idea behind both Potts model approaches is the same. The model gratifies existing edges inside the community and penalizes missing edges inside the community. However, by means of the parameter $\gamma$ in APM directly we can adjust the weight applied to missing edges. In the RBCM, $\gamma$ enables to adjusts the weight applied to the null model. The energy is minimized by sequentially placing single vertices to the communities, which best lowers the overall system's energy, till stop condition is reached.

The presented method uses a new partition stability criterion introduced a year later by the same authors in a successive paper [Ronhovde 10]. A new approach computes similarity of partitions obtained for the same γ and different initial conditions and then plotting the similarity as a function of the resolution parameter γ. Most replicas delivered by the robust partition in a given range of γ values will be very similar. On contrary, if a region between two strong partitions is considered, the replicas will deliver one or many partitions and the individual replicas will be, on average, not so similar to each other. In the chart created in this way, peaks in the similarity spectrum correspond to stable/relevant partitions.

### 3.3.6  Clique Percolation Method

The clique percolation method (CFinder) proposed by Palla et al. [Derenyi 05], [Palla 05] is the most widely used algorithm for extracting overlapping communities. The CFinder method works locally and its basic idea assumes that the internal edges of the group have a tendency to form cliques as a result of high density among them. Oppositely, the edges connecting different communities are unlikely to form cliques. A complete graph with *k* members is called *k*-clique. Two *k*-cliques are treated as a disjoining if the number of shared members equals *k*–1. Lastly, a *k*-clique community is the graph achieved by the union of all adjoining *k*-cliques [Adamcsek 06]. Such an assumption is made to reflect the fact that it is a crucial feature of the group that its nodes can be attained through densely joint subsets of nodes. The algorithm works as follows:

---

**CFinder Algorithm**

1. Find all cliques for different values of *k*.
2. Create a square matrix $M = n \times n$, where *n* is the number of cliques found. Each cell [*i, j*] contains the number of nodes shared by cliques *i* and *j*.
3. Select all cliques of size equal or greater than *k* and between the cliques of the same size find connections in order to create a *k*-clique chain.

---

Palla et al. proposing their method aim for algorithm which is (i) not too rigorous, (ii) takes into account the density of edges, (iii) works locally, and (iv) allows nodes to be a part of several groups. All of this were fulfilled.

### 3.4 Evaluation of Community Detection

There is a large number of measures which are able to assess the similarity of structure between two networks [Meila 07], [Kuncheva 04]: Rand measure [Rand 71], adjusted Rand measure [Steinley 04], Fowlkes–Mallows measure [Fowlkes 83], Jaccard measure [Ben Hur 02], Wallace measure [Wallace 83], Mirkin measure [Mirkin 96], van Dongen measure [Dongen 00a], etc.

In this thesis for assessing similarity between clustering results and model group structure for the analysed networks, the *normalized mutual information measure* (*NMI*) [Fred 03] will be used. The reason behind this decision is that the measure is proven to be very good for comparing community structure [Kuncheva 04], [Danon 05] and in fact for past few years it was most often used by researchers to assess the community detection algorithms, so the results will be easy to underspend and relate to other methods (at least for *SSN*).

Normalized mutual information measure comes from the information theory. It is based on defining a confusion matrix $N$, where the rows reflects model communities $A=\{G_1, G_2, ..., G_a\}$, and columns correspond to extracted communities $B=\{G_1, G_2, ..., G_b\}$. The elements inside the matrix $N$ describe the similarity between groups. The element $n_{ij}$ ($i^{th}$ row, $j^{th}$ column) is the number of nodes in the model community $i$ that occur in the extracted community $j$. Normalized mutual information measure for two networks $A$ and $B$ is defined as:

$$NMI(A,B) = \frac{-2\sum_{i=1}^{a}\sum_{j=1}^{b} n_{ij} \log\left(\frac{n_{ij}n}{n_i n_j}\right)}{\sum_{i=1}^{c_A} n_i \log\left(\frac{n_i}{n}\right) + \sum_{j=1}^{c_B} n_j \log\left(\frac{n_j}{n}\right)},$$

(3.3)

where $a$ is the number of model communities; $b$ is the number of extracted communities; $n_i$ is the sum of elements in row $i$; $n_j$ is the sum of elements in column $j$; and finally $n$ is a number of nodes in the analysed network [Fred 03]. The output is a real number from the range [0,1] where 0 means that groups do not have any mutual information i.e. the clustering is completely wrong, and 1 means that groups have 100% of mutual information i.e. the clustering is correct.

## 3.5 Group Extraction in Multi-layered Social Network

Only few researchers have tried to struggle with community detection in *MSN*s. The first method to find the community structure was presented in [Mucha 10]. The authors have proposed a general framework to detect community in time-dependent, multiscale, and multiplex networks using generalized Fast modularity optimization method (see Section 3.3.3). Unfortunately, the method description in the article is too general to be implement independently and the 1.0 version in MATLAB source code[10] was released on January 5, 2012 but it is available only to the limited number of people since it is still under development. Based on the information received from the authors the public version will be realised around July 2012. Additionally, the dataset on which authors were testing their solution, *Tastes, Ties, and Time*[11], is currently unavailable due to privacy concerns.

A similar idea, based also on the generalized Louvain method was presented in [Carchiolo 11]. However, in this case the work is only theoretical by now. Barigozzi at al. in [Barigozzi 11] struggled to find community structure in the multi-network of international trade by aggregating trade with commodity-specific trade and optimising modularity using a tabu search algorithm. The normalized information measure for the results vary from 0.05 to 0.46 depending on the year and selected commodity-specific trade type. Berlingerio et al. in [Berlingerio 11] have described the community detection problem and proposed two measures to quantify and disambiguate the density of the community, however, a multidimensional (multi-layered) community discovery algorithm was presented as the future work.

It is worth noticing that only one of the presented papers is more than one year old, so the problem of community extraction in multi-layered social network is quite new. Additionally the presented methods were designed for undirected and unweighted networks and extract separated communities (in the cease of the Louvain method communities are also hierarchical).

---

[10] http://netwiki.amath.unc.edu/GenLouvain/GenLouvain
[11] http://cyber.law.harvard.edu/node/4682

### 3.5.1 Cross Layered Edge Clustering Coefficient Method

The *Cross Layered Edge Clustering Coefficient method* (*CLECC* method) is a novel attempt to group extraction in multi-layered social networks introduced in [**Bródka** 11a] It has been developed based on the idea of cross layered edge clustering coefficient measure (see Section 2.4.), adjusted to the case of multi-layered networks. *CLECC* method, just like *CLECC* measure is inspired by edge clustering coefficient for single-layered social network and associated clustering method presented in [Radicchi 04].

Applying a multi-layered neighbourhood allowed the *CLECC* measure to consider all layers simultaneously. The adjustable parameter $\alpha$ is responsible for restrictiveness of the algorithm. This is especially relevant, if the multi-layered networks consist of layers with significantly different number of edges. When layers with high density exist along with sparse ones, the probability of two vertices to be in the multi-layered network for $\alpha$ equal |L| is small, thus the method outcomes will present only the strongest communities. However, by lowering $\alpha$, the method limitation enforced on *CLECC* measure decreases, unveiling more of underlying community structures. Thus, the parameter $\alpha$ sets robustness of communities delivered by the *CLECC* method.

---

**The *CLECC* Algorithm**

**Input:**    The multi-layered social network *MSN*, parameter $\alpha$

**Output**:   The list of groups within the *MSN*

1. Calculate the $CLECC(x,y,\alpha)$ for each pair $(x,y)$ where $x \in MN(y,\alpha)$ and for a given $\alpha$
2. Remove all edges between par $(x,y)$ for which $CLECC(x,y,\alpha)$ is the lowest.
3. Recalculate *CLECC* for all affected edges i.e. the $CLECC(x,z,\alpha)$ and $CLECC(y,z,\alpha)$ for all $z:z \in MN(x,\alpha) \cup MN(y,\alpha)$ and for a given $\alpha$.
4. If the deletion of edges will lead to the separation of the network into subgraphs, validate them against the selected condition for the group existence (in the original *MSN*). If the subgraph is a group do not remove any more edges.
5. Repeat from step 2 until there are only groups or single nodes.

---

*CLECC* is a divisive algorithm i.e. extracts separated communities and the group structure emerges by continuous edge removal. However, when the edge removal leads to the split of the graph, it is necessary to evaluate whether the disjoined subgraphs are communities. In this dissertation, the *CLECC* method, uses the concept of weak community (Equation 3.2). Of course, any other group evaluation indicator can be used (see Section 3.2). Additionally the *CLECC* method is designed for unweight and undirected network and for such networks it has been tested. However the cross layered edge clustering coefficient measure (see Section 2.4.4) can be adjusted to take into consideration the weight and the direction of edges during nodes neighbourhoods calculation, e.g. using different type of the multi-layered neighbourhood (see Section 2.4.1).

### 3.5.2 Cross Layered Edge Clustering Coefficient Method – the Computational Complexity

The computational complexity for the worst case scenario is polynominal and is equal to $O(n^4)$, where $n = |V|$, or $O(m^2)$, where $m$ is a number of pairs $(x,y)$ where $x,y \in V$ and there exist at least one edge $<x,y,l>$ $l \in L$ between $x,y$. For the worst case scenario, the method will remove one edge in one iteration until it removes all edges ($m$ times) and for each iteration the *CLECC* measure will have to be calculated for all remaining edges ($\frac{m}{2}$ times), so the complexity is $O(m^2)$.

If we would like to present complexity using nodes, then for complete multi-graph i.e. for a multi-graph where there is at least one edge between each pair of nodes the number of pairs $(x,y)$ is $m = \frac{|V|(|V|-1)}{2} = \frac{n^2-n}{2}$ then: $\frac{m^2}{2} = \frac{\left(\frac{n^2-n}{2}\right)^2}{2} = \frac{n^4 - 2n^3 + n^2}{2} \Rightarrow O(n^4)$.

However if we take into consideration that social networks are sparse networks the real computational complexity will almost never reach $O(n^4)$. The comparison of *CLECC* computational complexity with other, popular community detection methods are presented in Table 3.1.

| Author | Referefce | Label | Computational complexity |
|---|---|---|---|
| **Bagrow & Bollt** | [Bagrow 05] | BB | $O(n^3)$ |
| **Blondel et al.** | [Blondel 08] | Blondel et al | $O(m)$ |
| **Bródka et al.** | [**Bródka** 11a] | *CLECC* | $O(n^4); O(m^2)$ |
| **Capocci et al.** | [Capocci 05] | CSCC | $O(n^2)$ |
| **Clauset et al.** | [Clauset 04] | Clauset et al. | $O(n \log^2 n)$ |

| Author | Reference | Label | Computational complexity |
|---|---|---|---|
| **Donetti & Munoz** | [Donetti 04] | DM | $O(n^3)$ |
| **van Dongen** | [Dongen 00a] | MCL | $O(nk^2), k < n$ parameter |
| **Duch & Arenas** | [Duch 05] | DA | $O(n^2 \log n)$ |
| **Eckmann & Moses** | [Eckmann 02] | EM | $O(m\langle k \rangle^2)$ |
| **Fortunato et al.** | [Fortunato 04] | FLM | $O(m^3 n)$ |
| **Girvan & Newman** | [Girvan 02] | GN | $O(nm^2)$ |
| **Guimera et al.** | [Guimera 05] | Sim. Ann. | parameter dependent |
| **Latapy & Pons** | [Latapy 05] | LP | $O(n^3)$ |
| **Newman & Girvan** | [Newman 04b] | NG | $O(nm^2)$ |
| **Newman & Leicht** | [Newman 07] | EM | parameter dependent |
| **Palla et al.** | [Palla 05] | CFinder | $O(e^n)$ |
| **Radicchi et al.** | [Radicchi 04] | Radicchi et al. | $O\left(\dfrac{m^2}{n^2}\right)$ |
| **Reichardt & Bornholdt** | [Reichardt 04] | RB | parameter dependent |
| **Ronhovde & Nussinov** | [Ronhovde 09] | RN | $O(m^\beta \log n), \beta \approx 1.3$ |
| **Rosvall & Bergstrom** | [Rosvall 07] | Infomod | parameter dependent |
| **Rosvall & Bergstrom** | [Rosvall 08] | Infomap | $O(m)$ |
| **Wu & Huberman** | [Wu 04] | WH | $O(n+m)$ |
| **Zhou & Lipowsky** | [Zhou 04] | ZL | $O(n^3)$ |

Table 3.1 The compilation of computational complexities analysed in [Danon 05] and [Lancichinetti 09a] expressed by the number of nodes *n*, the number of links *m* or the average degree *<k>*.

### 3.5.3 Cross Layered Edge Clustering Coefficient Method – the Example

The example about how the *CLECC* method works, was done using the well-known *karate club* dataset[12] [Zachary 77]. This single-layered network was utilized because of simplicity of presentation and the well know status as a benchmarking network for the clustering methods. This dataset represents a karate club, which was observed for a period of three years. The political situation in the club was informal and despite the fact the club had four officers, all decisions were made by consensus at club meetings. The karate instructor at the club was Mr Hi. At the beginning of the study, there was a small conflict between club's chief administrator John A. and Mr Hi over the lessons prices. Mr Hi wanted to raise the prices but John A. wouldn't allowed it. While the time passed the club became divided over the issue. Mr Hi supporters saw in him fatherly figure who was their spiritual and physical mentor, and Mr Hi opponent saw him as disobedient employee who only care about money. After few very sharp confrontation Mr Hi was fired for attempt to raise the prices. The supporters of Mr Hi left with him and form new karate club. The karate club social network is presented in Table 3.2 and Figure 3.2.

---
[12] Network data http://www-personal.umich.edu/~mejn/netdata/karate.zip

| Edge | *CLECC* 1st iteration | *CLECC* 2nd iteration | *CLECC* 3rd iteration |
|---|---|---|---|
| <2,1>   | 0.4118 | 0.5000 | 0.5385 |
| <3,1>   | 0.2500 | 0.3333 | 0.2857 |
| <3,2>   | 0.2857 | 0.4000 | 0.4444 |
| <4,1>   | 0.3125 | 0.3571 | 0.3846 |
| <4,2>   | 0.4000 | 0.4444 | 0.4444 |
| <4,3>   | 0.3636 | 0.5000 | 0.5714 |
| <5,1>   | 0.1250 | 0.1429 | 0.1538 |
| <6,1>   | 0.1176 | 0.1333 | 0.1429 |
| <7,1>   | 0.1176 | 0.1333 | 0.1429 |
| <7,5>   | 0.2000 | 0.2000 | 0.2000 |
| <7,6>   | 0.4000 | 0.4000 | 0.4000 |
| <8,1>   | 0.1875 | 0.2143 | 0.2308 |
| <8,2>   | 0.3333 | 0.3750 | 0.3750 |
| <8,3>   | 0.3000 | 0.4286 | 0.5000 |
| <8,4>   | 0.5000 | 0.5000 | 0.5000 |
| <9,1>   | 0.0526 | 0.0588 | - |
| <9,3>   | 0.1667 | 0.2222 | 0.0000 |
| <10,3>  | 0.0000 | - | - |
| <11,1>  | 0.1250 | 0.1429 | 0.1538 |
| <11,5>  | 0.2500 | 0.2500 | 0.2500 |
| <11,6>  | 0.2000 | 0.2000 | 0.2000 |
| <12,1>  | 0.0000 | - | - |
| <13,1>  | 0.0625 | 0.0714 | 0.0769 |
| <13,4>  | 0.1667 | 0.1667 | 0.1667 |
| <14,1>  | 0.1765 | 0.2143 | 0.2308 |
| <14,2>  | 0.3000 | 0.3750 | 0.3750 |
| <14,3>  | 0.2727 | 0.4286 | 0.5000 |
| <14,4>  | 0.4286 | 0.5000 | 0.5000 |
| <17,6>  | 0.2500 | 0.2500 | 0.2500 |
| <17,7>  | 0.2500 | 0.2500 | 0.2500 |
| <18,1>  | 0.0625 | 0.0714 | 0.0769 |
| <18,2>  | 0.1111 | 0.1250 | 0.1250 |
| <20,1>  | 0.0588 | 0.0714 | 0.0769 |
| <20,2>  | 0.1000 | 0.1250 | 0.1250 |
| <22,1>  | 0.0625 | 0.0714 | 0.0769 |
| <22,2>  | 0.1111 | 0.1250 | 0.1250 |
| <26,24> | 0.0000 | - | - |
| <26,25> | 0.2500 | 0.5000 | 0.5000 |

| Edge | CLECC 1st iteration | CLECC 2nd iteration | CLECC 3rd iteration |
|---|---|---|---|
| <28,3> | 0.0000 | - | - |
| <28,24> | 0.1429 | 0.2500 | 0.2500 |
| <28,25> | 0.0000 | - | - |
| <29,3> | 0.0000 | - | - |
| <30,24> | 0.3333 | 0.4000 | 0.4000 |
| <30,27> | 0.2500 | 0.2500 | 0.2500 |
| <31,2> | 0.0000 | - | - |
| <31,9> | 0.3333 | 0.4000 | 0.5000 |
| <32,1> | 0.0000 | - | - |
| <32,25> | 0.1429 | 0.2000 | 0.2000 |
| <32,26> | 0.1429 | 0.2000 | 0.2000 |
| <32,29> | 0.1429 | 0.2000 | 0.2000 |
| <33,3> | 0.0500 | 0.0588 | - |
| <33,9> | 0.2308 | 0.2308 | 0.1667 |
| <33,15> | 0.0833 | 0.0833 | 0.0909 |
| <33,16> | 0.0833 | 0.0833 | 0.0909 |
| <33,19> | 0.0833 | 0.0833 | 0.0909 |
| <33,21> | 0.0833 | 0.0833 | 0.0909 |
| <33,23> | 0.0833 | 0.0833 | 0.0909 |
| <33,24> | 0.1429 | 0.1538 | 0.1667 |
| <33,30> | 0.1538 | 0.1538 | 0.1667 |
| <33,31> | 0.1538 | 0.1667 | 0.1818 |
| <33,32> | 0.0625 | 0.0667 | 0.0714 |
| <34,9> | 0.1053 | 0.1250 | 0.1333 |
| <34,10> | 0.0000 | - | - |
| <34,14> | 0.0000 | - | - |
| <34,15> | 0.0588 | 0.0714 | 0.0714 |
| <34,16> | 0.0588 | 0.0714 | 0.0714 |
| <34,19> | 0.0588 | 0.0714 | 0.0714 |
| <34,20> | 0.0000 | - | - |
| <34,21> | 0.0588 | 0.0714 | 0.0714 |
| <34,23> | 0.0588 | 0.0714 | 0.0714 |
| <34,24> | 0.1667 | 0.2143 | 0.2143 |
| <34,27> | 0.0588 | 0.0714 | 0.0714 |
| <34,28> | 0.0526 | 0.0714 | 0.0714 |
| <34,29> | 0.0556 | 0.0714 | 0.0714 |
| <34,30> | 0.1765 | 0.2143 | 0.2143 |
| <34,31> | 0.1111 | 0.1429 | 0.1429 |
| <34,32> | 0.1000 | 0.1250 | 0.1250 |
| <34,33> | 0.5556 | 0.6667 | 0.7143 |

Table 3.2 The values of *CLECC* measure for each edge in all three iterations. Edges removed in the first iteration are marked in blue, in the second one are marked in orange and in the last one in orange. *CLECC* values which have not been recalculated are marked in yellow.

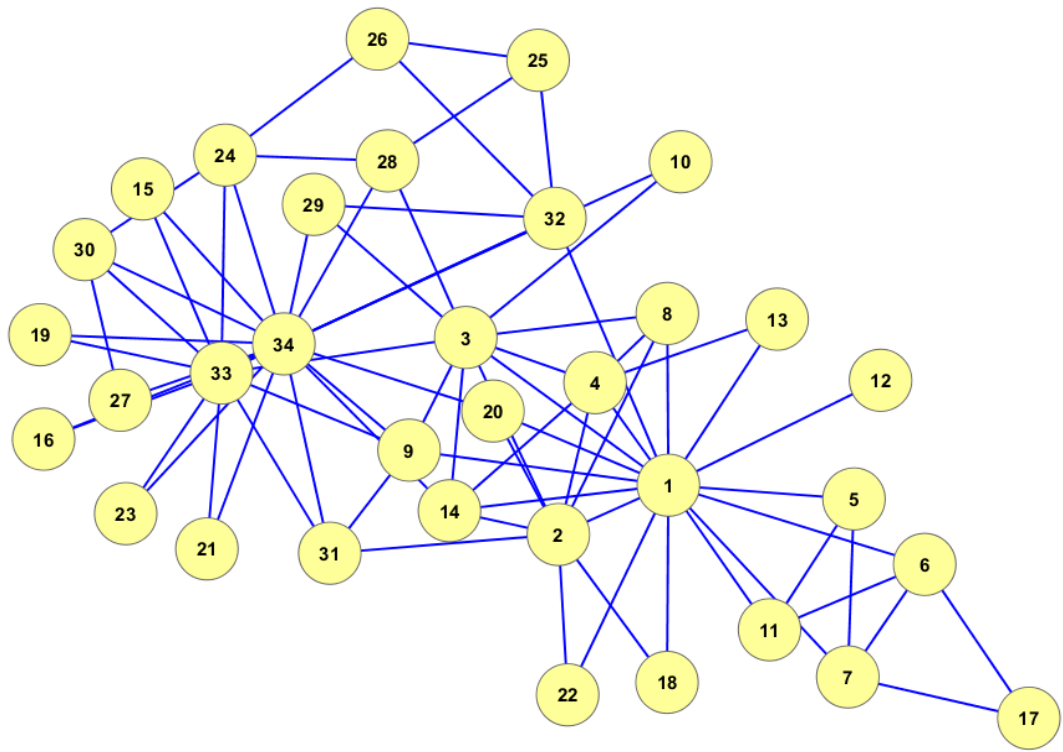

Figure 3.2 The karate club social network.

The *CLECC* method has needed 3 iteration to extract two groups. The values of *CLECC* measure for each edge for all iteration are presented in Table 3.2. In the 1$^{st}$ iteration 11 edges were removed and the *CLECC* measure was 0 for all of them (see Figure 3.3). In the 2$^{nd}$ iteration, next two edges were removed (*CLECC* measure = 0.0588). As it can be noticed in Figure 3.4, the first two iteration have created a bridge, the edge <9,3>, which connects two subgraphs and this relation was removed in the 3$^{rd}$ iteration. When edge <9,3> was removed, two separated subgraphs appeared. Both of them were validated according to condition for the week community existence (see Section 3.2.3) and both subgraphs meet the condition so they are groups. Since only separated groups or single nodes have left, the method terminates. It is worth noticing that the *CLECC* measure was recalculated only if necessary. Edges not affected by the removal process (yellow cells Table 3.2) can be skipped in the *CLECC* measure recalculation, what significantly decreases the method complexity.

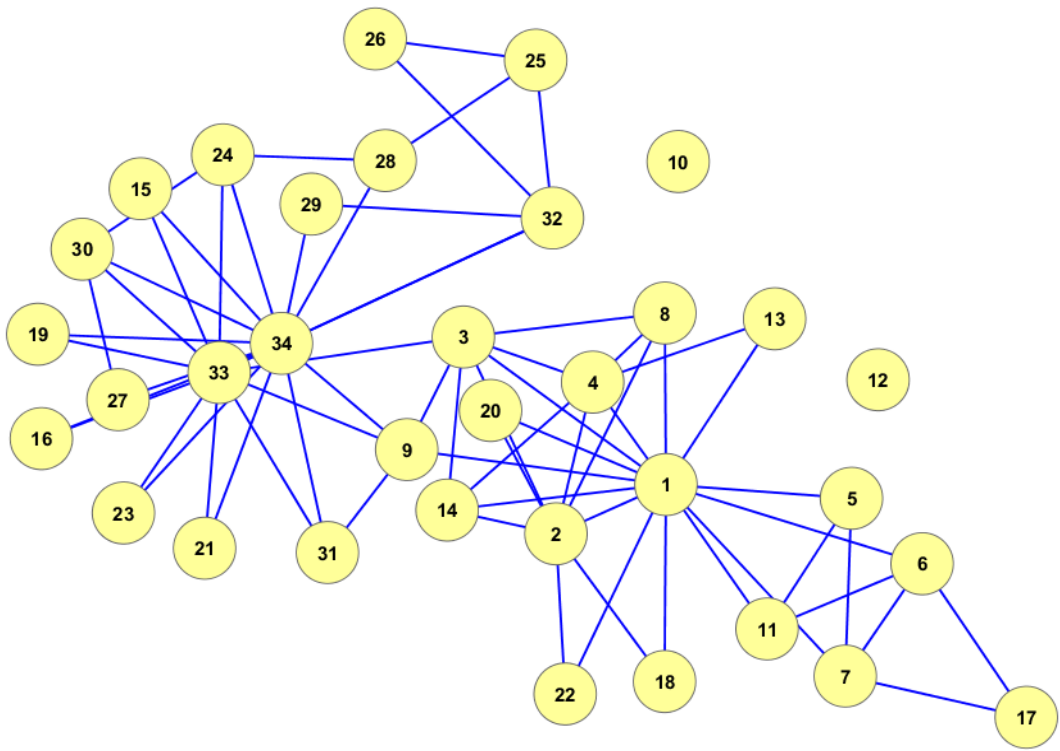

Figure 3.3 The network after the 1st iteration. Removed edges: <10,3>; <12,1>; <26,24>; <28,3>; <28,25>; <29,3>; <31,2>; <32,1>; <34,10>; <34,14>; <34,20>.

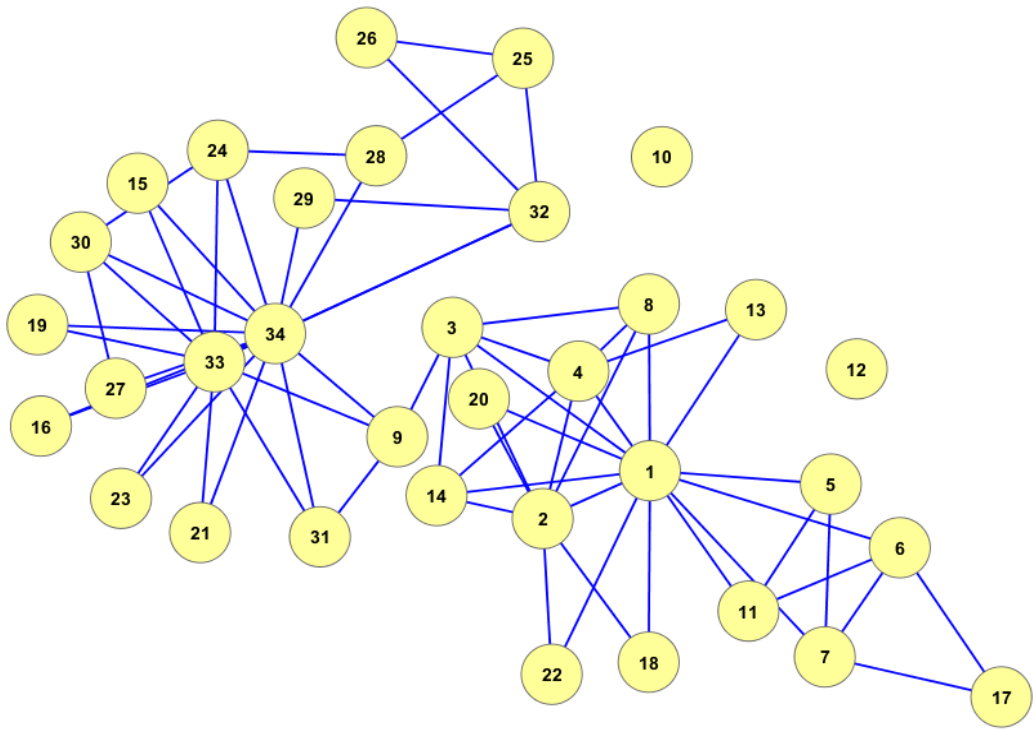

Figure 3.4 The network after the 2nd iteration. Removed edges: <9,1>; <33,3>.

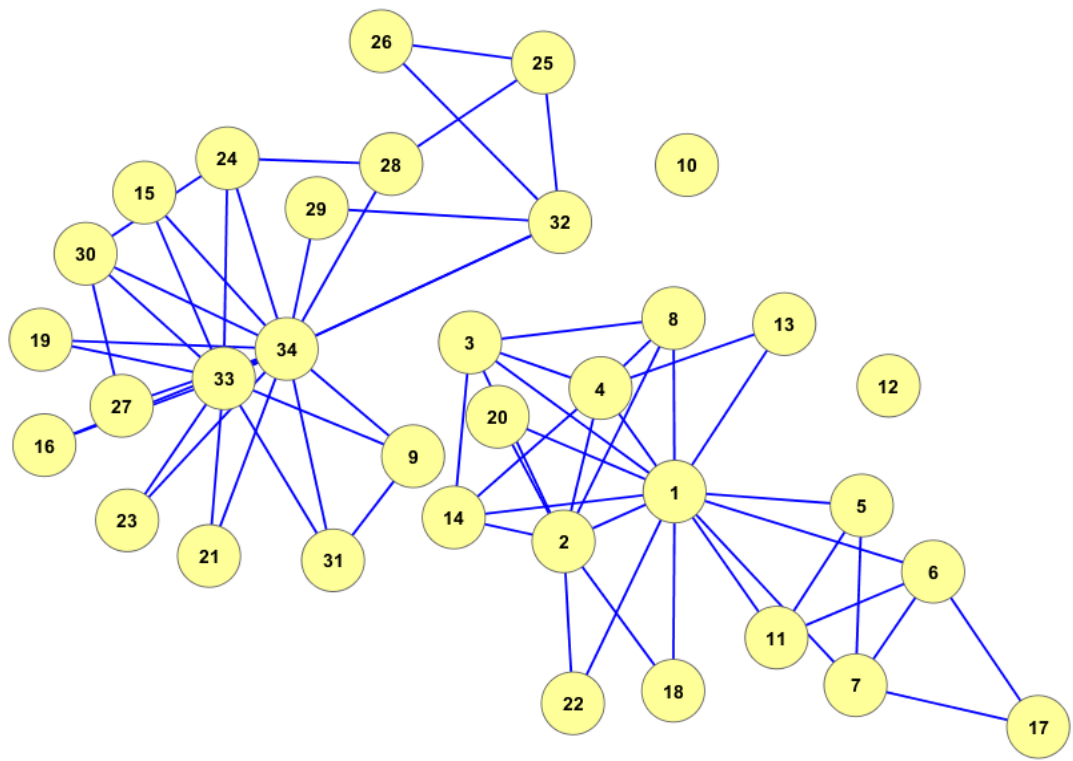

Figure 3.5 The network after the 3[rd] iteration. Removed edges: <9,3>.

The final result and its comparison with the original data [Zachary 77] is presented in Table 3.3 and Figure 3.6.

| Karate club member | Faction | Club After Fission | Club After Fission as Modelled in [Zachary 77] | Group assigned by *CLECC* method |
|---|---|---|---|---|
| 1 | Mr Hi – Strong | Mr Hi's | Mr Hi's | Mr Hi's |
| 2 | Mr Hi – Strong | Mr Hi's | Mr Hi's | Mr Hi's |
| 3 | Mr Hi – Strong | Mr Hi's | Mr Hi's | Mr Hi's |
| 4 | Mr Hi – Strong | Mr Hi's | Mr Hi's | Mr Hi's |
| 5 | Mr Hi – Strong | Mr Hi's | Mr Hi's | Mr Hi's |
| 6 | Mr Hi – Strong | Mr Hi's | Mr Hi's | Mr Hi's |
| 7 | Mr Hi – Strong | Mr Hi's | Mr Hi's | Mr Hi's |
| 8 | Mr Hi – Strong | Mr Hi's | Mr Hi's | Mr Hi's |
| 9 | John – Weak | Mr Hi's | John's | John's |
| 10 | None | John's | John's | Not assigned |
| 11 | Mr Hi – Strong | Mr Hi's | Mr Hi's | Mr Hi's |
| 12 | Mr Hi – Strong | Mr Hi's | Mr Hi's | Not assigned |
| 13 | Mr Hi – Weak | Mr Hi's | Mr Hi's | Mr Hi's |
| 14 | Mr Hi – Weak | Mr Hi's | Mr Hi's | Mr Hi's |
| 15 | John – Strong | John's | John's | John's |
| 16 | John – Weak | John's | John's | John's |
| 17 | None | Mr Hi's | Mr Hi's | Mr Hi's |
| 18 | Mr Hi – Weak | Mr Hi's | Mr Hi's | Mr Hi's |

| Karate club member | Faction | Club After Fission | Club After Fission as Modelled in [Zachary 77] | Group assigned by *CLECC* method |
|---|---|---|---|---|
| 19 | None | John's | John's | John's |
| 20 | Mr Hi – Weak | Mr Hi's | Mr Hi's | Mr Hi's |
| 21 | John – Strong | John's | John's | John's |
| 22 | Mr Hi – Weak | Mr Hi's | Mr Hi's | Mr Hi's |
| 23 | John – Strong | John's | John's | John's |
| 24 | John – Weak | John's | John's | John's |
| 25 | John – Weak | John's | John's | John's |
| 26 | John – Strong | John's | John's | John's |
| 27 | John – Strong | John's | John's | John's |
| 28 | John – Strong | John's | John's | John's |
| 29 | John – Strong | John's | John's | John's |
| 30 | John – Strong | John's | John's | John's |
| 31 | John – Strong | John's | John's | John's |
| 32 | John – Strong | John's | John's | John's |
| 33 | John – Strong | John's | John's | John's |
| 34 | John – Strong | John's | John's | John's |

Table 3.3 The final result of the *CLECC* method and theirs comparison with the original clustering results [Zachary 77].

*Karate club member* is an unique identifier which refers to a given karate club member. *Faction* gives the factional affiliation of the club members either to Mr Hi, John A. or none, the *Strong/Week* designation indicates whether individual was strong or weak supporter. *Club After Fission* indicated the club which member has chosen. *Club After Fission as Modelled in [Zachary 77]* describes Zachary's prediction based on original data. *Group assigned by CLECC* method indicate group assigned by *CLECC* method.

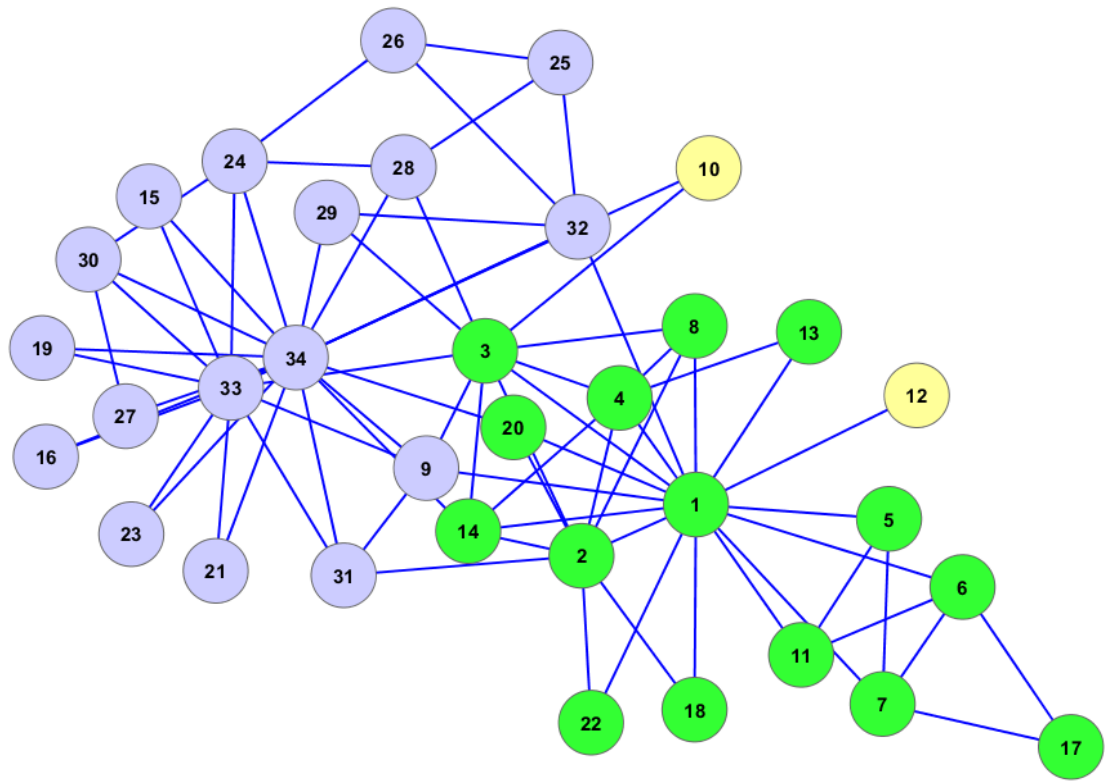

Figure 3.6 The final result of the *CLECC* method.

As it is presented in the final results, 31 members out of 34 are assigned to correct groups. Unfortunately, some members were allocated incorrectly.

Member no. 9 was a week supporter of John but at the end he joined Mr Hi's club. This is an unavoidable situation since people sometimes undertake decisions unsupported by their past actions. The *CLECC* method assigned member no. 9 according to his past behaviour so method acted just like Zachary's model.

Member no.10 according to the data did not support any side of the conflict, additionally, he has the same number of connections to both groups, that is why the *CLECC* method has not assigned this member to neither of two groups.

Member no. 12 was not assigned to any group because it is connected only to one member, i.e. member no. 1 and the *CLECC* measure for this connection is 0, so this tie was removed in the first iteration. This is the main weakness of the *CLECC* method – it is not able to assign leafs in the network. This problem can be easily resolved by adding a new step, which will allocate all nodes without assigned group to the group containing the highest number of their neighbours. If a node can be assigned to more than one group it serves as a bridge between groups. Since such a node does not belong clearly to any group it should be left without any group. The upgraded *CLECC+* method is presented below.

---

**The *CLECC+* Algorithm**

**Input:** The multi-layered social network *MSN*, parameter $\alpha$

**Output**: The list of groups within the *MSN*

1. Calculate the *CLECC*(*x,y,α*) for each pair (*x,y*) where $x \in MN(y)$ and for a given $\alpha$
2. Remove all edges between par (*x,y*) for which the *CLECC*(*x,y,α*) is the lowest.
3. Recalculate *CLECC* for all affected edges i.e. the *CLECC*(*x,z,α*) and *CLECC*(y,*z,α*) for all $z: z \in MN(x) \cup MN(y)$ and for a given $\alpha$.
4. If the deletion of edges will lead to the separation of the network into subgraphs, validate them against the selected condition for the group existence (in the original *MSN*). If the subgraph is a multi-layered group do not remove any more edges.
5. Repeat from step 2 until there are only groups or single nodes.
6. Allocate each node without an assigned group to the group with the highest number of its neighbours. If a node may be assigned to more than one group, remain it unallocated.

---

The results of the *CLECC+* method are presented in table Table 3.4 and Figure 3.7. Now, the member no. 12 is correctly assigned to Mr Hi's group. Member no. 10 has one neighbour from both groups so he remains unassigned.

| Karate club member | Faction | Club After Fission | Club After Fission as Modelled in [Zachary 77] | Group assigned by *CLECC* method | Group assigned by *CLECC+* method |
|---|---|---|---|---|---|
| 1 | Mr Hi - Strong | Mr Hi's | Mr Hi's | Mr Hi's | Mr Hi's |
| 2 | Mr Hi - Strong | Mr Hi's | Mr Hi's | Mr Hi's | Mr Hi's |
| 3 | Mr Hi - Strong | Mr Hi's | Mr Hi's | Mr Hi's | Mr Hi's |
| 4 | Mr Hi - Strong | Mr Hi's | Mr Hi's | Mr Hi's | Mr Hi's |
| 5 | Mr Hi - Strong | Mr Hi's | Mr Hi's | Mr Hi's | Mr Hi's |
| 6 | Mr Hi - Strong | Mr Hi's | Mr Hi's | Mr Hi's | Mr Hi's |
| 7 | Mr Hi - Strong | Mr Hi's | Mr Hi's | Mr Hi's | Mr Hi's |
| 8 | Mr Hi - Strong | Mr Hi's | Mr Hi's | Mr Hi's | Mr Hi's |
| 9 | John - Weak | Mr Hi's | John's | John's | Mr Hi's |

| Karate club member | Faction | Club After Fission | Club After Fission as Modelled in [Zachary 77] | Group assigned by *CLECC* method | Group assigned by *CLECC+* method |
|---|---|---|---|---|---|
| 10 | None | John's | John's | Not assigned | Not assigned |
| 11 | Mr Hi - Strong | Mr Hi's | Mr Hi's | Mr Hi's | Mr Hi's |
| 12 | Mr Hi - Strong | Mr Hi's | Mr Hi's | Not assigned | Mr Hi's |
| 13 | Mr Hi - Weak | Mr Hi's | Mr Hi's | Mr Hi's | Mr Hi's |
| 14 | Mr Hi - Weak | Mr Hi's | Mr Hi's | Mr Hi's | Mr Hi's |
| 15 | John - Strong | John's | John's | John's | John's |
| 16 | John - Weak | John's | John's | John's | John's |
| 17 | None | Mr Hi's | Mr Hi's | Mr Hi's | Mr Hi's |
| 18 | Mr Hi - Weak | Mr Hi's | Mr Hi's | Mr Hi's | Mr Hi's |
| 19 | None | John's | John's | John's | John's |
| 20 | Mr Hi - Weak | Mr Hi's | Mr Hi's | Mr Hi's | Mr Hi's |
| 21 | John - Strong | John's | John's | John's | John's |
| 22 | Mr Hi - Weak | Mr Hi's | Mr Hi's | Mr Hi's | Mr Hi's |
| 23 | John - Strong | John's | John's | John's | John's |
| 24 | John - Weak | John's | John's | John's | John's |
| 25 | John - Weak | John's | John's | John's | John's |
| 26 | John - Strong | John's | John's | John's | John's |
| 27 | John - Strong | John's | John's | John's | John's |
| 28 | John - Strong | John's | John's | John's | John's |
| 29 | John - Strong | John's | John's | John's | John's |
| 30 | John - Strong | John's | John's | John's | John's |
| 31 | John - Strong | John's | John's | John's | John's |
| 32 | John - Strong | John's | John's | John's | John's |
| 33 | John - Strong | John's | John's | John's | John's |
| 34 | John - Strong | John's | John's | John's | John's |

Table 3.4 The final result of the *CLECC* and *CLECC+* method and their comparison with the original clustering results [Zachary 77].

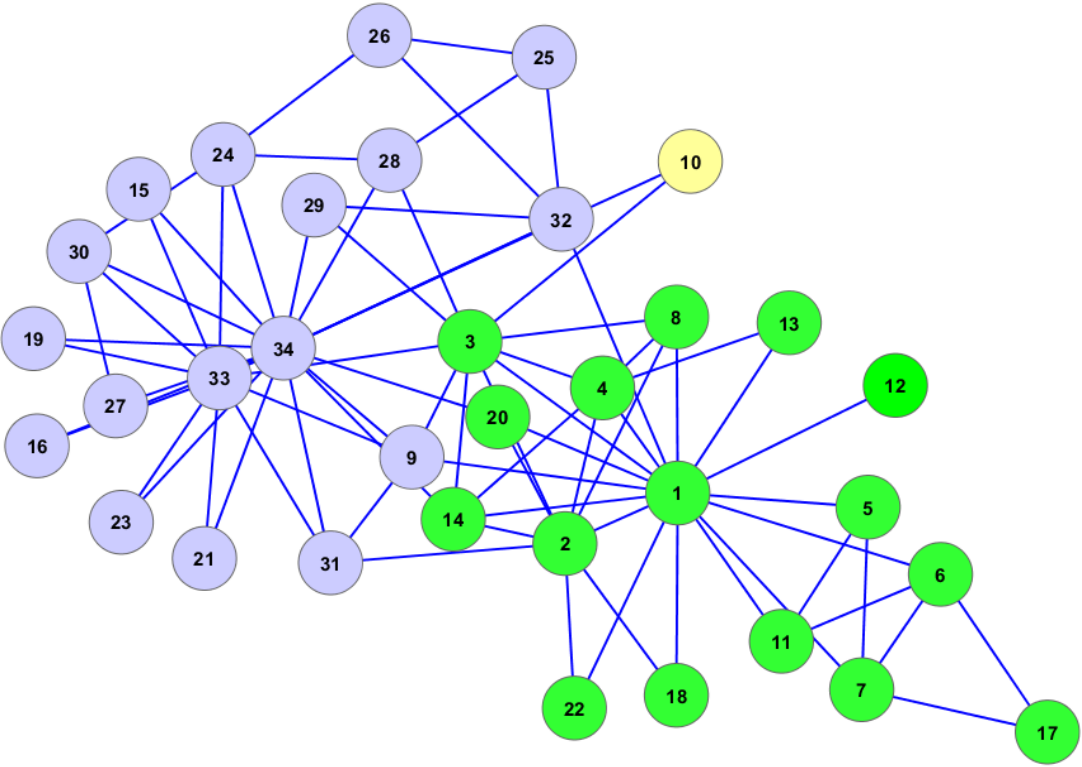

Figure 3.7 The final result of the *CLECC+* method.

### 3.6 Experiments and Results – Single-layered Social Network

The *CLECC* method was designed to work on multi-layered social networks. The most common *MSN*s are still single-layered social networks, so the method has to extract communities from such networks as well. Thus, the experiments were divided into two parts. In this one, the method will be analysed for the special case of *MSN*s which are *SSN*. There is a number of reference network for such a network, like the analysed before karate club [Zachary 77], and benchmarking systems like *GN* Benchmark [Girvan 02] or the recently best *LFR* Benchmark [Lancichinetti 08]. Therefore, it is much easier to evaluate the method and compare it at least to the methods specialized for *SSN*.

#### 3.6.1  Reference Networks

Apart from the presented karate club, two more datasets were utilized. The first one was *American College football*[13] [Girvan 02] – the network of American football games between Division I-A colleges during the regular season Fall 2000. Nodes in the network represent teams and edges represent games between the two teams they connect. What makes this network interesting is that it incorporates a known community structure. The teams are divided into conferences containing around 8–12 teams each. Games are more frequent between members of the same conference than between members of different conferences, with teams playing in average about seven games with the other teams of their own conference and four games with the teams form other conferences [Girvan 02].

The second dataset is a social network of the community of 62 bottlenose dolphins[14] living in Doubtful Sound, New Zealand. The network was derived by 7 years of observations of statistically significant frequent association between dolphin pairs [Lusseau 03], [Lusseau 04].

The mutual information measure for the analysed networks is presented in Table 3.5.

| Network name | *CLECC* | *CLECC+* | FFO [Lancichinetti 09c] | CFinder [Palla 05] |
|---|---|---|---|---|
| **Karate** | 0.723 | 0.840 | 0.690 | 0.170 |
| **Football** | 0.741 | 0.741 | 0.754 | 0.697 |
| **Dolphins** | 0.629 | 0.841 | 0.781 | 0.254 |

Table 3.5 The comparison of normalized mutual information measure for clustering results generated by *CLECC*, *CLECC+*, FFO [Lancichinetti 09c] and CFinder [Palla 05].

---

[13] Network data http://www-personal.umich.edu/~mejn/netdata/football.zip
[14] Network data http://www-personal.umich.edu/~mejn/netdata/dolphins.zip

The results of *CLECC* and *CLECC+* method are very similar to other existing methods and prove that *CLECC* method works correctly for real reference datasets. For the complete results for Football dataset see Appendix II.

### 3.6.2 *GN* Benchmark

The *GN* Benchmark was the first widely accepted and used benchmark for evaluation of group extraction methods proposed by Newman and Girvan in [Girvan 02]. The *GN* Benchmark is in fact the recipe for computer generated networks. The network should have 128 nodes divided into 4 communities 32 nodes in each group. The degree centrality of all nodes should be equal to 16 ($k_i = 16$), i.e. each node needs to have exactly 16 neighbours. Next $k_i^{(in)}$ members from the same community are randomly connected to node $i$ and remaining $k_i^{(out)} = k_i - k_i^{(in)}$ connections are randomly made to nodes from outside of the $i^{th}$ node community.

For the purpose of the experiment, ten networks have been generated for eight different values of ratio $\frac{k_i^{(out)}}{k_i}$ (called also mixing parameter see Section 3.6.3). The *CLECC* method has extracted the communities for each network and it was compared to 5 different well known methods (the data for comparison was provided by Mr Andrea Lancichinetti[15]). The results are presented in Table 3.6 and Figure 3.8.

| $k_i^{(out)} / k_i$ | **0.1** | **0.2** | **0.3** | **0.35** | **0.4** | **0.45** | **0.5** | **0.6** |
|---|---|---|---|---|---|---|---|---|
| **Blondel et al.** | 1.00 | 1.00 | 0.97 | 0.95 | 0.78 | 0.61 | 0.31 | 0.01 |
| **Clauset et al.** | 1.00 | 1.00 | 0.91 | 0.89 | 0.71 | 0.56 | 0.38 | 0.11 |
| *CLECC* | 1.00 | 1.00 | 1.00 | 0.79 | 0.61 | 0.42 | 0.35 | 0.21 |
| **Cfinder** | 0.89 | 0.58 | 0.41 | 0.37 | 0.24 | 0.20 | 0.11 | 0.03 |
| **GN** | 1.00 | 1.00 | 0.98 | 0.80 | 0.42 | 0.33 | 0.17 | 0.04 |
| **Radicchi et al.** | 1.00 | 1.00 | 0.91 | 0.52 | 0.09 | 0.01 | 0.00 | 0.00 |
| | | | | | | | | |
| **Standard deviation for *CLECC*** | 0.00 | 0.00 | 0.00 | 0.08 | 0.13 | 0.05 | 0.05 | 0.05 |

Table 3.6 Normalized mutual information measure comparison for *CLECC* method, Blondel et al., Clauset et al., Cfinder, GN and Radicchi et al. on *GN* Benchmark.

---

[15] https://sites.google.com/site/andrealancichinetti/

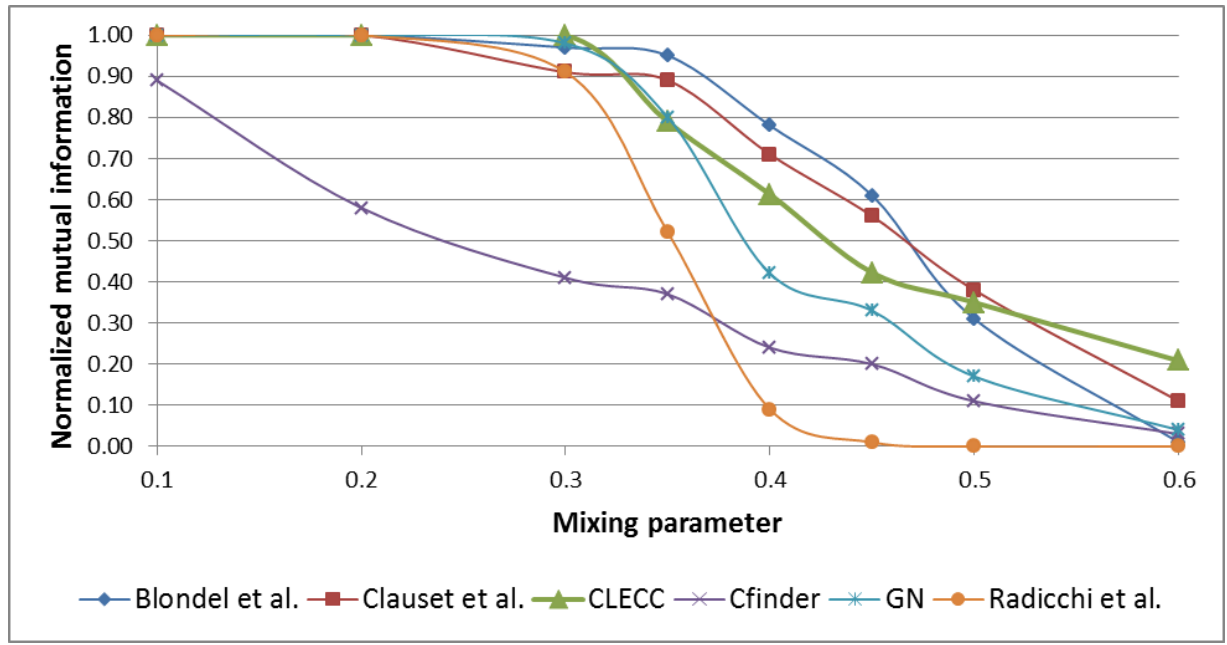

Figure 3.8 Normalized mutual information measure comparison of the *CLECC* method with Blondel et al., Clauset et al., Cfinder, GN, Radicchi et al. on *GN* Benchmark.

Just like for the real reference networks, *CLECC* method works properly for networks from *GN* Benchmark. Unfortunately, as it can be seen in Figure 3.8 there are methods which are better than the *CLECC* measure on *GN* Benchmark. However, the method does not deviate significantly from the best algorithms, especially for high values of $k_i^{(out)}/k_i$. Additionally, despite the fact that *CLECC* method was inspired by Radicchi et al. the results produced by those two methods are quite different.

*CLECC+* method had no effect on the clustering results since all the nodes were grouped by the *CLECC* method.

### 3.6.3  *LFR* Benchmark

*LFR* Benchmark (L – Lancichinetti, S – Fortunato, R – Radicchi) was proposed in [Lancichinetti 08]. *LFR* Benchmark is a special case of planted *l*-partition model [Condon 01] with different groups size and vertices degree. Both parameters follow typical distribution of real networks – the power law distribution, with exponents $\tau_1$, for nodes degree. and $\tau_2$, for community sizes. In addition, the procedure of graph creation in *LFR* Benchmark, allows generation of graphs with different sizes. Thus, the *LFR* Benchmark does not share main imperfections of the *GN* Benchmark [Girvan 02], which is constant size. The *LFR* become the new standard benchmark for generating artificial networks with the community structure.

Moreover, in [Lancichinetti 09c] authors have extended capabilities of *LFR* and provided the possibility to generate weighted and directed networks, with overlapping groups.

Except, exponents for power law distributions $\tau_1$ and $\tau_2$, the *LFR* Benchmark introduces several other parameters. The most important is mixing parameter $\mu$ which expresses the proportion between the external degree of a node $k_i^{(out)}$ and the total degree $k_i$ of the node *i*. The mixing parameter $\mu$ values can range from 0 to 1 inclusive, where setting the value of $\mu$ between 0 and 0.5 yields a community structure and for values above existence of communities is less likely. Setting the mixing parameter $\mu$ to 1 would cause a generation of completely random graph while selection 0 will lead to number of unconnected communities. Additional parameters are number of nodes in the network, average degree, maximal degree, maximal size of community and minimal size of community.

Now short explanation how the benchmark generate the undirected and unweighted networks will be presented, for complete description of all mechanisms and math behind the benchmark please see [Lancichinetti 08], [Lancichinetti 09c].

At the beginning the sizes of the communities are selected, by randomly picking the numbers from range [minimal size of community; maximal size of community] and with power law distribution where exponent is $\tau_2$. The sum of the sizes of the communities has to be equal to the number of nodes |V| in the *SSN*=<V,E> with exemption when the *SSN* with overlapping communities are generated. Next, each community is treated as an isolated graph and for each node the internal degree within community is assigned $k_i^{(in)} = (1-\mu)k_i$ where $k_i$ is node degree picked with power law distribution and exponent $\tau_1$. Of course, $k_i$ cannot exceed maximal degree for network and average degree for all nodes *i* in generated network need to be as close to average degree parameter, set for the network, as possible. In this way, each node *i* has the number of neighbours within community equal to $(1-\mu)k_i$ which need to be selected and connected to the node *i*. This is done according to the configuration model [Molloy 95], i.e., by randomly attaching pairs of randomly selected nodes to each other until there are no more "free" nodes in the community. Finally all communities within the network have to be connected to each other. In order to do this for each node *i* a number of external neighbours, equal to $k_i^{(out)} = \mu k_i$, need to be connected to this node according to the configuration model [Molloy 95]. Note that $k_i = k_i^{(in)} + k_i^{(out)} = (1-\mu)k_i + \mu k_i$, so in this way, the final graph satisfies the conditions which was set at the beginning on the distributions of degree and community size.

During experiments 10 networks for nine different values of mixing parameter {0.1, 0.2, 0.3, 0.4, 0.5, 0.6, 0.7, 0.8, 0.9} was generated using *LFR* Benchmark[16] and parameters used by Lancichinetti in [Lancichinetti 09a] namely: 1000 nodes, community size is between 10 and 50 nodes, the average degree is 20, the maximum degree is 50, the exponent of the degree distribution is 2, and that of the community size distribution is 1. For those networks the *CLECC* method was calculated and compared with 12 other methods analysed in [Lancichinetti 09a], i.e. Blondel et al. [Blondel 08], Cfinder [Palla 05], Clauset et al. [Clauset 04], DM [Donetti 04], EM [Eckmann 02], GN [Girvan 02], Infomap [Rosvall 08], Infomod [Rosvall 07], Radicchi et al. [Radicchi 04], RN [Ronhovde 09], MCL [Dongen 00a], Sim. Ann. [Guimera 05]. Once again the data for comparison was provided by Mr Andrea Lancichinetti. The results are presented in Table 3.7 and in Figures 3.9, 3.10, 3.11.

| Mixing parameter | 0.1 | 0.2 | 0.3 | 0.4 | 0.5 | 0.6 | 0.7 | 0.8 | 0.9 |
|---|---|---|---|---|---|---|---|---|---|
| *CLECC* | 0.89 | 0.88 | 0.88 | 0.91 | 0.93 | 0.80 | 0.64 | 0.52 | 0.43 |
| **Blondel et al.** | 1.00 | 1.00 | 1.00 | 0.99 | 0.97 | 0.93 | 0.54 | 0.02 | 0.00 |
| **Cfinder** | 1.00 | 0.99 | 0.98 | 0.93 | 0.76 | 0.48 | 0.25 | 0.07 | 0.00 |
| **Clauset et al.** | 0.83 | 0.60 | 0.43 | 0.29 | 0.20 | 0.12 | 0.01 | 0.00 | 0.00 |
| **DM** | 0.99 | 0.97 | 0.97 | 0.94 | 0.97 | 0.85 | 0.01 | 0.00 | 0.00 |
| **EM** | 0.80 | 0.60 | 0.37 | 0.13 | 0.01 | 0.00 | 0.00 | 0.00 | 0.00 |
| **GN** | 1.00 | 1.00 | 1.00 | 1.00 | 0.54 | 0.23 | 0.05 | 0.00 | 0.00 |
| **Infomap** | 1.00 | 1.00 | 1.00 | 1.00 | 1.00 | 1.00 | 0.72 | 0.00 | 0.00 |
| **Infomod** | 0.78 | 0.75 | 0.73 | 0.70 | 0.53 | 0.29 | 0.00 | 0.00 | 0.00 |
| **Radicchi et al.** | 1.00 | 1.00 | 1.00 | 1.00 | 0.16 | 0.00 | 0.00 | 0.00 | 0.00 |
| **RN** | 1.00 | 1.00 | 1.00 | 1.00 | 1.00 | 1.00 | 1.00 | 1.00 | 0.00 |
| **MCL** | 1.00 | 1.00 | 1.00 | 0.84 | 0.43 | 0.20 | 0.03 | 0.00 | 0.00 |
| **Sim. Ann.** | 0.96 | 0.94 | 0.90 | 0.84 | 0.76 | 0.65 | 0.47 | 0.00 | 0.00 |
| | | | | | | | | | |
| **Std dev for *CLECC*** | 0.039 | 0.045 | 0.047 | 0.062 | 0.016 | 0.028 | 0.022 | 0.010 | 0.015 |

Table 3.7 Normalized mutual information measure comparison on *LFR* Benchmark

---

[16] https://sites.google.com/site/andrealancichinetti/files

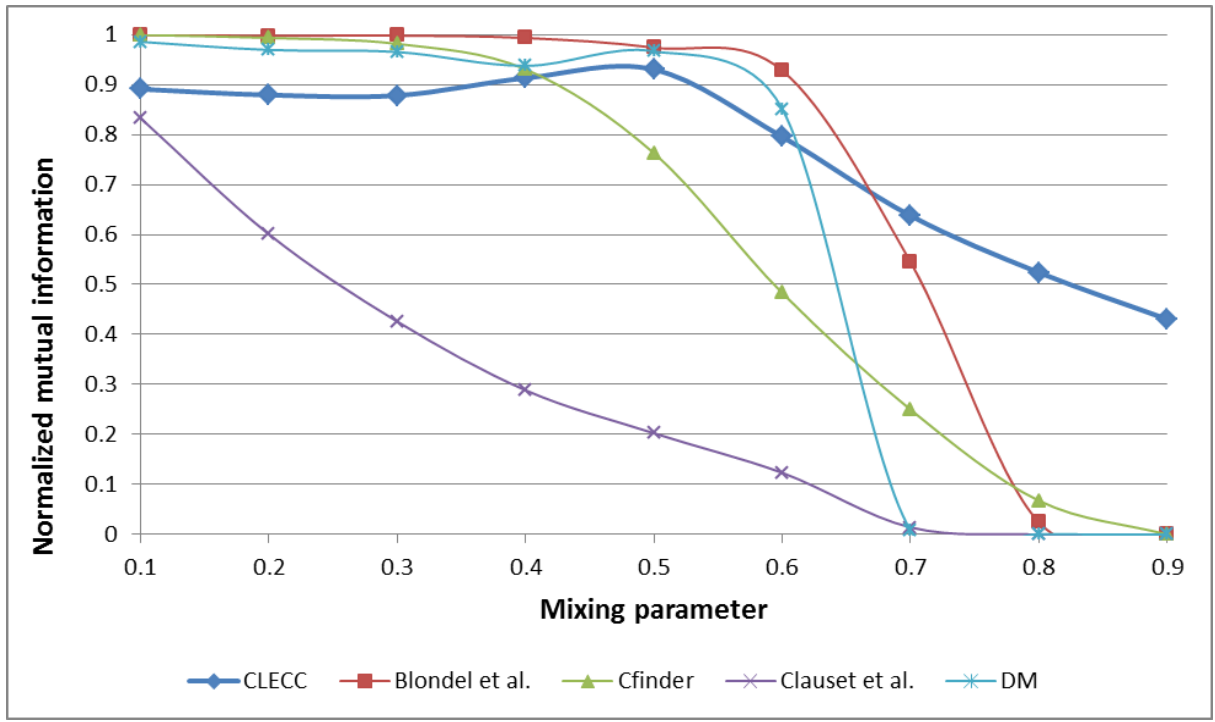

Figure 3.9 Normalized mutual information measure comparison for *CLECC*, Blondel et al., Cfinder, Clauset et al. and DM on *LFR* Benchmark.

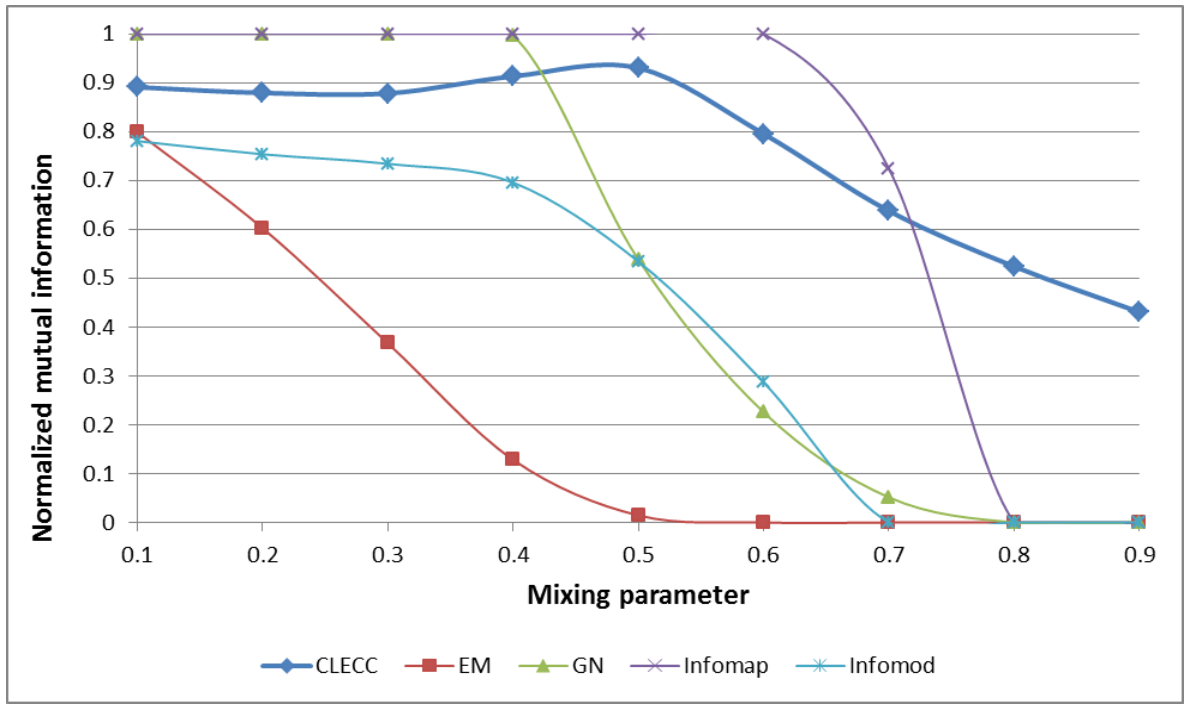

Figure 3.10 Normalized mutual information measure comparison for *CLECC*, EM, GN, Infomap and Infomod on *LFR* Benchmark.

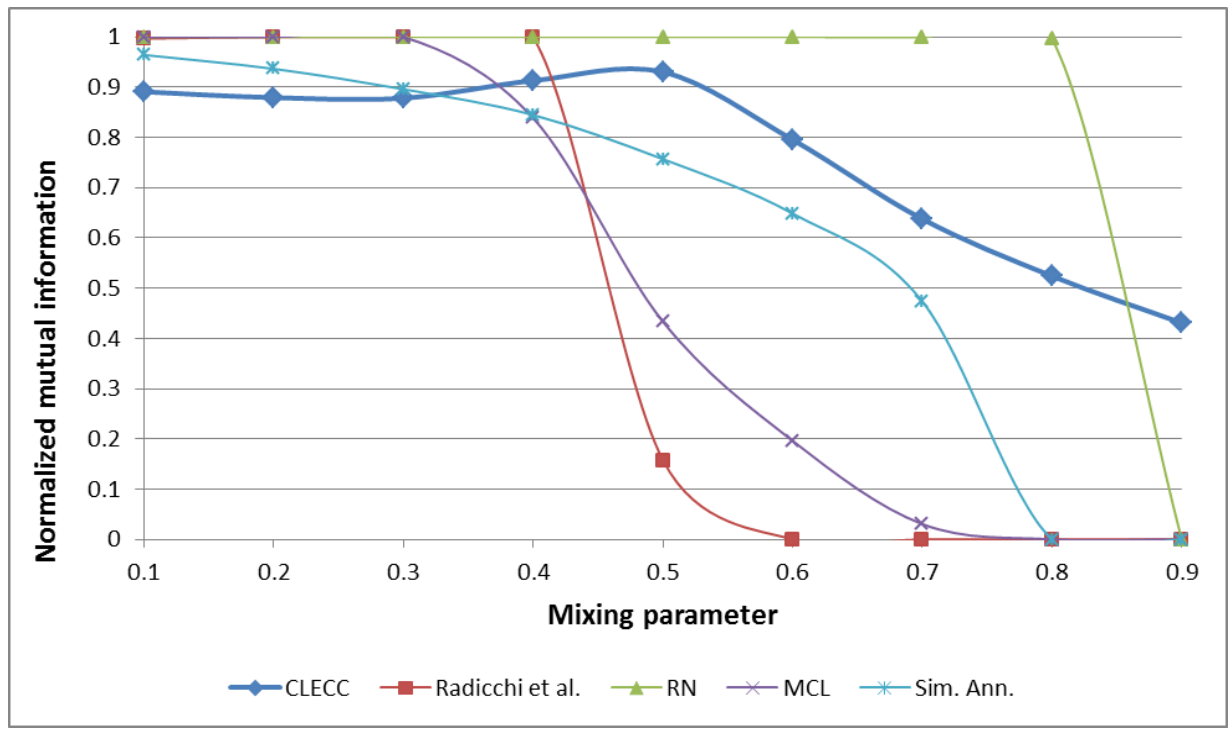

Figure 3.11 Normalized mutual information measure comparison for *CLECC*, Radicchi et al., RN, MCL and Sim. Ann. on *LFR* Benchmark.

Once again *CLECC* method is not the best method for *SSN* but it performs very well, especially for high values of mixing parameter when the group structure is weakly defined. Additionally, just like in case of *GN* Benchmark, despite the fact that *CLECC* method was inspired by Radicchi et al. the results produced by those two methods are quite different.

The *CLECC+* method had no effect on the clustering results.

## 3.7 Experiments and Results – Multi-layered Social Network

The second part of the experiments was performed utilizing networks extracted from virtual world and networks generated by the new, extended version of *LFR* Benchmark which is able to produce multi-layered networks.

### 3.7.1 Virtual world networks

The virtual world networks were extracted from polish social game Timik.pl. Timik is a virtual world where the users creates their avatars and rooms (home for each avatar). However, the most important future are public rooms created by admins where avatars can interact. Public room can represent almost everything: disco club, restaurant, school, hospital, street, park etc. Like in the real world, in the Timik people can talk, share ideas, sell items, borrow money, make friend etc. In the Figure 3.12 the exemplary room is presented, where users talk and protest together against ACTA (Anti-Counterfeiting Trade Agreement).

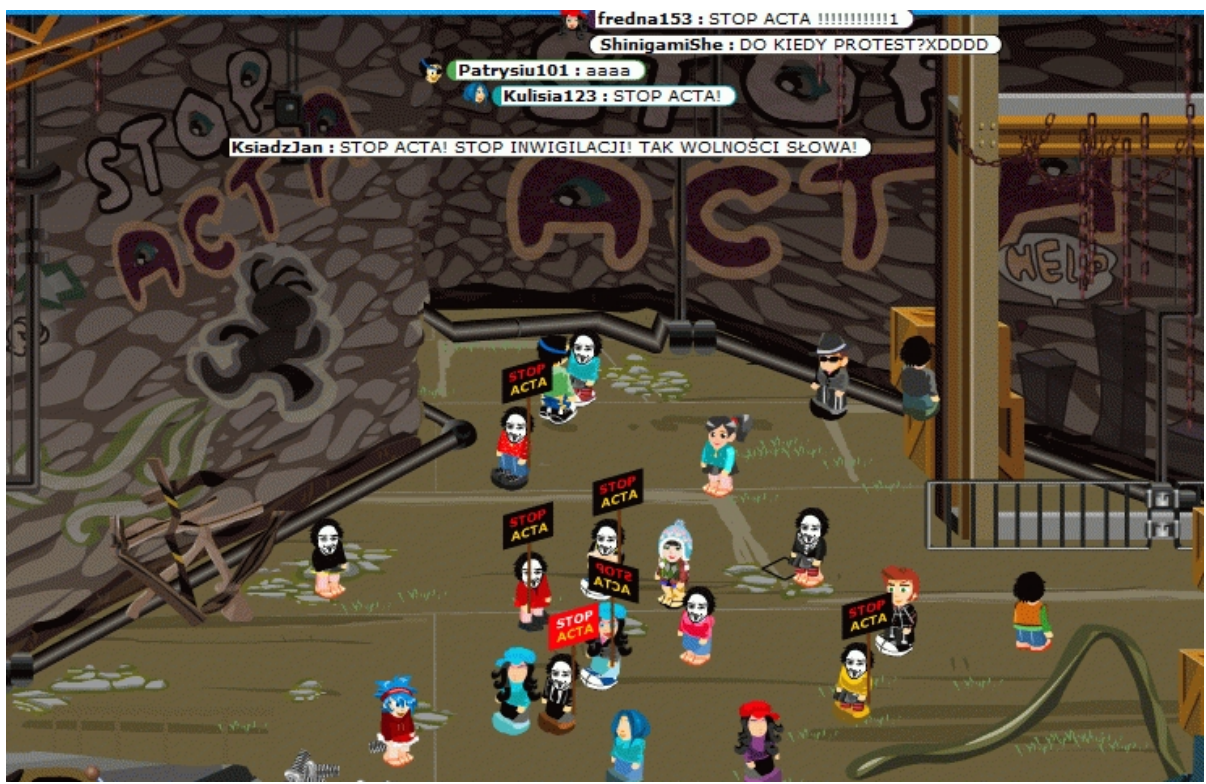

Figure 3.12 Protest against ACTA in Timik.pl.

The first network was extracted from virtual world during the *Clans Event*. During two month users were able to found the clan and invite other users to this clan or/and join clans created by other users because everyone could join many clans. Three layers has been extracted. The first one with 36,938 connections based on the private messages users have sent to each other. The second one with 2,824 connections based on virtual money transfers, and the third one with 178,193 connections based on friend lists. Overall the number of users participating in the event was 2,583. *CLECC* method was run three times for different values of $\alpha$ parameter $\alpha=1$, $\alpha=2$ and $\alpha=3$. The results are presented in Table 3.8.

|  | $\alpha = 3$ | $\alpha = 2$ | $\alpha = 1$ |
|---|---|---|---|
| **NMI** | 0.4592 | 0.4157 | 0.4406 |
| **Time [s]** | 192 | 2520 | 9000 |

Table 3.8 The results of normalized mutual information measure (NMI) for the multi-layered virtual world network.

The results are not impressive, this could be explained by the fact that users could join many clans, thus the groups overlap each other and *CLECC* method is designed to extract disjointed group. To check if using the *CLECC* method gives any new information it was applied to each layer separately, the results are presented below.

|  | private messages $l_1$ | virtual money transfers $l_2$ | friend lists $l_3$ |
|---|---|---|---|
| **NMI** | 0.4026 | 0.4022 | 0.3999 |
| **Time [s]** | 1,576 | 81 | 8,263 |
| **No. of edges** | 36,938 | 2,824 | 178,193 |

Table 3.9 The results of normalized mutual information measure (NMI) for each layer of the virtual world network.

The results of normalized information measure clearly indicates that using multi-layered network provides better results than analysing each layer separately. The best result is for the first layer and it is poorer than the worst result for $\alpha=2$. Moreover the computational time for the best multi-layered result ($\alpha=1$) is ten times better than the best result for single layer (layer no. 1). This indicate that *CLECC* method fulfils its main purpose i.e. provides new and better information about the analysed network (see Figure 3.13).

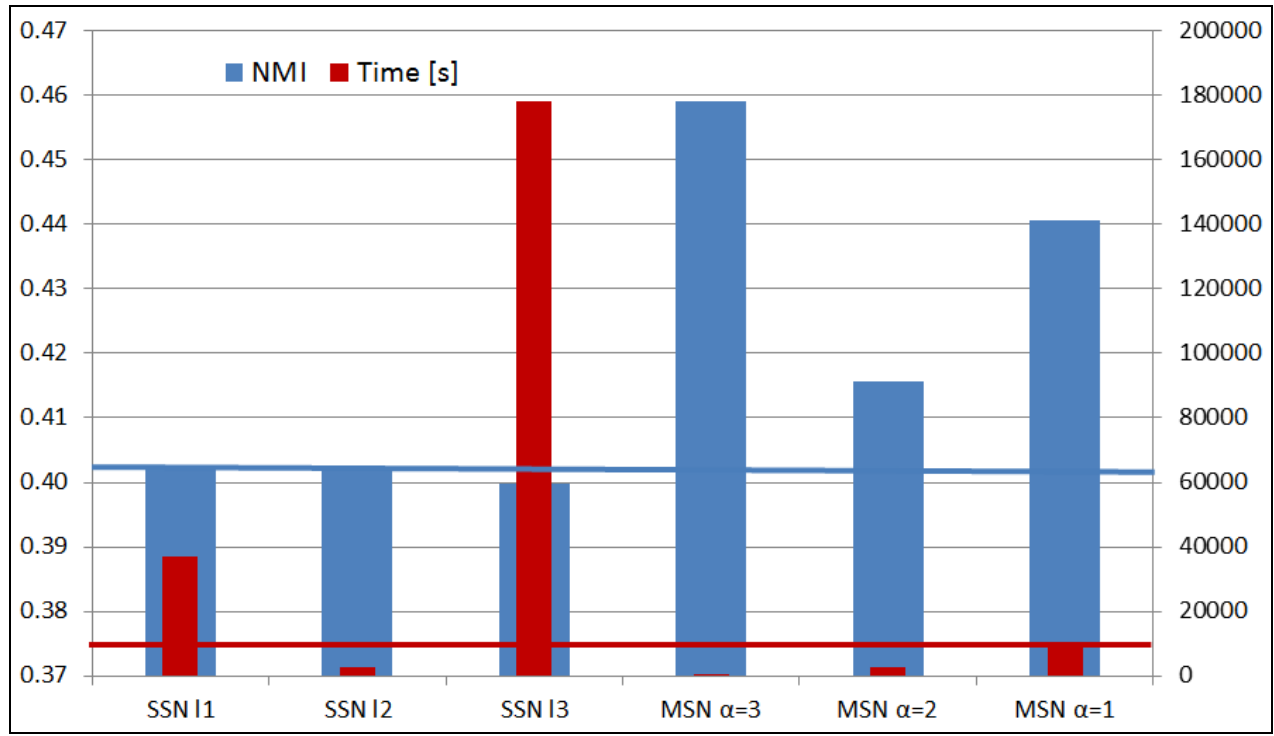

Figure 3.13 The comparison between the results for *SSNs* and *MSN*.

The second network from the Timik.pl virtual world was extracted during *Jelly Clans Event*. For the period of two months, users were able to join pre-defined jelly clans (each clan had different colour). The main difference between the *Jelly Clans Event* and the *Clans Event* is that, one person can join only one jelly clan. What is more, in order to get full membership in the clan, the five clan members with full membership had to agree in voting (except first few days when the game moderator has decided who can get full membership). Thus, except three layers from the previous cease (private messages, virtual money transfers and friend lists), there is a new layer created based on voting preferences. The number of participants was 10,034 and the number of edges for each layer is presented in Table 3.10. *CLECC* method was run four times for different values of $\alpha$ parameter $\alpha=1$, $\alpha=2$, $\alpha=3$ and $\alpha=4$. The results are presented in Table 3.10.

|          | $\alpha = 4$ | $\alpha = 3$ | $\alpha = 2$ | $\alpha = 1$ |
|----------|--------------|--------------|--------------|--------------|
| **NMI**  | 0.8178       | 0.4171       | 0.4235       | 0.4665       |
| **Time [s]** | 40       | 1349         | 7184         | 18187        |

Table 3.10 The Results of normalized mutual information measure (NMI) for the multi-layered virtual world network.

This time, the results are much better, probably thanks to voting layer on which members are connected only within the clan. To check if this assumption is correct the *CLECC* method was calculated on each layer separately. The results are presented in Table 3.11.

|  | private messages $l_1$ | virtual money transfers $l_2$ | friend lists $l_3$ | voting $l_4$ |
|---|---|---|---|---|
| **NMI** | 0.1662 | 0.2978 | 0.08420 | 0.5119 |
| **Time [s]** | 6,984 | 1,294 | 14,298 | 409 |
| **No. of edges** | 98,072 | 10,691 | 268,785 | 5,784 |

Table 3.11 The results of normalized mutual information measure (NMI) for each layer of the virtual world network.

Despite the fact that voting layer has the lowest number of edges it produces the best results because of its diversifying nature. However, once again using *MSN* creates better results than analysing one *SSN* (see Figure 3.14), but the difference is smaller than in *Clans* network.

For both networks *CLECC+* method has not improved the results.

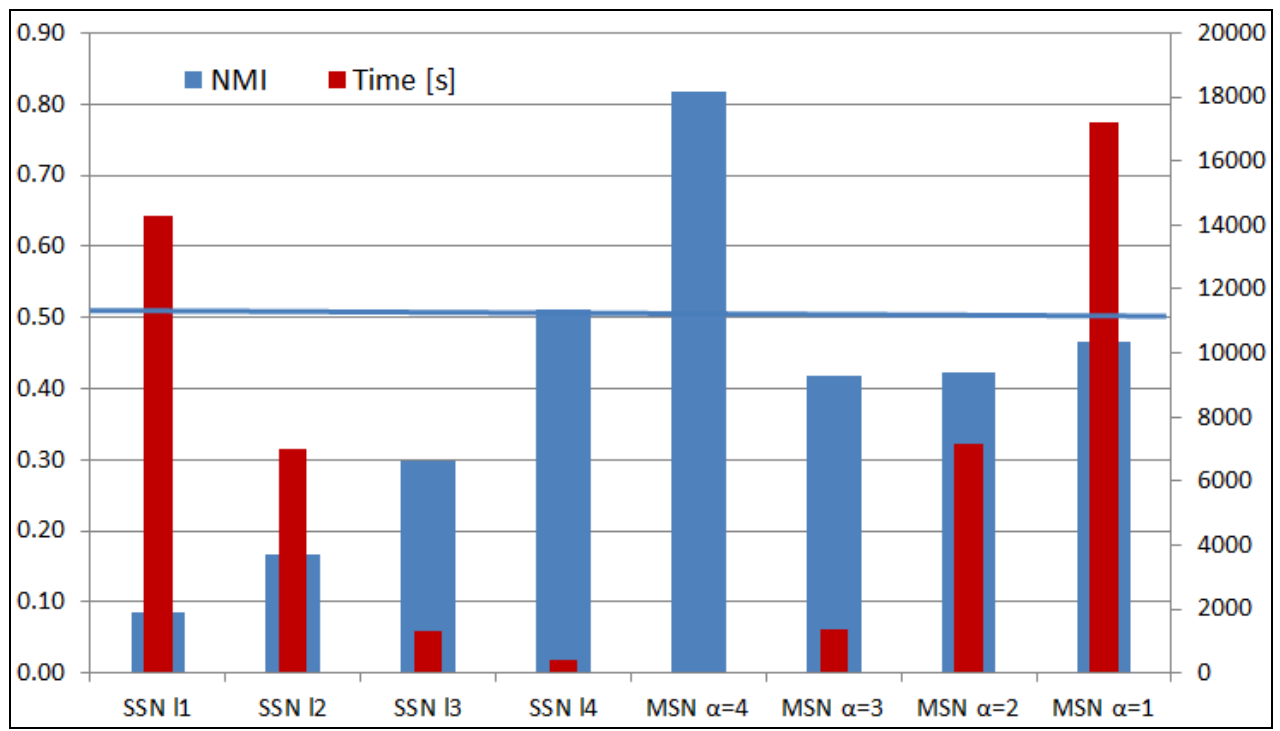

Figure 3.14 The comparison between *SSN* and *MSN* for *Jelly Clans* network.

### 3.7.2 m*LFR* Benchmark

For single-layered social network, there is a number of reference dataset like karate club [Zachary 77] or football league [Girvan 02], plus widely accepted and well tested benchmarks like *LFR* Benchmark [Lancichinetti 08], [Lancichinetti 09b]. Unfortunately for multi-layered social networks there is no reference data sets or benchmarks and because of that m*LFR* Benchmark has been created. The m*LFR* Benchmark is the extended *LFR* Benchmark which is able to generate multi-layered social networks (multi-layered *LFR* Benchmark).

When considering differences between edge distribution of vertex on a different layers, the internal degree $k_i^{(in)}$ of a vertex $i$ approaching to the community size $k_i^{(in)} \to S_G$ do not leave many possibilities when it comes to varying edges distribution between layers. In case of undirected networks it does not only affect the vertex with high internal degree, but $S_G - (k_i^{(in)} + 1)$ other vertices which are obligated to have vertex $i$ as their neighbour, thus further lowering possibility of effectuating changes in edge distribution of other vertices. The natural solution seems to be varying the internal degree distribution in community.

However, since the internal degree of the vertex $i$ is always a fixed fraction of its degree, far-reaching changes in internal degree distribution in community should affect the degrees of all vertices belonging to this community. The necessary changes for one community could be done, however, the algorithm would be restricted by the *LFR* Benchmark input parameters, i.e. average degree, maximum degree and, computed on the basis of previous two, minimal degree. Keeping the value of average degree set by user as an input parameter would made necessary to mirror changes in all other communities but with inversed direction, i.e. increasing degree of chosen vertices have to be followed by lowering degrees of other. This could be potentially dangerous for the community integrity.

This all makes benchmark networks generated by the *LFR* for the topology mixing parameter values approaching 0 potentially hard to vary between layers, when obeying the *LFR* Benchmark parameters. In order to cope with this task one could follow several directions. Due to the fact that layers of multi-layered social network considered separately are single-layered social networks it is tempting to just use *LFR* Benchmark several times, one for each layer of the multi-layered social network. However, even for the same input parameters generated networks can differ extremely, which would not be acceptable for the multi-layered social network. Therefore, each output would require a tuning to make layers similar to the satisfactory extent.

One of the *LFR* Benchmark achievements is linear computational complexity, but this could only be done if the network generation is not bound by the inner restrictions how the output network should look like. To be precise, the *LFR* Benchmark is restricted by the set of input parameters, but in general the process of the network generation is characterised by the high level of randomness. For example, the maximum, minimal and average degrees are influencing the power law distribution creation, but vertex degree is assigned by drawing a random number from the distribution. Similar situation is for the communities' sizes distribution. As a result assigning vertices to the communities is fitting randomly chosen vertices (having appropriate degree, but assigning degree is a random process) to communities. What is more, each community is in fact a random graph, thus even if the same vertices were assigned to the same communities through layers, the edges between vertices would be chosen on a random basis. The number of changes necessary to perform, makes the tuning process unrealistic or at least having objectionable computational costs.

The approach chosen for the created extension was following the *LFR* input parameter regardless the restrictions introduced by them, which significantly reduces the room for manoeuvre when designing extension. Still, it seems natural that any extension ought to obey restrictions of its base. Simultaneously, the changes effectuated were accepted whenever they do not increase the computational costs to the not acceptable level. Following this aims the extension considers the network generated by the *LFR* Benchmark as a base layer of multi-layered network. Then, the other layers are created with accordance to the base layer. Naturally, several ways of differentiating the layers were analysed.

The most obvious is different distribution of edges through the layers. Conforming to the aim of creating solution which provides the widest range of possibilities, algorithm can allocate edges correspondingly to the power law distribution. Algorithm, before inserting an edge joining vertex $x$ and vertex $y$ on a layer $l$, checks how many connections exist between vertices $x$ and $y$ on all layers. Then for given number of connections asks for a value of cumulative distribution function tuned to return 0 for no edges existing and 1 for connection on all layers. The new connection is accepted only if the randomly chosen value from 0 to 1, inclusively, obtained from random function is equal or greater than the value from cumulative distribution function.

Due to the aforementioned features of the m*LFR* Benchmark, instructing the extension to organize edge distribution following power law distribution does not necessarily guarantee that task can be successfully performed. In order to increase the odds of favourable solution and further increase dissimilarities between layers, the *LFR* Benchmark extension introduces two additional methods, one changing degrees of vertices through the layers and the second changing membership of vertex on layers.

The degree changing method swaps internal degrees of vertices belonging to the same community. It allows to avoid formerly mentioned restrictions given by the input parameters of *LFR* Benchmark, i.e. the vertices that consolidate community are still within the community and since, the changes are restricted to one community no changes has to be done to a graph as a whole or other communities. The vertices qualified to the transposing degree are chosen on a random basis, but the probability of triggering change can be set by the user. The value can be any real value from 0 to 1, inclusively, where 0.1 is a 10% chance of swapping degree. However, overlapping vertices are not considered by the algorithm and will not be changed. The overlapping vertex has their internal degree equally distributed between all communities it belongs to. Thus, any change in the internal degree of one community should be mirrored in all others communities to which overlapping vertex belong to. Even the minor change could destroy the balance between the sum if internal degrees and external degree of a vertex with overlapping memberships. Moreover, due to possibly different communities' sizes the change is not always possible. Even in case of vertices with single membership, before accepting transpose between vertex *x* with internal degree $k_x^{(in)}$ and vertex *y* with internal degree $k_y^{(in)}$ additional conditions have to be met. Firstly, external degree of any transposed vertex cannot exceed the internal degree of the other, i.e. $k_x^{(in)} < k_x^{(ext)}$ and $k_y^{(in)} < k_y^{(ext)}$. Next, the second vertex to transpose is chosen on a random basis, the transpose will not be accepted if *x=y*. Transposing only the internal degree of a vertex, breaks the rule of internal degree being a fixed fraction of the vertex degree. However, this is done because the *LFR* Benchmark considers the connections between communities as in fact another subgraph (each community is an individually considered subgraph). The degree transposition of external degrees could be done by considering it as one of the subgraphs. Moreover, since the changes are done within one community and situation where the external degree of a vertex exceeds its internal degree is prohibited it does not effectuate the integrity of the community.

Second method varies the membership of a vertex through the layers. As in case of the former one, qualification to the transposing degree is done on a random basis. Moreover, the similarity extends to the possibility of setting any chance of triggering the transpose. However, in this case there is no difference between overlapping vertices and ones having single membership, except that the overlapping vertices having the number of memberships equal to the number of communities will not be moved. The method does not perform any changes in the communities' degree sequences, however the degrees of vertices being transposed will be switched. This is because the method in fact changes the identifiers of the vertices, not two vertices, between two communities on a certain layer, i.e. the vertex $x$ with internal degree $k_x^{(in)}$ and external degree $k_x^{(ext)}$ with memberships to $G_x$, when switched with vertex $y$, receives its memberships $G_y$, internal degree $k_x^{(in)}$ and external degree $k_x^{(ext)}$. The method design is a result of an approach not to overlap functionality of methods. In this case, the degree changes are performed only by the former method, thus their frequency depends only on the former method as well. Naturally, due to the method varying memberships design, it does not effectuate input parameters of the *LFR* Benchmark.

The core part of the algorithm is responsible for distributing edges between vertices of the same group on different layers. Each group has to be considered separately, because of possible existence of overlapping vertices with internal degree shared by two or more groups. The algorithm preserves taken from *LFR* Benchmark equal division of edges between different communities. Not considering each group as a detached case could possibly lead to the situation when, for example, vertex with two memberships and internal degree equal 10 would have 8 edges assigned to the vertices from the first group and 2 to the second one. In this case this vertex would lose the sense of overlap.

The algorithm uses a layer created by the *LFR* Benchmark as a template (base layer) and tries to assign connections on other layers with power-law distribution. Power-law distribution was chosen after analysis of real multi-layered social network performed in [**Bródka** 11b], [**Bródka** 12a], [Kazienko and **Bródka** 10a] where it is shown that the number of layers on which vertex $x$ and $y$ are connected usually fallow the power-law distribution.

The most important part of the algorithm is a collection which based on its behaviour could be named as a sorted list of pairs ($k_{x,l}^{(in)}$, $x$), where $k_{x,l}^{(in)}$ is the internal degree of user $x$ on layer *l*. At the beginning each $k_{x,l}^{(in)} = k_x^{(in)}$ i.e. *x*'s internal degree. For each layer, except the base one, such list of pairs is created, each describing the same group on a different layer. All lists are enclosed in a map with keys being an identifier of a layer and a values being mentioned lists for a specific layer. For the group presented in the Figure 3.15 before distributing any edge, the list would consist of elements: [($k_{2,l}^{(in)}$,2), ($k_{5,l}^{(in)}$,5), ($k_{3,l}^{(in)}$,3), ($k_{3,l}^{(in)}$,4), ($k_{6,l}^{(in)}$,6), ($k_{1,l}^{(in)}$,1)] and if we replace the degree with numbers it will be [(2,2), (2,5), (3,3), (3,4), (4,6), (5,1)].

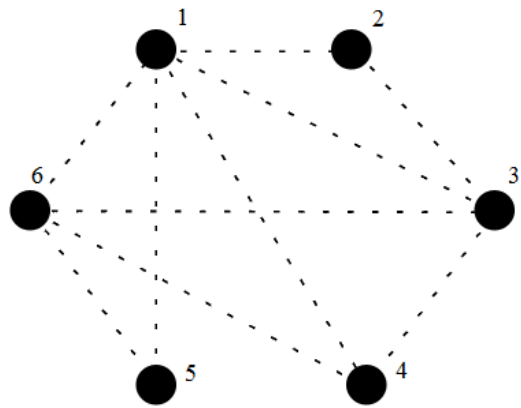

Figure 3.15 A community example presenting only internal edges. The number next to the vertex is its identifier.

The m*LFR* starts distributing edges with randomly choosing a layer, which the pair list is not empty. Then, it considers the last element of the pair list for a chosen layer and assigns number of neighbours equal to the internal degree of vertex *x*. Then the element describing vertex *x* is removed from the pair list and all neighbours assigned to the vertex *x* have their number of edges to distribute reduced by one. Then, the procedure is repeated for another randomly chosen layer. The changes of layers after each iteration preserves from skewed distribution. If the algorithm distributed all connections for one layer, then it would proceed to another, the first considered layer would have high probability of being similar to the base layer for all edges. Ideally, one should consider one edge at the time, then choose another layer. However, it can potentially block the distribution. The differences in approaches has been presented in the Table 3.12 and Table 3.13, which delineates the distribution of the same network in two cases: first, one edge at the time and second, fully distributing edges of the vertex, then proceeding to another vertex. It is worth noticing, that the network has only one possible distribution, where vertex 3 and vertex 4 are adjacent to all vertices in the network.

| Step | The pair list for chosen layer | Action |
|---|---|---|
| 1 | (2, 0), (3, 1), (3, 2), (4, 3), (4, 4) | Vertex 4 is connected with vertex 2 |
| 2 | (2, 0), (2, 2), (3, 1), (3, 4), (4, 3) | Vertex 3 is connected with vertex 4 |
| 3 | (2, 0), (2, 2), (2, 4), (3, 1), (3, 3) | Vertex 3 is connected with vertex 0 |
| 4 | (1, 0), (2, 2), (2, 4), (2, 3), (3, 1) | Vertex 1 is connected with vertex 0 |
| 5 | (2, 2), (2, 4), (2, 3), (2, 1) | Vertex 4 no longer can be joined with vertex 0 |

Table 3.12 The delineation of potential threats connected with distribution of edges considering one edge at the time. Algorithm fails to distribute all connections.

| Step | The pair list for chosen layer | Action |
|---|---|---|
| 1 | (2, 0), (3, 1), (3, 2), (4, 3), (4, 4) | Vertex 4 is connected with vertex 2<br>Vertex 4 is connected with vertex 3<br>Vertex 4 is connected with vertex 0<br>Vertex 4 is connected with vertex 1 |
| 2 | (1, 0), (2, 2), (2, 1), (3, 3) | Vertex 3 is connected with vertex 1<br>Vertex 3 is connected with vertex 0<br>Vertex 3 is connected with vertex 2 |
| 3 | (1, 1), (1, 2) | Vertex 2 is connected with vertex 1 |
| 4 | Fully connected network | |

Table 3.13 Presentation of the same example as in Table 5.1, but considering one vertex at the time, till full distribution of its "connections". As a result network is fully distributed.

Naturally, the presented distributions are only for demonstrative purpose. The edges for a considered vertex are not distributed on a random basis, but following chosen organization. This process can be divided into two parts. Firstly, converges the distribution to the one from base layer. This is done, only if the considered vertex $x$ belongs to the same community on both layers. The process of converging proposes randomly chosen neighbours of vertex $x$ on a base layer, to its equivalent on a considered layer. The new connection is accepted only if it is coincident with chosen statistical distribution. The base layer is a correctly connected community, thus, transmitting any of its edges to the considered layer does not carry the chance of blocking the algorithm. If any free connections are left after it, the second part, disposes yet not arranged connections for a considered vertex. The neighbours propositions are taken from the pair list, iterated from the element next to last, as it does not carry the chance of blocking the distribution, as shown in Table 3.13. However, such a solution would skew the organization of edges. In order to accommodate the both competing interests, the algorithm introduces the jumper variable, which holds information how many elements can be skipped while iterating the pair list to fully distribute connections and lower the chance of not possible allocation of edges. However, the situation is still possible, but occurs only for the vertices with internal degree approaching the community size. In this case, the algorithm will skip allocation of this edge. Moreover, the solution had to be implemented because the parameters describing the community created by the *LFR* Benchmark not always are valid and allows the full connection of a community. The similar solution, which in case of not possible distribution cuts the edge has been implemented in the *LFR* Benchmark. The pseudocode of m*LFR* Benchmark algorithm can be found in Appendix I.

### 3.7.3 Tests on m*LFR* Benchmark

At the beginning of the experiments, the networks with the number of layers equal to three, five, and seven were generated. For each number of layers, six different mixing parameters were selected (0.1, 0.3, 0.5, 0.7, 0.8, 0.9) and for each mixing parameter 10 networks were generated using m*LFR* Benchmark. Hence sixty 3-layers networks, sixty 5-layers networks and sixty 7-layers network were analysed. The benchmark parameters were the same like for single-layered network i.e. 1000 nodes, community size between 10 and 50 nodes, the average degree was 20, the maximum degree was 50, the exponent of the degree distribution was 2, and that of the community size distribution was 1. The new parameter namely exponent of the layers distribution was 2.

*CLECC* method was run for all possible values of $\alpha$ parameter for each network. The results for 3-layer social network are presented in Table 3.14, Figure 3.16 and Figure 3.17.

| The average | | Mixing parameter | | | | | |
|---|---|---|---|---|---|---|---|
| | | 0.1 | 0.3 | 0.5 | 0.7 | 0.8 | 0.9 |
| NMI | $\alpha=3$ | 0.81 | 0.81 | 0.72 | 0.63 | 0.58 | 0.59 |
| | $\alpha=2$ | 0.76 | 0.64 | 0.66 | 0.54 | 0.48 | 0.44 |
| | $\alpha=1$ | 0.80 | 0.76 | 0.74 | 0.59 | 0.27 | 0.28 |
| Time [s] | $\alpha=3$ | 52.2 | 336.6 | 241.8 | 97.8 | 19.8 | 16.8 |
| | $\alpha=2$ | 806.2 | 701.6 | 566.6 | 381.6 | 375.8 | 366.0 |
| | $\alpha=1$ | 1084.8 | 1293.4 | 1441.2 | 1443.4 | 1517.0 | 1493.2 |
| Standard deviation | | | | | | | |
| NMI | $\alpha=3$ | 0.0437 | 0.0370 | 0.0376 | 0.0167 | 0.0195 | 0.0037 |
| | $\alpha=2$ | 0.0465 | 0.0365 | 0.0136 | 0.0164 | 0.0226 | 0.0206 |
| | $\alpha=1$ | 0.0469 | 0.0196 | 0.0235 | 0.0132 | 0.0318 | 0.0208 |
| Time [s] | $\alpha=3$ | 20.60 | 44.60 | 51.63 | 34.01 | 2.71 | 1.16 |
| | $\alpha=2$ | 33.02 | 83.93 | 52.12 | 7.31 | 24.62 | 13.46 |
| | $\alpha=1$ | 79.76 | 116.93 | 101.79 | 53.15 | 46.32 | 35.22 |

Table 3.14 The average values and standard deviation of normalized mutual information measure (NMI) and executions time for different $\alpha$ parameter and mixing parameter values.

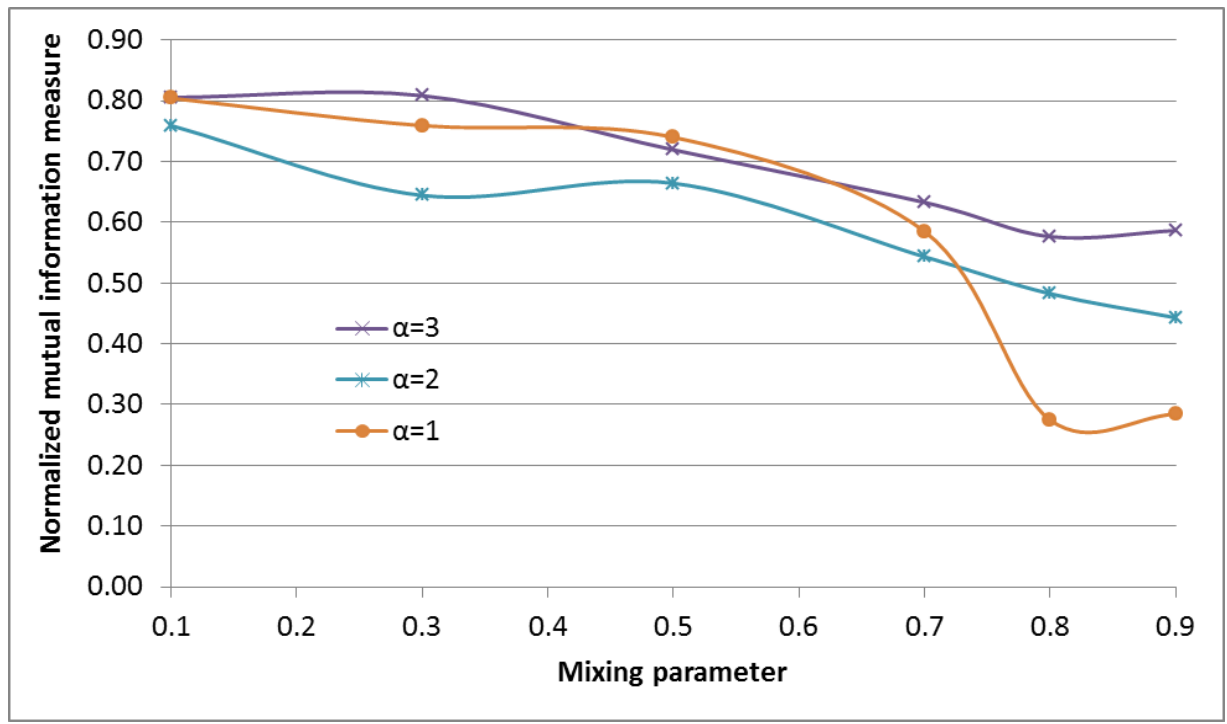

Figure 3.16 The average values of normalized mutual information measure (NMI).

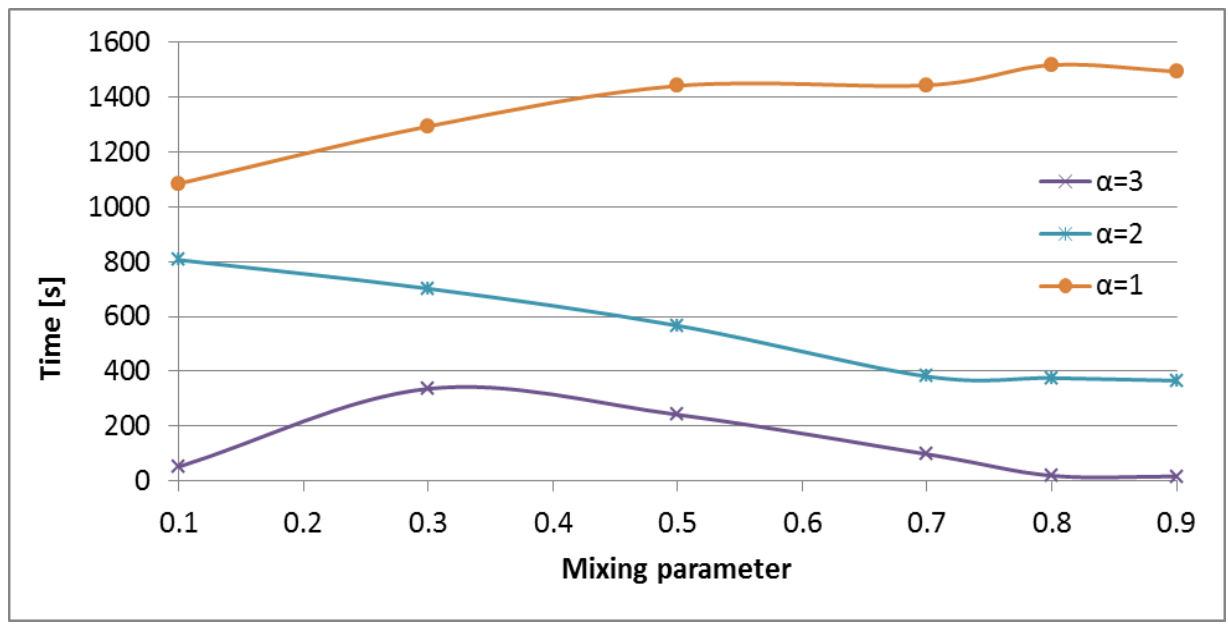

Figure 3.17 The average values of *CLECC* method executions time.

The best results are almost always for *α*=3 because of *CLECC* measure which for *α*>1 acts as a filter and removes week connection present only on one layer. If we consider for example a group of 50 people then even for mixing parameter equal to 0.7 there is much bigger chance that two users will be connected on many layers, than the probability that they are connected on many layers with some random person from the rest of network (950 people). Even though that each user has 70% relations going outside. Additionally the filtering produce sparser graph, so the *CLECC* method is very fast.

The Figure 3.17 and Figure 3.18. seems to confirm this. For *α*=3 and *α*=2 executions time are much lower than for *α*=1 and are decreasing with increasing mixing parameter i.e. when whole network structure approach the random graph model and more and more connections are filtered out. Additionally the results plot in Figure 3.17 behave quite similar to *CLECC* method plots in Figures 3.9 – 3.11 i.e. for *SSNs* generated using *LFR* Benchmark.

Different situation is with *α*=1. In this case the *CLECC* measure creates an union of all layers, producing denser graph and increasing the number of connections between communities, while the number of intercommunity connection remains almost the same. This is reflected in poorer results, especially for mixing parameter $\mu > 0.6$, and very long execution time which is the only one increasing. This might be explained by network structure approaching the random graph model, thus the union graph becomes denser.

However, very interesting thing seems to happened when mixing parameter is crossing the value of 0.8. The results for *α*=1 no longer deteriorate and even become a little better. This could be explained by the fact that from this value there is so few intercommunity connections that the union of layers also enhance a little the group density.

This conclusions was confirmed by the experiments on second set of networks namely 5 – layer social networks. The results are presented in Table 3.15, Figure 3.18 and Figure 3.19.

| The average | | Mixing parameter | | | | | |
|---|---|---|---|---|---|---|---|
| | | 0.1 | 0.3 | 0.5 | 0.7 | 0.8 | 0.9 |
| NMI | *α*=5 | 0.95 | 0.98 | 0.83 | 0.71 | 0.68 | 0.67 |
| | *α*=4 | 0.88 | 0.77 | 0.68 | 0.60 | 0.58 | 0.57 |
| | *α*=3 | 0.74 | 0.73 | 0.58 | 0.52 | 0.47 | 0.42 |
| | *α*=2 | 0.63 | 0.69 | 0.63 | 0.55 | 0.51 | 0.36 |
| | *α*=1 | 0.68 | 0.79 | 0.72 | 0.52 | 0.20 | 0.19 |
| Time [s] | *α*=5 | 41.6 | 52.0 | 43.0 | 17.2 | 9.2 | 8.4 |
| | *α*=4 | 607.6 | 693.2 | 429.2 | 172.8 | 68.0 | 25.0 |
| | *α*=3 | 901.6 | 971.0 | 600.0 | 398.2 | 316.8 | 283.4 |
| | *α*=2 | 1213.8 | 1539.6 | 1389.2 | 1216.0 | 1078.6 | 1051.0 |
| | *α*=1 | 1627.6 | 2063.6 | 2238.8 | 2440.8 | 2355.4 | 2367.8 |

| Standard deviation | | 0.1 | 0.3 | 0.5 | 0.7 | 0.8 | 0.9 |
|---|---|---|---|---|---|---|---|
| **NMI** | α=5 | 0.017 | 0.010 | 0.056 | 0.019 | 0.012 | 0.007 |
| | α=4 | 0.060 | 0.044 | 0.034 | 0.020 | 0.030 | 0.015 |
| | α=3 | 0.081 | 0.019 | 0.056 | 0.024 | 0.012 | 0.017 |
| **NMI** | α=2 | 0.055 | 0.020 | 0.036 | 0.027 | 0.040 | 0.065 |
| | α=1 | 0.041 | 0.018 | 0.027 | 0.031 | 0.069 | 0.038 |
| **Time [s]** | α=5 | 19.38 | 1.41 | 6.66 | 3.87 | 1.17 | 1.36 |
| | α=4 | 316.51 | 76.38 | 61.25 | 27.21 | 32.48 | 0.89 |
| | α=3 | 446.53 | 31.34 | 45.27 | 63.82 | 63.31 | 21.28 |
| | α=2 | 433.49 | 47.46 | 43.70 | 53.98 | 8.45 | 11.40 |
| | α=1 | 104.40 | 50.40 | 65.12 | 87.07 | 31.23 | 78.55 |

Table 3.15 The average values and standard deviation of normalized mutual information measure (NMI) and execution time in seconds for different *α* parameter and mixing parameter values.

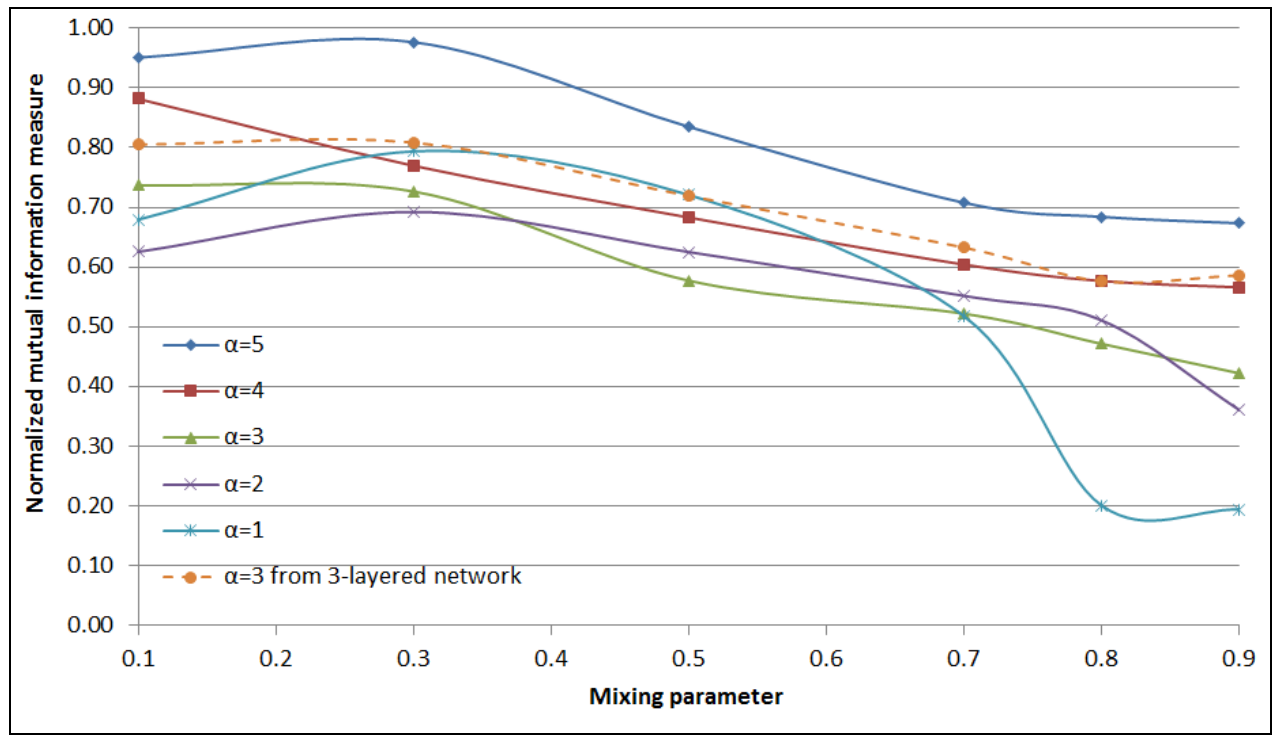

Figure 3.18 The average values of normalized mutual information measure (NMI).

As mentioned, the conclusion from 5-layer network analysis are the same like for previous 3-layer network. However, very interesting property of the *CLECC* method can be noticed while comparing the results from 5-layer network and 3-layer network. With the increasing number of layers in the network, the results of the method also become better. To visualise this, additional plot was added, in Figure 3.18, for results of the *CLECC* method for α=3 from the 3-layered network. The similar trend can be noticed for 7-layer network (Figure 3.20). Maybe it is not so impressive as the difference between 3 and 5-layered networks, but still it is noticeable.

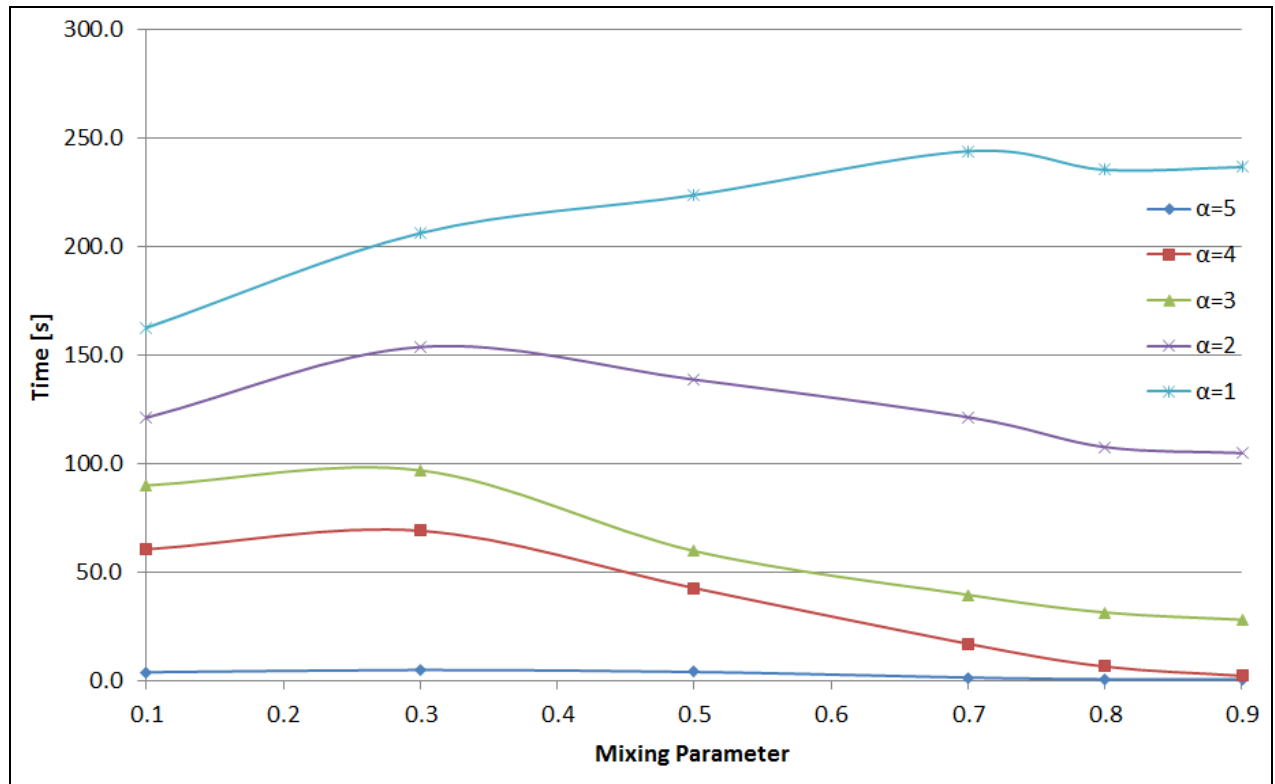

Figure 3.19 The average values of *CLECC* method execution time.

The rest of the results from 7-layered network analysis confirms the previous conclusions form analysis of the 3 and 5-layered network. The results for 7-layered network are in Table 3.16 and Figures 3.20 and 3.21.

| The average | | Mixing parameter | | | | | |
|---|---|---|---|---|---|---|---|
| | | 0.10 | 0.30 | 0.50 | 0.70 | 0.80 | 0.90 |
| NMI | α=7 | 0.99 | 0.98 | 0.83 | 0.72 | 0.70 | 0.68 |
| | α=6 | 0.94 | 0.86 | 0.78 | 0.65 | 0.68 | 0.66 |
| | α=5 | 0.83 | 0.79 | 0.65 | 0.61 | 0.58 | 0.58 |
| | α=4 | 0.78 | 0.69 | 0.65 | 0.55 | 0.48 | 0.44 |
| | α=3 | 0.74 | 0.64 | 0.59 | 0.59 | 0.54 | 0.44 |
| | α=2 | 0.74 | 0.68 | 0.66 | 0.59 | 0.54 | 0.37 |
| | α=1 | 0.79 | 0.75 | 0.72 | 0.37 | 0.14 | 0.12 |
| Time [s] | α=7 | 49 | 42 | 32 | 19 | 16 | 9 |
| | α=6 | 53 | 47 | 187 | 35 | 22 | 10 |
| | α=5 | 692 | 737 | 393 | 173 | 106 | 24 |
| | α=4 | 1118 | 1049 | 590 | 336 | 239 | 237 |
| | α=3 | 1486 | 1441 | 1345 | 877 | 717 | 678 |
| | α=2 | 1680 | 1872 | 1982 | 1746 | 1568 | 1566 |

|  |  | 2045 | 2477 | 2983 | 2864 | 2849 | 3614 |
|---|---|---|---|---|---|---|---|
|  | *α*=1 | | | | | | |
| **Standard deviation** | | **0.10** | **0.30** | **0.50** | **0.70** | **0.80** | **0.90** |
| NMI | *α*=7 | 0.008 | 0.015 | 0.022 | 0.021 | 0.019 | 0.012 |
|  | *α*=6 | 0.030 | 0.058 | 0.048 | 0.036 | 0.011 | 0.011 |
|  | *α*=5 | 0.039 | 0.017 | 0.027 | 0.034 | 0.018 | 0.008 |
|  | *α*=4 | 0.052 | 0.043 | 0.032 | 0.030 | 0.036 | 0.020 |
|  | *α*=3 | 0.100 | 0.040 | 0.021 | 0.021 | 0.048 | 0.028 |
|  | *α*=2 | 0.089 | 0.052 | 0.034 | 0.040 | 0.050 | 0.051 |
|  | *α*=1 | 0.081 | 0.057 | 0.015 | 0.181 | 0.023 | 0.021 |
| Time [s] | *α*=7 | 4.83 | 1.85 | 15.61 | 7.36 | 6.53 | 1.48 |
|  | *α*=6 | 4.21 | 1.96 | 81.54 | 9.09 | 10.59 | 0.89 |
|  | *α*=5 | 135.03 | 93.74 | 48.92 | 65.62 | 67.60 | 7.70 |
|  | *α*=4 | 77.67 | 124.99 | 37.73 | 19.48 | 78.62 | 75.36 |
|  | *α*=3 | 69.16 | 76.43 | 68.18 | 146.90 | 231.37 | 207.76 |
|  | *α*=2 | 66.26 | 94.30 | 72.97 | 177.80 | 404.74 | 411.77 |
|  | *α*=1 | 54.69 | 215.82 | 105.96 | 427.46 | 558.09 | 1737.36 |

Table 3.16 The average values and standard deviation of normalized mutual information measure (NMI) and execution time in seconds for different *α* parameter and mixing parameter values.

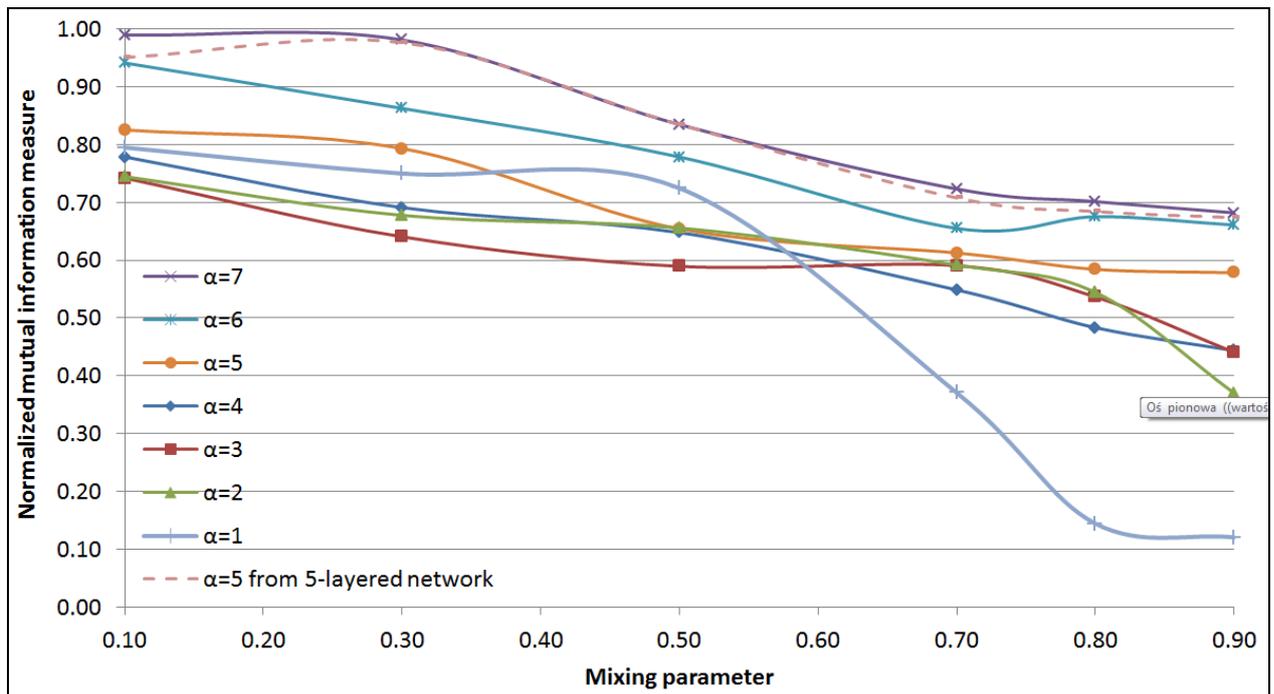

Figure 3.20 The average values of normalized mutual information measure (NMI).

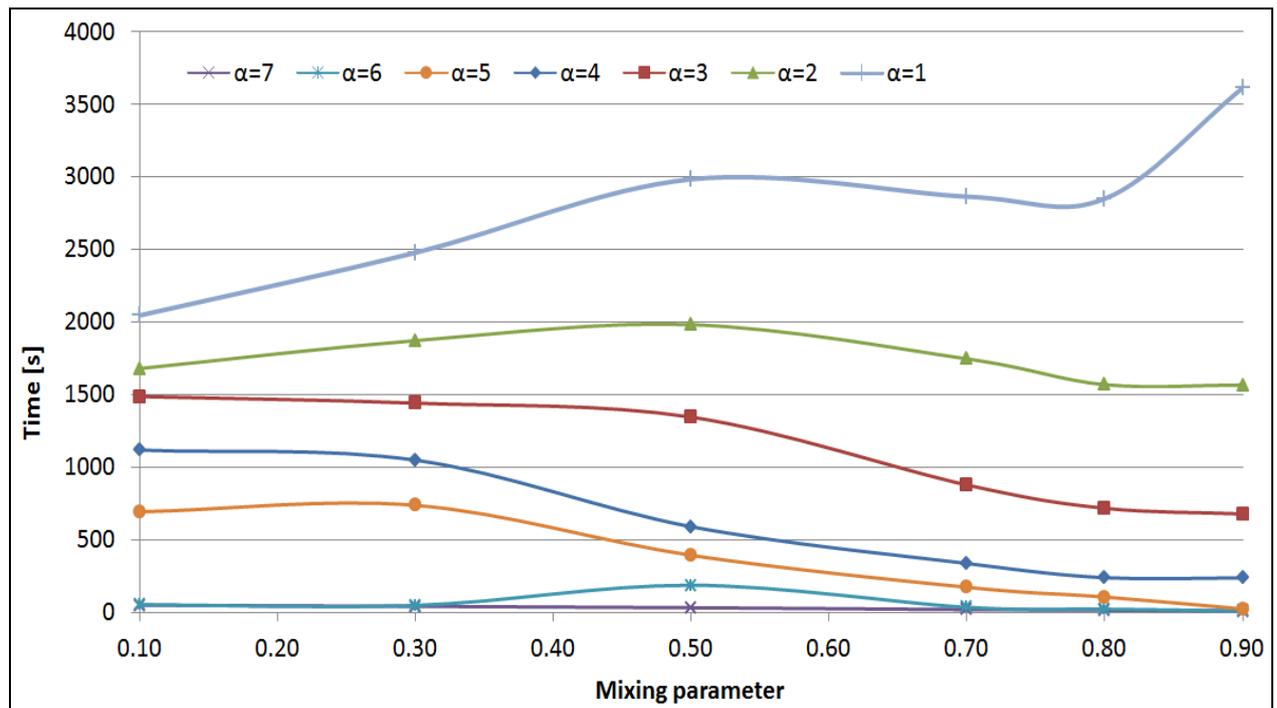

Figure 3.21 The average values of *CLECC* method execution time.

The *CLECC+* method had no effect on the clustering results for none of the analysed network.

## 4. Group Evolution in Social Networks

The continuous interest in the social network area contributes to the fast development of this field. The new possibilities of obtaining and storing data facilitate deeper analysis of the entire network, extracted social groups and single individuals as well. One of the most interesting research topic is the dynamics of social groups, it means analysis of group evolution over time. Additionally it seems to be the natural step after community extraction. Having extracted communities, appropriate knowledge and methods for dynamic analysis, one may attempt to predict the future of the group, and then manage it properly in order to achieve or change this predicted future according to specific needs. Such ability would be a powerful tool in the hands of human resource managers, personnel recruitment, marketing, telecommunication companies, etc.

As mentioned before the group extraction and their evolution are among the topics which arouse the greatest interest in the domain of social network analysis. However, while the group extraction methods for social networks are being developed very dynamically, what was described in the previous chapter, the methods of group evolution discovery and analysis are still 'uncharted territory' on the social network analysis map.

The ideas and research presented in this chapter was partially presented in [**Bródka** 11d], [**Bródka** 11e], [**Bródka** 12b].

### 4.1    Introduction to Group Evolution in Social Network

In recent years, several methods for tracking changes in social groups have been proposed. Sun et al. have introduced GraphScope [Sun 07], Chakrabarti et al. have presented another original approach in [Chakrabarti 06], Lin et al. have provided the framework called FacetNet [Lin 08] using evolutionary clustering, Kim and Han in [Kim 09] have introduced the concept of nono-communities, Hopcroft et al. have also investigated group evolution, but no method which can be implemented have been provided [Hopcroft 04].However two methods stand out from the rest Asur et al. [Asur 07] and Palla et al. [Palla 07].

#### 4.1.1   Asur et al. Method

Asur et al. in [Asur 07] have proposed a simple approach for investigating group evolution over time. At first, groups are extracted in each time frame, then comparing the size and overlapping of every possible pair of groups in consecutive time steps, the events involving those groups are assigned. When none of the nodes in the group from time step $T_i$ occurs in the following time frame $T_{i+1}$, Asur et al. have described this situation as dissolve of the group. In opposite to dissolve, if none of the nodes in the group from time frame Ti was present in the previous time frame $T_{i-1}$, a group is marked as new born. The group continues its existence when identical occurrence of the group in the consecutive time frames is found. Case, when two groups from time step $T_{i-1}$ joined together overlap or overlap each other with more than a given percentage of the single group in time frame $Ti$, is called merge. In the opposite case, when two groups from time frame $Ti$ joined together overlap greater than a given part of the single group in time frame $T_{i+1}$, the event is marked as split. Asur et al. did not specify what method has been used for group extraction or if the method works for overlapping groups.

### 4.1.2 Palla et al. Method

Palla et al. in [Palla 2007] have used clique percolation method [Palla 05] (Section 3.3.6), which allows groups to overlap. Thanks to this feature analysing changes in groups over time is very simple. Networks at two consecutive time frames $Ti$ and $T_{i+1}$ are merged into a single graph $Q(T_i, T_{i+1})$ and groups are extracted using the CFinder. Next, the communities from time frames $T_i$ and $T_{i+1}$, which are the part of the same group from the joined graph $Q(T_i, T_{i+1})$, are considered to be matching. It may happen that more than two communities are contained in the same group. Then, matching is performed based on the value of their relative overlap sorted in descending order. Possible events between groups are: growth, contraction, merging, splitting, birth and death. Using the CFinder allowed Palla et al. to investigate evolution in overlapping groups, which can be extracted from the directed as well as weighted network.

### 4.1.3 Dynamic Social Network

In this thesis a list of following time frames (time windows) $T$, where each time frame is in fact a single-layered social network $SSN(V,E_l,\{l\})$, or multi-layered social network $MSN(V,E,L)$ is called a dynamic social network $DSN$:

$$\begin{aligned} DSN &= <T_1, T_2, ...., T_t>, \quad t \in N \\ T_i &= MSN_i(V_i, E_i, L), \quad i = 1, 2, ..., t, \quad L - constatnt \\ E_i &= <x, y, l>: x, y \in V_i, \quad i = 1, 2, ..., t, l \in L \end{aligned} \quad (4.1)$$

An example of a dynamic social network *DSN* for simple *SSN* is presented in Figure 4.1. It consists of five time frames, and each time frame is a separate social network created from data gathered in the particular interval of time. In the simplest case, one interval starts when the previous interval ends, but based on author's needs the intervals may overlap by a set of time or can even contain full history of previous time frames in the aggregated form. The concept of social network spited into the list of successive time frames has been proposed for the first time by White in [White 92], and since then is most often used to present social network dynamic.

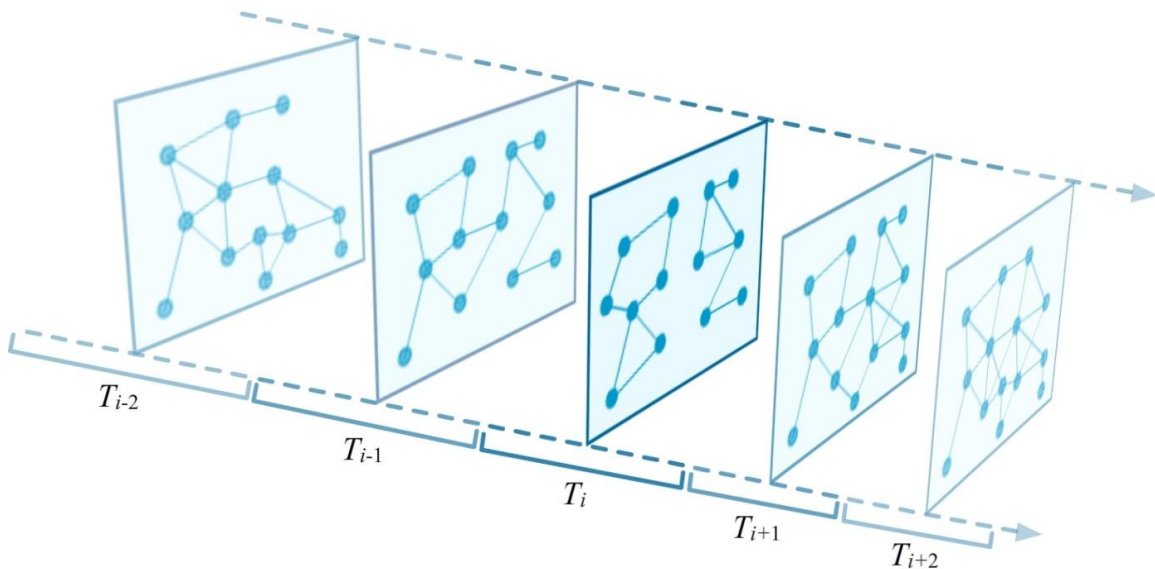

Figure 4.1 The example of dynamic social network consisting of five time frames.

### 4.1.4  Group Evolution

Methods for group evolution discovery are often dependent on clustering algorithm (Pala et. al. and CFinder) or type of communities (Asur et. al. and disjointed groups). But, we would like from the new method to work independently from clustering algorithms. Thus from now on the relaxed definition of group will be used i.e. A group *G* extracted from the multi-layered social network *MSN*(*V*,*E*,*L*) is a subset of vertices from *V* (*G*⊆*V*), extracted using *any community extraction method* (clustering algorithm).

Group evolution is a sequence of events (changes) succeeding each other in the consecutive time windows (time frames) within the social network. Palla et al. in [Palla 07] and Asur et al. in [Asur 07] have proposed some types of events but their lists were incomplete. Thus, in this paper, the possible list of events in social group evolution was extended. Seven independent types of events have been identified changing the state of a group or groups between two following time windows (see Figure 4.2):

1. *Continuing* (stagnation) – a group continues its existence, when two groups in the consecutive time windows are identical or when two groups differ only by few nodes but their size remains the same.

2. *Shrinking* – a group shrinks when some nodes have left the group, making its size smaller than in the previous time window. A group can shrink slightly, i.e. by a few nodes or greatly losing most of its members.

3. *Growing* (opposite to shrinking) – a group grows when some new nodes have joined the group, making its size bigger than in the previous time window. A group can grow slightly as well as significantly, doubling or even tripling its size.

4. *Splitting* – a group splits into two or more groups in the next time window $T_{i+1}$, when some groups from time frame $T_{i+1}$ consist of members of one group from the previous time frame $T_i$. We can distinguish two types of splitting: (1) *equal split*, which means the contribution of all resulting groups in the splitting group is almost the same and (2) *unequal split* when one of the final groups has much greater contribution in the splitting group, what in turn for this greater group might be similar to shrinking.

5. *Merging*, (reverse to splitting) – a group has been created by merging several other groups when one group from time frame $T_{i+1}$ consists of two or more groups from the previous time frame $T_i$. Merge, just like the split, might be (1) *equal*, when the contribution of all source groups in the merged, target group is almost the same, or (2) *unequal*, if one of the groups has much greater contribution into the merged group. In second case, for the biggest group the merging might be similar to growing.

6. *Dissolving* happens when a group ends its life and does not occur in the next time window at all, i.e. its members have vanished or stopped communicating with each other and are scattered among the rest of the groups.

7. *Forming* of the new group (opposite to dissolving) occurs when a group, which did not exist in the previous time window $T_i$, appears in next time window $T_{i+1}$. When a group remains inactive over several time frames; such case is treated as dissolving of the first group and forming again of the second, new one.

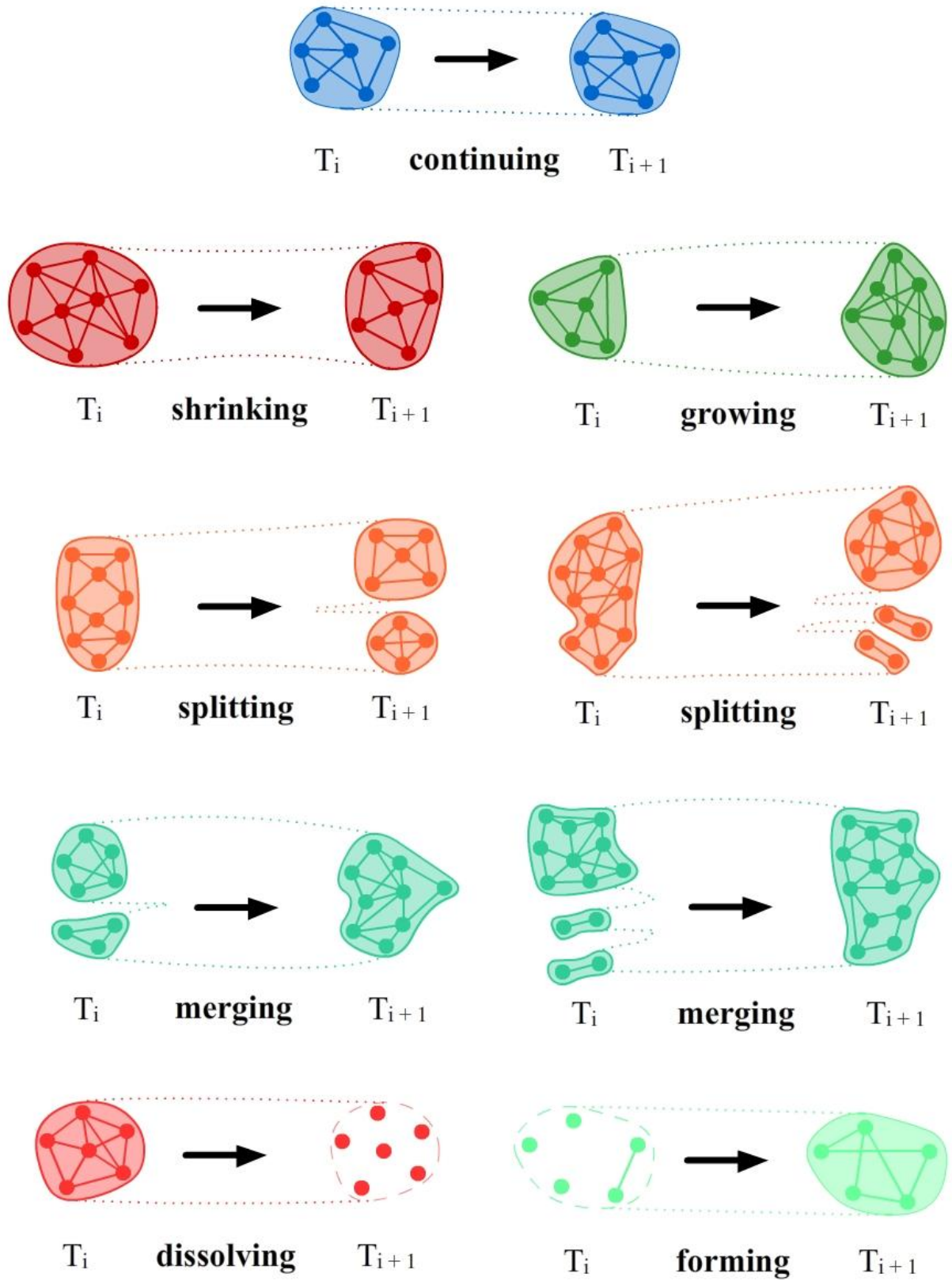

Figure 4.2 Seven possible types of events in the group evolution.

## 4.2 Method of Group Evolution Discovery

To discover group evolution in the social network a new method called *GED* (Group Evolution Discovery) was developed. The most important component of this method is a new measure called *inclusion*. This measure allows to evaluate the inclusion of one group in another. Therefore, inclusion $I(G_1,G_2)$ of group $G_1$ in group $G_2$ is calculated as follows:

$$I(G_1,G_2) = \overbrace{\frac{|G_1 \cap G_2|}{|G_1|}}^{group\ quantity} \cdot \underbrace{\frac{\sum_{x \in (G_1 \cap G_2)} NI_{G_1}(x)}{\sum_{x \in (G_1)} NI_{G_1}(x)}}_{group\ quality}, \qquad (4.2)$$

where $NI_{G_1}(x)$ is the value reflecting importance of the node $x$ in group $G_1$.

As a node importance $NI_{G_1}(x)$ measure any metric which indicate member position within the community can be used, e.g. centrality degree, betweenness degree, page rank, social position etc. (see Section 2.2.1). The second factor in Equation 4.2 would have to be adapted accordingly to selected measure. For example, if social position measure (Equation 2.9) is utilized, the Equation 4.2 will be as follows:

$$I(G_1,G_2) = \overbrace{\frac{|G_1 \cap G_2|}{|G_1|}}^{group\ quantity} \cdot \underbrace{\frac{\sum_{x \in (G_1 \cap G_2)} SP_{G_1}(x)}{\sum_{x \in (G_1)} SP_{G_1}(x)}}_{group\ quality}.$$

The *GED* method, used to discover group evolution, respects both the quantity and quality of the group members. The *quantity* is reflected by the first part of the *inclusion* measure, i.e. what portion of members from group $G_1$ is in group $G_2$, whereas the *quality* is expressed by the second part of the *inclusion* measure, namely what contribution of important members from group $G_1$ is in $G_2$. It provides a balance between the groups that contain many of the less important members and groups with only few but key members.

One might say that inclusion measure is "unfair" for not identical groups, because if the community differs even by only one member, inclusion is reduced through not having all nodes and also through not having social position of those nodes. Indeed, it is slightly "unfair" (or rather strict), but using member position within the community calculated on the basis of users relations, makes *inclusion* to focus not only on nodes (members) but also on edges (relations) giving great advantage over methods, which are using only members' overlapping for event identification (group quantity factor in inclusion measure).

It is assumed that only one event may occur for two groups ($G_1$, $G_2$) in the consecutive time frames, however, one group in time frame $T_i$ may be involved in several events with different groups in $T_{i+1}$.

The procedure for the Group Evolution Method (*GED*) is as follows:

---

### *The GED Method*

**Input:** Dynamic social network *TSN*, in which groups are extracted by any community detection algorithm separately for each time frame $T_i$ and any node importance measure is calculated for each group.

1. For each pair of groups <$G_1$, $G_2$> in consecutive time frames $T_i$ and $T_{i+1}$ inclusion $I(G_1,G_2)$ for $G_1$ in $G_2$ and $I(G_2,G_1)$ for $G_2$ in $G_1$ is computed according to Equations 4.2.

2. Based on both inclusions $I(G_1,G_2)$, $I(G_2,G_1)$ and sizes of both groups only one type of event may be identified:

    a. *Continuing*: $I(G_1,G_2) \geq \alpha$ and $I(G_2,G_1) \geq \beta$ and $|G_1| = |G_2|$

    b. *Shrinking*: $I(G_1,G_2) \geq \alpha$ and $I(G_2,G_1) \geq \beta$ and $|G_1| > |G_2|$ OR $I(G_1,G_2) < \alpha$ and $I(G_2,G_1) \geq \beta$ and $|G_1| \geq |G_2|$ and there is only one match (matching event) between $G_2$ and all groups in the previous time window $T_i$

    c. *Growing*: $I(G_1,G_2) \geq \alpha$ and $I(G_2,G_1) \geq \beta$ and $|G_1|<|G_2|$ OR $I(G_1,G_2) \geq \alpha$ and $I(G_2,G_1) < \beta$ and $|G_1| \leq |G_2|$ and there is only one match (matching event) between $G_1$ and all groups in the next time window $T_{i+1}$

    d. *Splitting*: $I(G_1,G_2) < \alpha$ and $I(G_2,G_1) \geq \beta$ and $|G_1| \geq |G_2|$ and there is more than one match (matching event) between $G_2$ and all groups in the previous time window $T_i$

    e. *Merging*: $I(G_1,G_2) \geq \alpha$ and $I(G_2,G_1) < \beta$ and $|G_1| \leq |G_2|$ and there is more than one match (matching event) between $G_1$ and all groups in the next time window $T_{i+1}$

f.  *Dissolving*: for $G_1$ in $T_i$ and each group $G_2$ in $T_{i+1}$  $I(G_1,G_2) < 10\%$ and $I(G_2,G_1) < 10\%$

g.  *Forming*: for $G_2$ in $T_{i+1}$ and each group $G_1$ in $T_i$  $I(G_1,G_2) < 10\%$ and $I(G_2,G_1) < 10\%$

The scheme, which facilitates understanding of the event selection (identification) for the pair of groups in the *GED* method is presented in Figure 4.3.

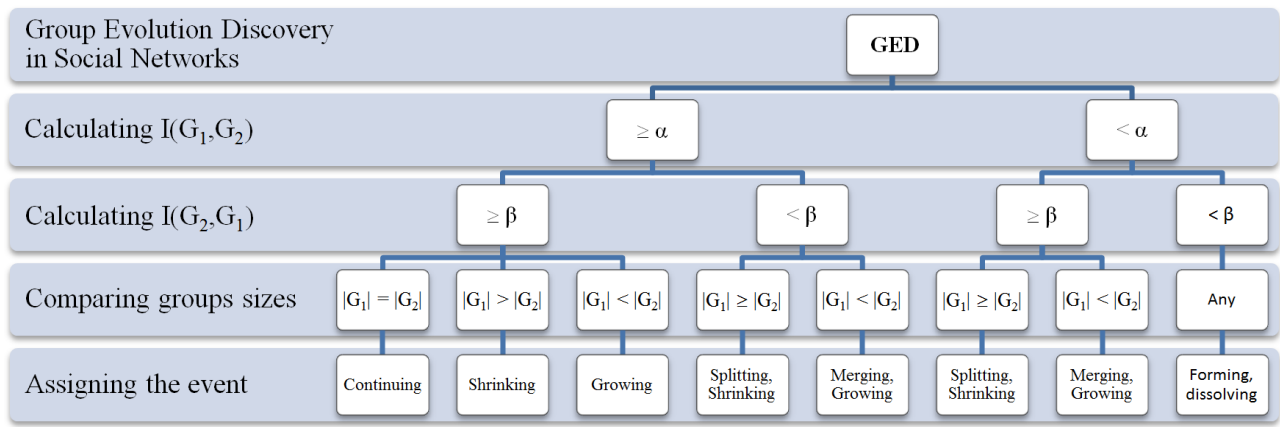

Figure 4.3 The decision tree for assigning the event type to a pair of groups.

$\alpha$ and $\beta$ are the *GED* method parameters, which can be used to adjust the method to the particular social network and community detection method. According to experimental analysis (see section 4.3) the values of $\alpha$ and $\beta$ from the range [50%;100%] are recommended.

Based on the list of extracted events, which have occurred for the selected group between each two successive time frames, the whole group evolution process may be created. In the sample social network in Figure 4.4 and Table 4.1, its lifetime consists of eight time windows. The group forms in $T_2$, then it grows in $T_3$ by gaining some new nodes, next it splits into two groups in $T_4$, afterwards the bigger group is shrinking in $T_5$ by losing one node, both groups continue over $T_6$ and they both merge with the third group in $T_7$, finally the group dissolves in $T_8$.

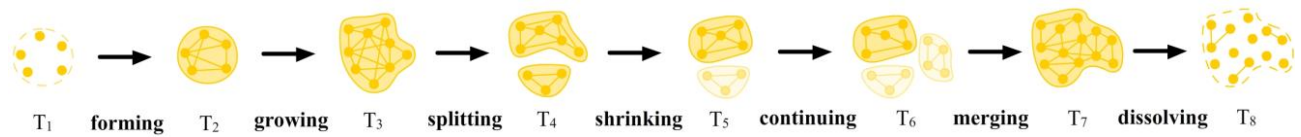

Figure 4.4 Changes over time for the single group.

| Event type | Group in $T_2$ | Event type | Group in $T_3$ | Event type | Group in $T_4$ | Event type | Group in $T_5$ | Event type | Group in $T_6$ | Event type | Group in $T_7$ | Event type |
|---|---|---|---|---|---|---|---|---|---|---|---|---|
| form | $G_1$ | grow | $G_1$- | split | $G_2$ | shrink | $G_2$ | continue | $G_2$ | merge | $G_5$ | dissolve |
| - | - | - |  | - | $G_3$ | continue | $G_3$ | continue | $G_3$ |  |  | - |
| - | - | - | - | - | - | - | - | form | $G_4$ |  |  | - |

Table 4.1 Changes over time for the single group.

The development of the *GED* method was also presented in [**Bródka** 11d], [**Bródka** 11e], [**Bródka** 12b].

## 4.3  Experiments and Results

The experiments were conducted on the data gathered from Wroclaw University of Technology email communication. The whole data set was collected within period of February 2006 – October 2007 and consists of 5,845 nodes (university distinct email addresses) and 149,344 edges (emails send from one address to another).

The dynamic social network consisted of fourteen 90-day time frame extracted from this source data. Timeframes have the 45-day overlap, i.e., the first time frame begins on the 1st day and ends on the 90th day, the second begins on the 46th day and ends on the 135th day and so on.

### 4.3.1  Experiment Based on Overlapping Groups Extracted by CFinder

In the first experiment, as a method for group extraction, *CFinder* was utilized (http://www.cfinder.org/). The groups were discovered for *k*=6 and for the directed and unweighted social network. The *CFinder* algorithm has extracted from 80 to 136 groups for different time windows (avg. 112 per time window). The average size of the group was 19 nodes. The smallest group had size of 6, because of the *k* parameter and the biggest one was of 613 in time window 10.

#### The Asur et al. Method

This method has been implemented in T-SQL language. The authors have suggested to set 30% or 50% as an overlapping threshold for merge and split. In the experiment, the threshold was set to 50%. It took more than 5.5 hours to calculate events between groups in all fourteen time frames. The total number of events found by Asur et al. method was 1,526, out of which 90 were continuation, 18 – forming, 29 – dissolving, 703 – merging and 686 were splitting.

Such a small number of continuing events is caused by the very rigorous condition, which requires the groups to remain unchanged. Small amount of forming (dissolving) events came from another strong condition, which states that none of the nodes from the considered group can exist in the network in previous (following) time window. A huge number of merging (splitting) events is a result of low overlapping threshold for merge (split).

However, it has to be noticed that these numbers are slightly overestimated. The method by Asur et al. allows one pair of groups to assign more than one type of events. This leads to anomalies when e.g. the group no. 1 in time window no. 1 ($T_1$) is continuing in group no. 2 in $T_2$ and simultaneously merging with group no. 13 from $T_1$ into group no. 2 in $T_2$. This should not happen if the condition for continuing is so rigorous.

The total number of anomalies is 128 cases, 8% of all results. More than a half of these cases are groups with *split* and *merge* event into another group at the same time. The rest of the cases are even worse, because one group has *continue* and *split* or *merge* event into another group simultaneously. Therefore, the total number of "distinct" events found by Asur et al. was as many as 1,398. All these unexpected cases revealed a significant weakness of the method by Asur et al.

### The Palla et al. Method

The method by Palla et al. has been implemented in T-SQL, but it required much more preparations. Apart from extracting groups in all time windows, yet another group extraction was needed. The data from two consecutive time windows were merged into a single graph, from which groups were extracted by means of the CFinder method (see Section 3.3.6). As easy to count, the group extraction had to be performed additional thirteen times, some of them took only five minutes to calculate, but there were also some lasting up to two days.

Palla et al. have designed their method in order to find all matching pairs of groups, even if they overlap in the slightest way, sharing only one node. The great advantage of this approach is that no event will be ignored. However, if one takes into account the fact, that Palla et al. only showed which event types may occur (and did not provide the algorithm to assign them), analysis of the group evolution during its life is very difficult and cumbersome. Each case of assigning event must be considered individually over a huge number of possibilities. As a result, it is very hard to find the key match. Moreover, Palla et al. did not explain how to choose the best match for the analysed groups or how to assign the event type. The authors only defined the case when there is the single highest overlapping for each group.

The total number of matched pairs found by Palla's et al. method was 9,797, out of which 4,183 pairs (42.7%) had an overlap higher than 0%. The authors did not specify how to interpret the rest of the groups that matched with the overlap equal 0%, but the intuition suggests to omit these cases. There were 90 cases when matched pairs had overlap equal 100%, which corresponds to *continuation* event in the Asur et al. method.

### The *GED* Method.

The *GED* method has also been implemented in T-SQL language. The method has been run frequently with different values of $\alpha$ and $\beta$ thresholds to analyse the influence of these parameters on the method, see Table 4.2. The time needed for a single run was about 6 minutes. The lowest checked value for the thresholds was set to 50%, which guarantees that at least a half of the considered groups are included in the matched group. The highest possible value is of course 100% which means the studied group is identical to the matched group. The thresholds for the *forming* and *dissolving* event were set to 10% based on average group size and intuition.

While analysing Table 4.2 and Figures 4.5-4.10, it can be observed that with the increase of $\alpha$ and $\beta$ thresholds, the total number of events is decreasing, when $\alpha$ and $\beta$ equal 50% this number is 1,734, and with thresholds equal 100% the number is only 1,091. It means that the parameters $\alpha$ and $\beta$ can be used to filter results, preserving from events where groups are highly overlapped. Another advantage of having parameters is possibility to adjust the results to one's needs. The linear increase of threshold $\alpha$ causes close to linear reduction in the number of merging events. In contrast, with linear increase of threshold $\beta$, the number of splitting events decreases in the almost linear way. As a consequence of the algorithm structure, raising the thresholds makes it difficult to match the groups (see Figure 4.3). Furthermore, dissolving events occur more frequently than forming events. The main reason is the fact that the last time window covers only the period of summer holidays, and as a result the email exchange is very low for that time. This causes the groups to be small and have low density.

Overall, the *GED* method found 90 continue events when both inclusions of groups ($\alpha$ and $\beta$) are equal to 100%.

| Threshold | | Number of events | | | | | | | |
|---|---|---|---|---|---|---|---|---|---|
| $\alpha$ % | $\beta$ % | form | dissolve | shrink | growth | continue | split | merge | total |
| 50 | 50 | 122 | 186 | 204 | 180 | 127 | 517 | 398 | 1,734 |
| 50 | 60 | 122 | 186 | 204 | 173 | 124 | 464 | 405 | 1,678 |
| 50 | 70 | 122 | 186 | 202 | 157 | 124 | 400 | 421 | 1,612 |
| 50 | 80 | 122 | 186 | 203 | 149 | 122 | 311 | 429 | 1,522 |
| 50 | 90 | 122 | 186 | 199 | 154 | 122 | 279 | 424 | 1,486 |
| 50 | 100 | 122 | 186 | 199 | 156 | 122 | 261 | 422 | 1,468 |
| 60 | 50 | 122 | 186 | 190 | 177 | 124 | 531 | 359 | 1,689 |
| 60 | 60 | 122 | 186 | 191 | 170 | 120 | 475 | 366 | 1,630 |
| 60 | 70 | 122 | 186 | 187 | 152 | 119 | 409 | 384 | 1,559 |
| 60 | 80 | 122 | 186 | 187 | 144 | 117 | 314 | 392 | 1,462 |
| 60 | 90 | 122 | 186 | 181 | 148 | 117 | 277 | 388 | 1,419 |
| 60 | 100 | 122 | 186 | 179 | 149 | 117 | 259 | 387 | 1,399 |
| 70 | 50 | 122 | 186 | 179 | 176 | 123 | 543 | 284 | 1,613 |
| 70 | 60 | 122 | 186 | 180 | 170 | 119 | 486 | 286 | 1,549 |
| 70 | 70 | 122 | 186 | 177 | 156 | 113 | 418 | 298 | 1,470 |
| 70 | 80 | 122 | 186 | 174 | 149 | 111 | 317 | 305 | 1,364 |
| 70 | 90 | 122 | 186 | 165 | 150 | 111 | 277 | 304 | 1,315 |
| 70 | 100 | 122 | 186 | 161 | 152 | 111 | 259 | 302 | 1,293 |
| 80 | 50 | 122 | 186 | 172 | 169 | 120 | 553 | 233 | 1,555 |
| 80 | 60 | 122 | 186 | 173 | 154 | 117 | 495 | 235 | 1,482 |
| 80 | 70 | 122 | 186 | 170 | 137 | 111 | 426 | 244 | 1,396 |
| 80 | 80 | 122 | 186 | 165 | 127 | 97 | 324 | 251 | 1,272 |
| 80 | 90 | 122 | 186 | 157 | 128 | 96 | 276 | 250 | 1,215 |
| 80 | 100 | 122 | 186 | 152 | 129 | 96 | 257 | 249 | 1,191 |
| 90 | 50 | 122 | 186 | 172 | 169 | 120 | 553 | 199 | 1,521 |
| 90 | 60 | 122 | 186 | 174 | 152 | 117 | 494 | 198 | 1,443 |
| 90 | 70 | 122 | 186 | 171 | 132 | 111 | 425 | 199 | 1,346 |
| 90 | 80 | 122 | 186 | 165 | 121 | 96 | 324 | 203 | 1,217 |
| 90 | 90 | 122 | 186 | 154 | 123 | 91 | 276 | 199 | 1,151 |
| 90 | 100 | 122 | 186 | 148 | 123 | 91 | 257 | 199 | 1,126 |
| 100 | 50 | 122 | 186 | 176 | 167 | 120 | 549 | 185 | 1,505 |
| 100 | 60 | 122 | 186 | 177 | 149 | 117 | 491 | 183 | 1,425 |
| 100 | 70 | 122 | 186 | 173 | 127 | 111 | 423 | 180 | 1,322 |
| 100 | 80 | 122 | 186 | 166 | 116 | 96 | 323 | 179 | 1,188 |
| 100 | 90 | 122 | 186 | 154 | 117 | 91 | 276 | 173 | 1,119 |
| 100 | 100 | 122 | 186 | 148 | 115 | 90 | 257 | 173 | 1,091 |

Table 4.2 The results of the *GED* computation on overlapping groups extracted by CFinder.

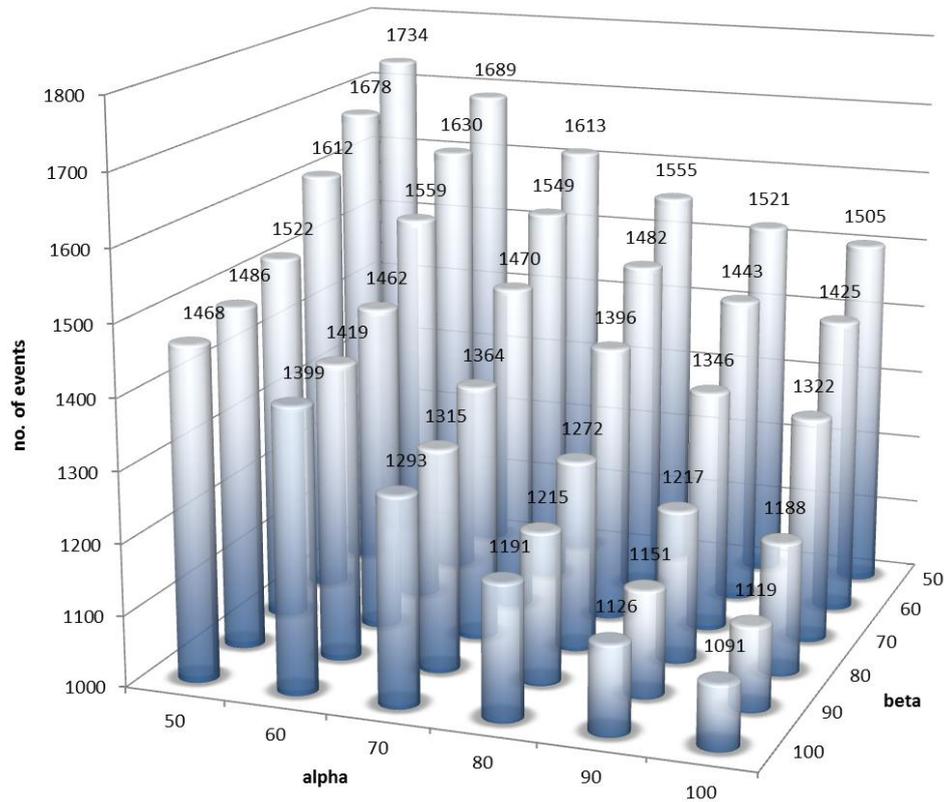

Figure 4.5 Alpha and beta influence on the number of events.

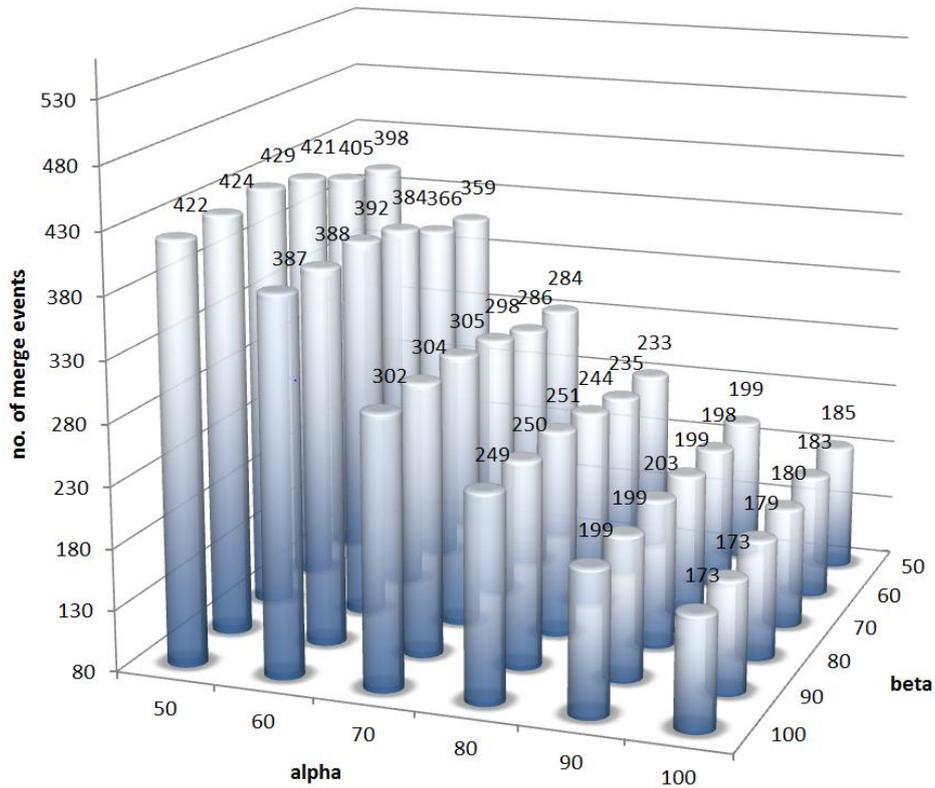

Figure 4.6 Alpha and beta influence on the number of merge events.

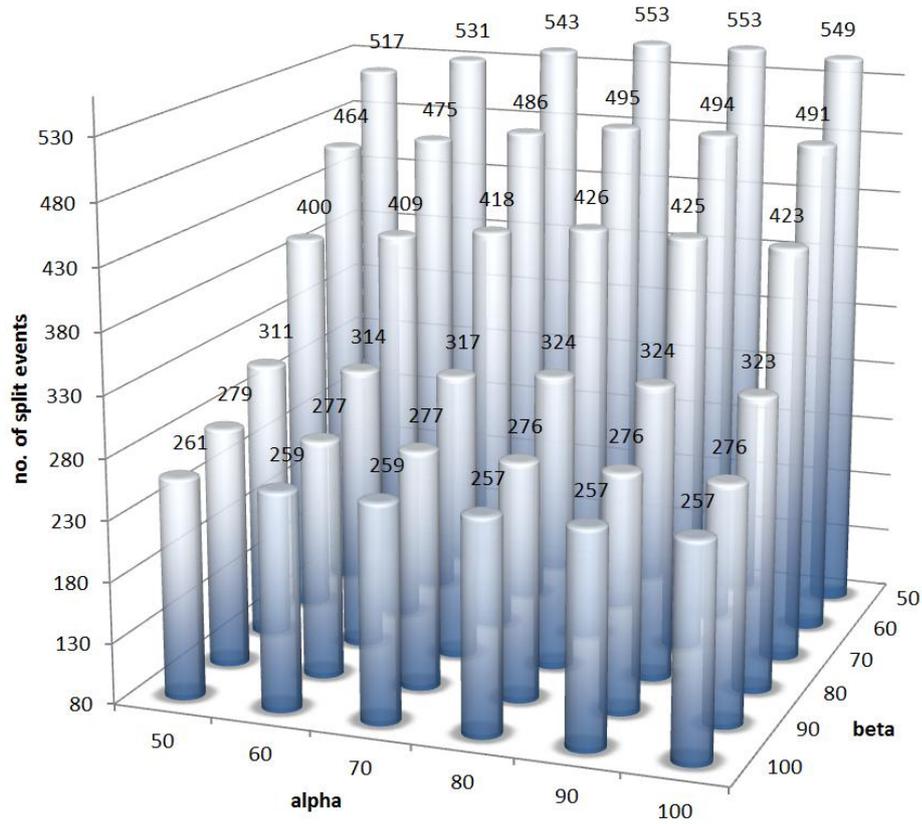

Figure 4.7 Alpha and beta influence on the number of split events.

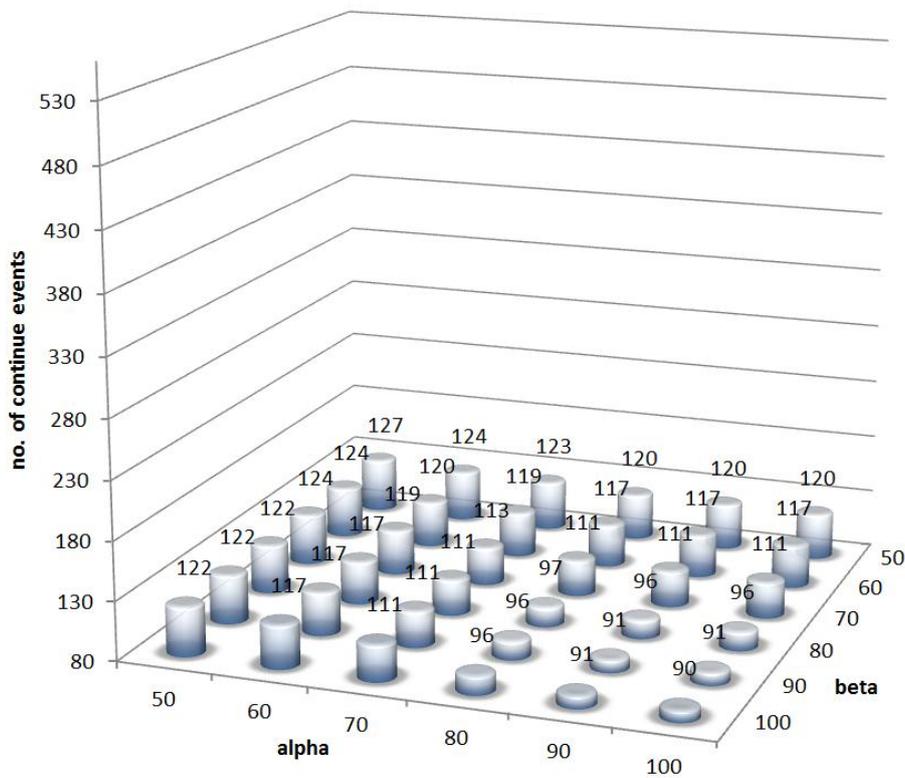

Figure 4.8 Alpha and beta influence on the number of continue events.

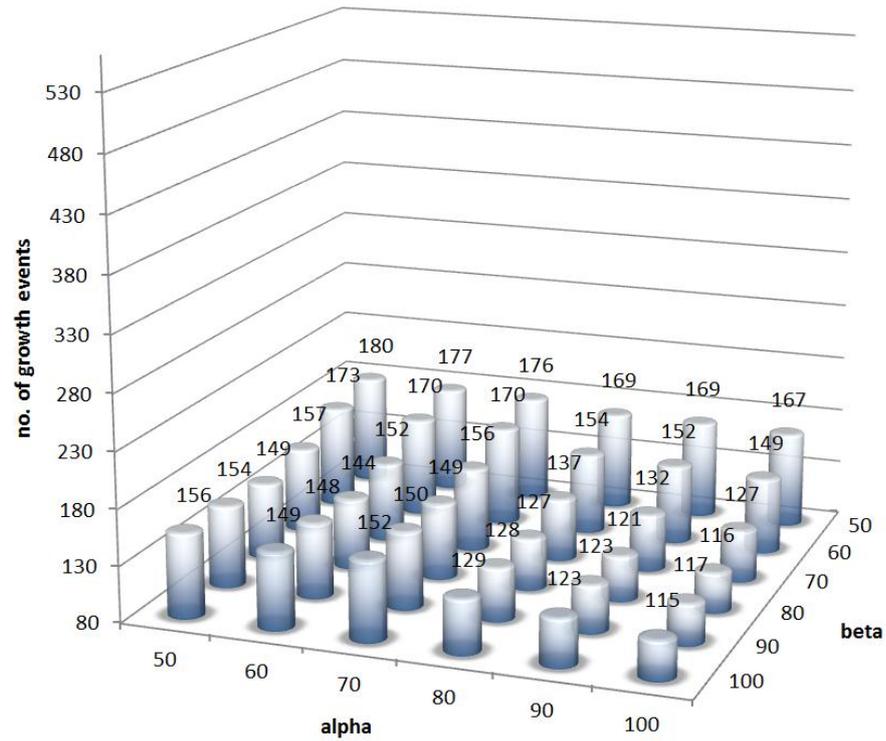

Figure 4.9 Alpha and beta influence on the number of growth events.

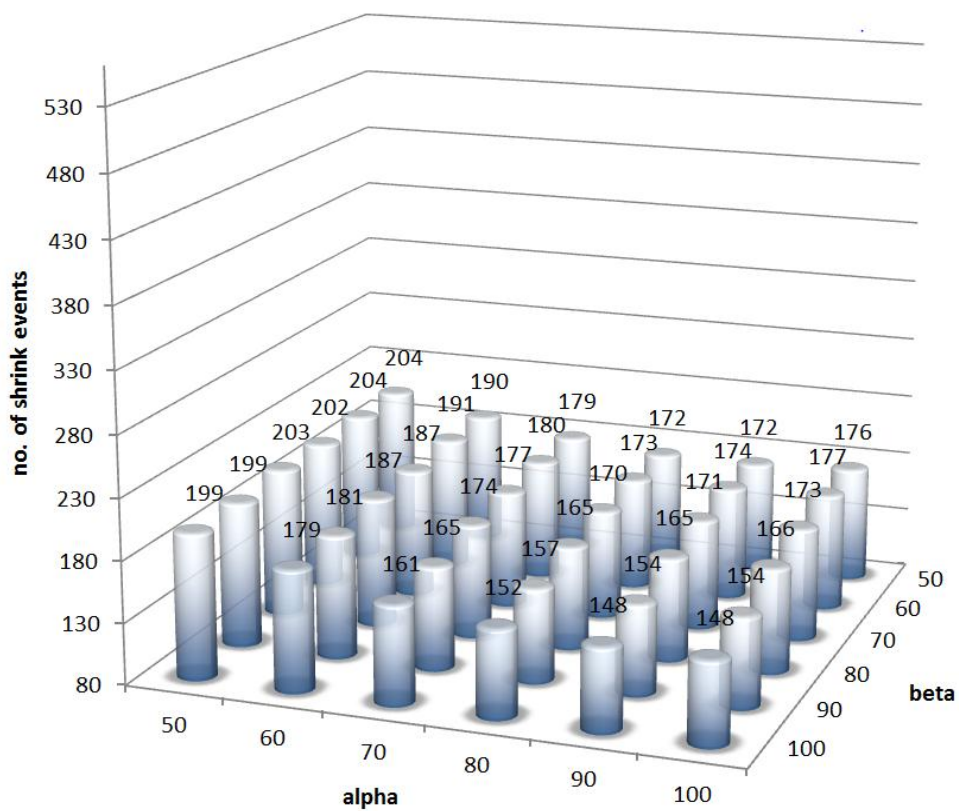

Figure 4.10 Alpha and beta influence on the number of shrink events.

### Differences between the *GED* Method and the Method by Asur et al.

As already mentioned, the computation time for Asur et al. method was more than 5.5 hours, while for *GED* it took less than 4 hours to calculate the whole Table 4.2. The single run of the *GED* method lasted less than 6 minutes, so it is over 50 times faster than the method by Asur et al.

The *GED* method with thresholds equal 50% has found 721 events which the method by Asur et al. has not discovered at all. Such a big lack in results obtained with Asur et al. method is caused mostly by its rigorous conditions for assigning events and almost no flexibility of the method. On the other hand, Asur et al. method has found 399 events which the *GED* method with thresholds 50% has not. However, it is not treated as a defect in *GED's* results because all these events had both inclusions below 50%, therefore, the *GED* algorithm skipped them on purpose (because of thresholds' values). To prove this, the *GED* method was run with thresholds equal 10% and this time none of events found by Asur et al. method were omitted by the *GED* method.

Furthermore, Asur et al. did not introduce the shrinking and growing events, what effects in assigning splitting and merging events or, in the worst case, missing the event. If two groups in the successive time windows differ only by one node, they will not be treated as continuation (since the overlapping is below 100%) and it might not be treated as merging (splitting), if there is no other group fulfilling the requirements for merging (splitting). Such a case is not possible in *GED*, which through the change of inclusion thresholds allows to adjust the results to user's needs.

The above analysis proves that the *GED* method is not only faster but also more accurate and much more flexible than method by Asur et al.

### Differences between the *GED* Method and the Method by Palla et al.

As noted before, the method by Palla et al. needs additional preparations to run the experiment, which lasted almost a week, therefore the *GED* method is faster, despite additional calculations of user importance measures required.

The great advantage of the method by Palla et al. is catching all matching pairs of groups. As in the case of comparing the *GED* method with the algorithm by Asur et al., Palla et al. method found more matched pairs than the *GED* method with thresholds at the level of 50%. Again, it is not treated as a defect in *GED's* results since all these events had both inclusions below 50%. To confirm that, the results obtained with *GED* on thresholds equal 10% have been compared, and this time all matched pairs found by Palla et al. and not found by the *GED* method had inclusions below 10%.

Another problem with Palla et al. method is the lack of the algorithm for assigning events. It is very difficult and time consuming to identify an event for the group in the next time window, not even to mention for all fourteen slots. So, the *GED* method with its automatic event assignment is much more useful and convenient.

Summing up, the *GED* method is not comparable when it comes to execution time. It is also definitely more specific in assigning events and therefore much more effective and accurate for tracking group evolution. The method by Palla et al. was helpful only in checking if the *GED* method found all events between the groups.

### 4.3.2   Experiment Based on Disjoint Groups Extracted using Blondel et. al.

For the second experiment the fast modularity optimization was used (see Section 3.3.3).

**The Asur et al. Method**

The method provided by Asur et al. needed almost 6 hours to calculate events between groups for all fourteen time windows. The overlapping threshold for merging and splitting events was set to 50%. The total number of events found by Asur et al. method was 747, out of which 120 were continuation, 23 were forming, 16 were dissolving, 255 were merging and 333 were splitting.

Again, the small number of continuing, forming, and dissolving events is caused by the too rigorous conditions. In turn, the great number of merging (splitting) events is a result of low overlapping threshold for merge (split).

As in case of CFinder grouping method, the number of events found on data grouped by the *Blondel* method is also overestimated. The number of anomalies this time is 40 cases, 5% of all results. Therefore, the total number of "distinct" events was 707. This mean that Asur et al. method works better for disjoint groups.

**The *GED* Method.**

As previously, for the data grouped with the CFinder method, the *GED* method have been run with different values of α and β thresholds and the results are presented in Table 4.3. The time needed for a single run was about 13 minutes. The thresholds for the *forming* and *dissolving* event was again set to 10%.

The total number of events found with thresholds equal to 50% was 1,231 but with thresholds equal to 100% only 663. This indicates that parameters α and β influence the number of events even more than in case of the CFinder method. The linear relation between the increase of threshold α or β and the reduction of the number of merging (splitting) is preserved (Table 4.3 and Figures 4.11-4.16 ).

The *GED* method found 120 continue events when both inclusions of groups (α and β) are equal to 100%, which correspond to *continuation* event in Asur et al. method.

In general, the *GED* method can be successfully used for both, overlapping or disjoint groups. If overlapping groups for a small network are needed then CFinder can be used, however, if one needs to extract groups very fast and for a big network then the method proposed by *Blondel* can be utilized. This flexibility and adaptability of the *GED* method is its big advantage because most methods can be used only for either overlapping or disjoint groups.

| Threshold | | Number of events | | | | | | | |
|---|---|---|---|---|---|---|---|---|---|
| α % | β % | form | dissolve | shrink | growth | continue | split | merge | total |
| 50 | 50 | 39 | 23 | 187 | 167 | 135 | 411 | 269 | 1231 |
| 50 | 60 | 39 | 23 | 181 | 161 | 135 | 378 | 275 | 1192 |
| 50 | 70 | 39 | 23 | 179 | 156 | 135 | 338 | 280 | 1150 |
| 50 | 80 | 39 | 23 | 178 | 153 | 135 | 294 | 283 | 1105 |
| 50 | 90 | 39 | 23 | 164 | 143 | 134 | 250 | 293 | 1046 |
| 50 | 100 | 39 | 23 | 154 | 143 | 134 | 224 | 293 | 1010 |
| 60 | 50 | 39 | 23 | 181 | 166 | 135 | 417 | 237 | 1198 |
| 60 | 60 | 39 | 23 | 176 | 159 | 134 | 383 | 244 | 1158 |
| 60 | 70 | 39 | 23 | 174 | 155 | 134 | 338 | 247 | 1110 |
| 60 | 80 | 39 | 23 | 171 | 151 | 134 | 294 | 251 | 1063 |
| 60 | 90 | 39 | 23 | 156 | 140 | 133 | 250 | 262 | 1003 |
| 60 | 100 | 39 | 23 | 148 | 140 | 133 | 218 | 262 | 963 |
| 70 | 50 | 39 | 23 | 169 | 164 | 134 | 429 | 216 | 1174 |
| 70 | 60 | 39 | 23 | 163 | 158 | 131 | 396 | 219 | 1129 |
| 70 | 70 | 39 | 23 | 164 | 154 | 130 | 345 | 221 | 1076 |
| 70 | 80 | 39 | 23 | 159 | 150 | 130 | 299 | 225 | 1025 |
| 70 | 90 | 39 | 23 | 144 | 139 | 129 | 245 | 236 | 955 |

| Threshold | | Number of events | | | | | | | |
|---|---|---|---|---|---|---|---|---|---|
| α % | β % | form | dissolve | shrink | growth | continue | split | merge | total |
| **70** | **100** | 39 | 23 | 137 | 138 | 129 | 204 | 237 | 907 |
| **80** | **50** | 39 | 23 | 162 | 165 | 134 | 436 | 180 | 1139 |
| **80** | **60** | 39 | 23 | 157 | 158 | 130 | 402 | 178 | 1087 |
| **80** | **70** | 39 | 23 | 156 | 152 | 129 | 350 | 176 | 1025 |
| **80** | **80** | 39 | 23 | 151 | 147 | 127 | 304 | 177 | 968 |
| **80** | **90** | 39 | 23 | 138 | 140 | 126 | 235 | 184 | 885 |
| **80** | **100** | 39 | 23 | 128 | 140 | 126 | 191 | 184 | 831 |
| **90** | **50** | 39 | 23 | 157 | 172 | 133 | 442 | 126 | 1092 |
| **90** | **60** | 39 | 23 | 153 | 161 | 129 | 407 | 124 | 1036 |
| **90** | **70** | 39 | 23 | 152 | 152 | 128 | 355 | 118 | 967 |
| **90** | **80** | 39 | 23 | 146 | 139 | 126 | 310 | 116 | 899 |
| **90** | **90** | 39 | 23 | 133 | 130 | 121 | 228 | 114 | 788 |
| **90** | **100** | 39 | 23 | 116 | 131 | 121 | 178 | 113 | 721 |
| **100** | **50** | 39 | 23 | 160 | 168 | 133 | 439 | 106 | 1068 |
| **100** | **60** | 39 | 23 | 156 | 154 | 129 | 404 | 104 | 1009 |
| **100** | **70** | 39 | 23 | 155 | 144 | 128 | 352 | 97 | 938 |
| **100** | **80** | 39 | 23 | 149 | 129 | 126 | 307 | 95 | 868 |
| **100** | **90** | 39 | 23 | 133 | 110 | 121 | 228 | 83 | 737 |
| **100** | **100** | 39 | 23 | 114 | 109 | 120 | 178 | 80 | 663 |

Table 4.3 The results of the *GED* identification process for disjoint groups.

**Differences between the *GED* Method and the Method by Asur et al.**

The *GED* method needed less than 8 hours to calculate results for the whole Table 4.3, while a single run of Asur et al. method lasted almost 6 hours. A single run of the *GED* method was only 13 minutes, so it is still much faster than the method by Asur et al.

The *GED* method run with thresholds equals 50% found as many as 613 events, which the method by Asur et al. did not recognize at all. Again, the big gap in results obtained with Asur et al. method is caused mostly by its rigorous conditions for assigning events and almost no flexibility of the method. Like in case of the CFinder method, Asur et al. method found events, which the *GED* method skipped because of threshold values. Reducing the thresholds effected in not omitting the mentioned groups.

The above considerations confirm that the *GED* method is better than Asur et al. method for overlapping as well as for disjoint methods of grouping.

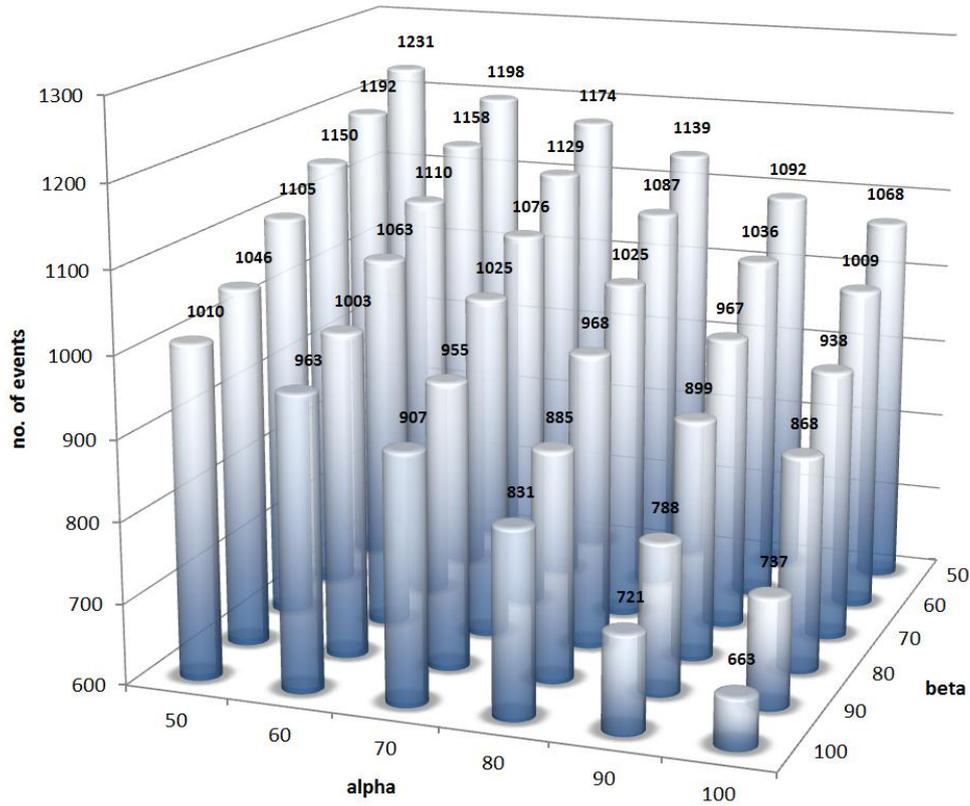

Figure 4.11 Alpha and beta influence on the number of events.

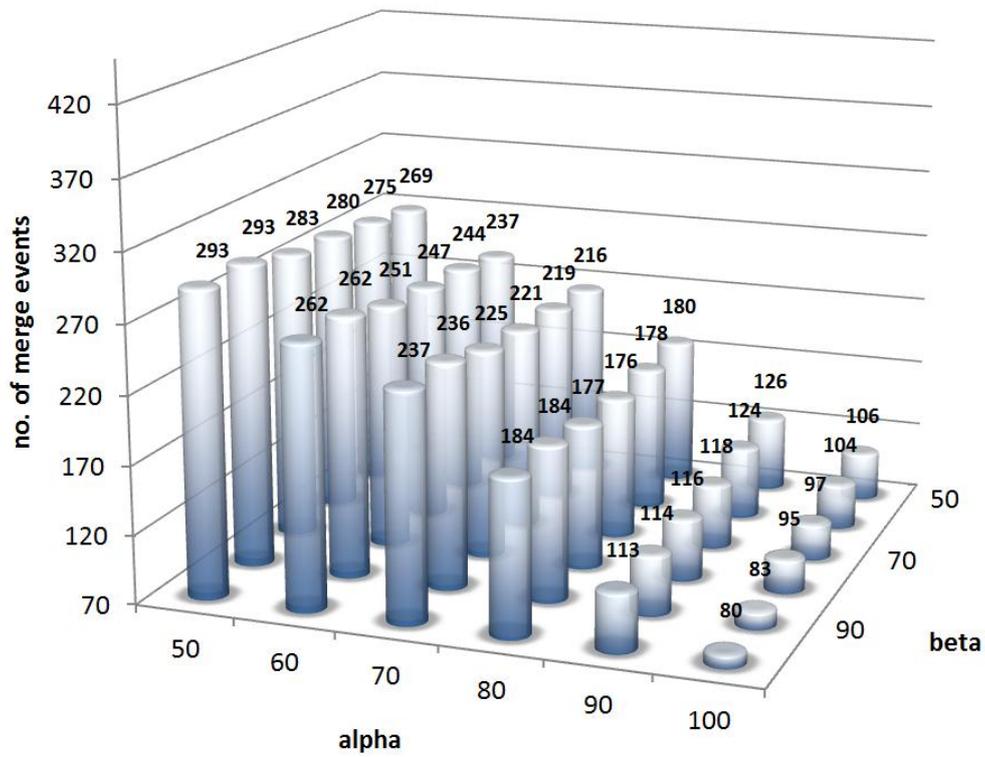

Figure 4.12 Alpha and beta influence on the number of merge events.

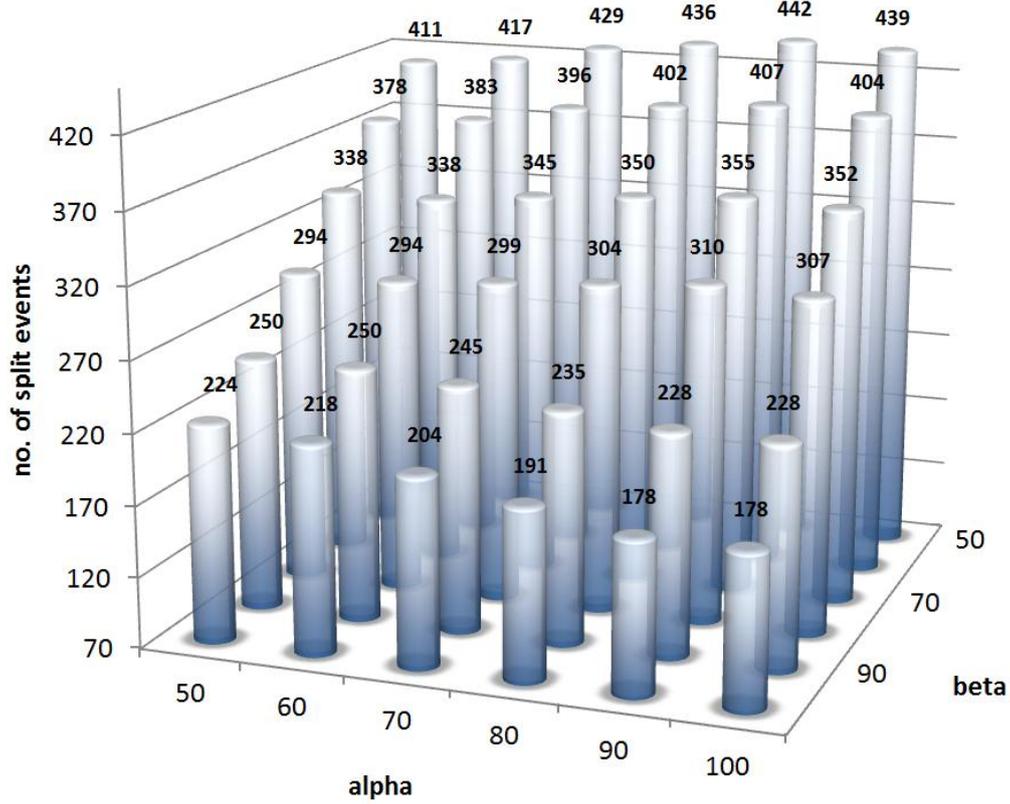

Figure 4.13 Alpha and beta influence on the number of split events.

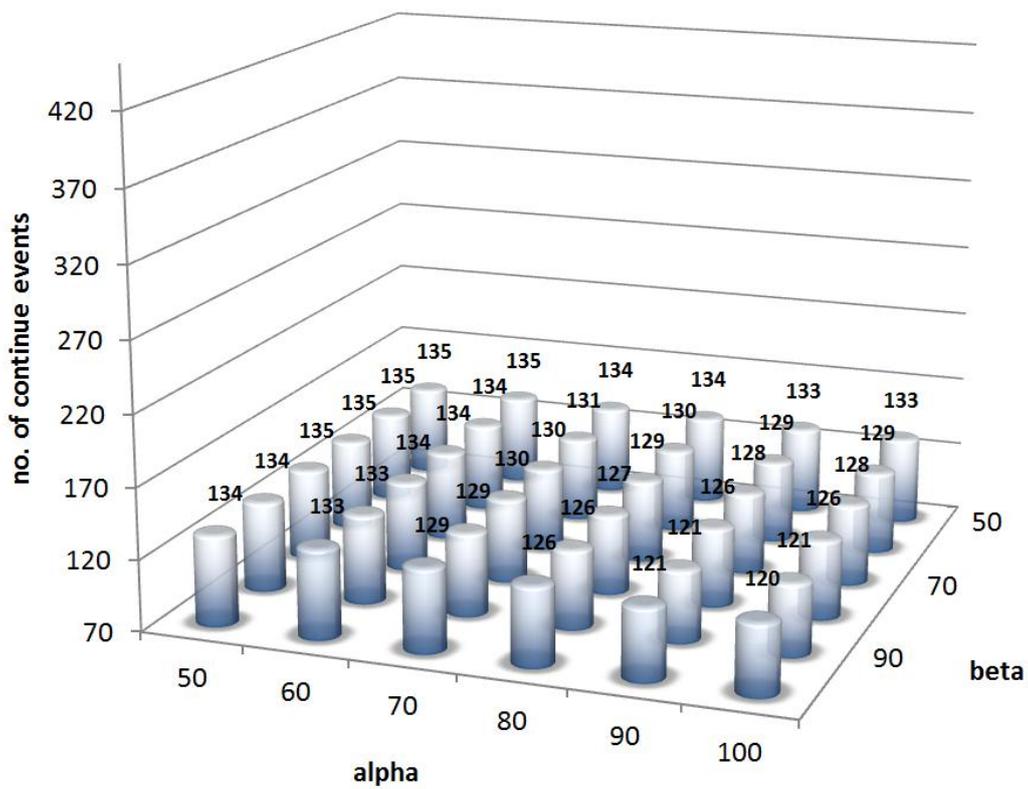

Figure 4.14 Alpha and beta influence on the number of continue events.

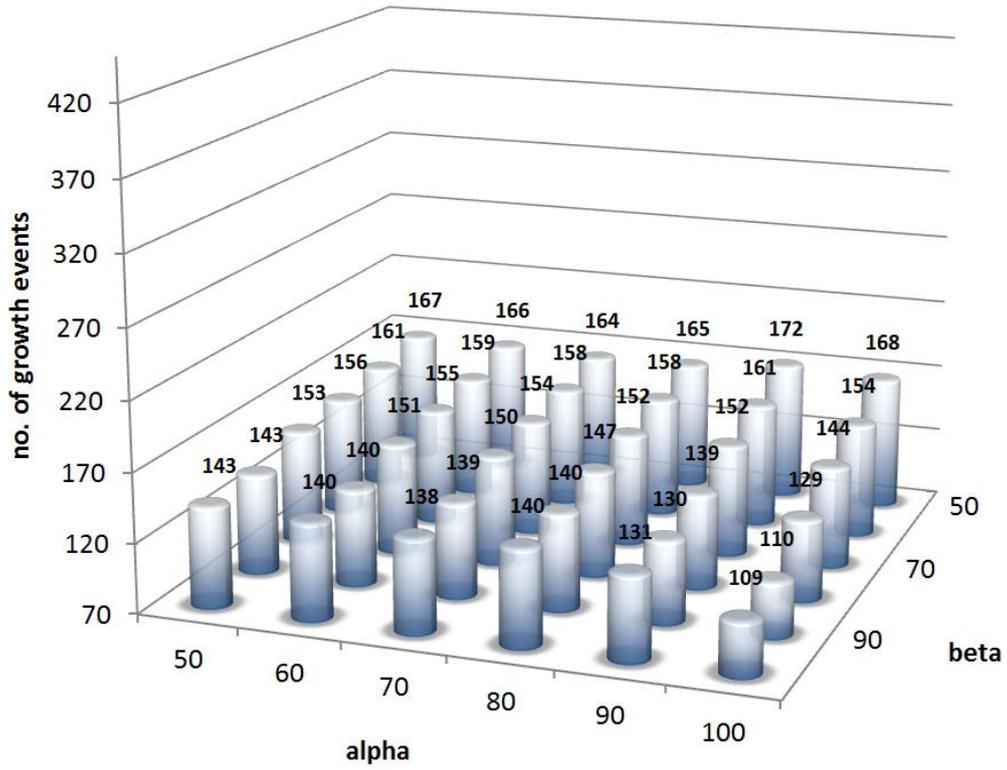

Figure 4.15 Alpha and beta influence on the number of growth events.

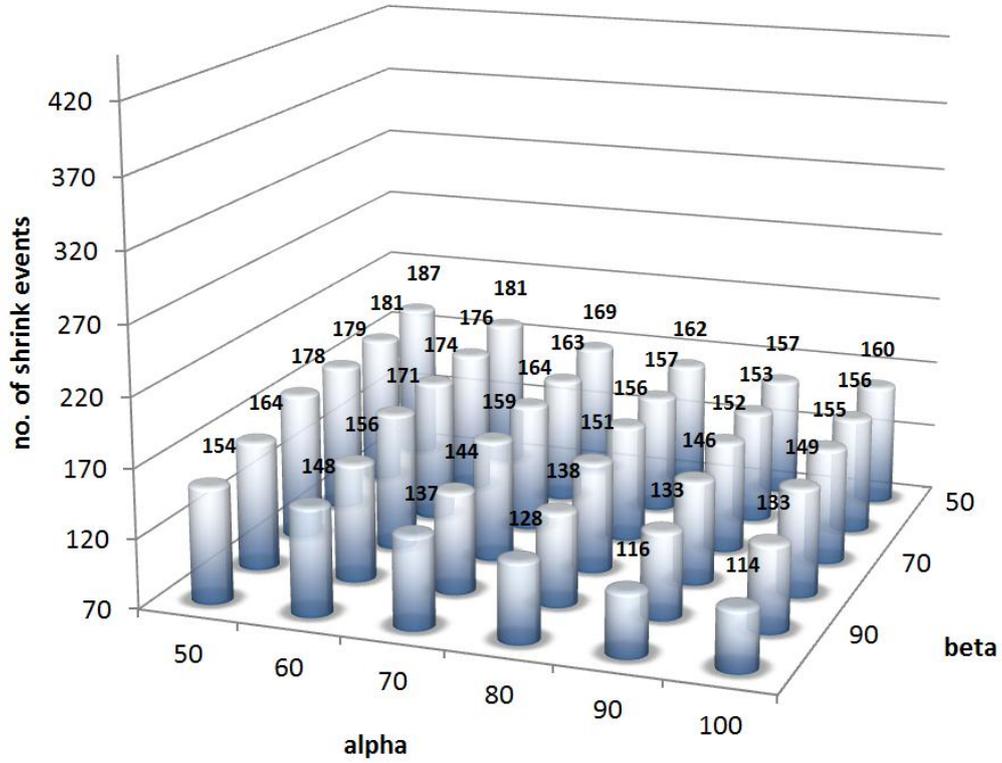

Figure 4.16 Alpha and beta influence on the number of shrink events.

### 4.3.3 Experiment Based on Different User Importance Measures

In the last experiment *GED* method was run: (i) with degree centrality measure instead of social position measure and (ii) without any measure, in order to investigate influence of the measure on calculations of inclusion values and also on results of the method. Like in case of the first experiment, overlapping groups extracted with CFinder were used.

The results obtained with degree centrality as a measure of user importance and results derived without any measure are very similar to the results obtained with social position measure, Table 4.4.

| Measure | Execution time [min] | Events found | Threshold | |
|---|---|---|---|---|
| | | | $\alpha$ | $\beta$ |
| **Social Position** | 6:00 | 1,470 | 70 | 70 |
| **Degree Centrality** | 5:55 | 1,447 | 70 | 70 |
| **No measure** | 5:30 | 1,483 | 70 | 70 |

Table 4.4. Results of the GED Method with different user importance measures.

Execution time for the *GED* method with degree centrality was slightly better than for the *GED* with social position, because degree centrality value was an integer, while the type for social position value is float. The degree centrality was not normalized, as the inclusion measure do not require normalized values and since summing the integers is faster than summing floats, the execution time do degree centrality is shorter. Of course the best execution time was for *GED* without any user importance indicate. Although, the number of events found in all three cases is more or less the same, it can be observed that *GED* without user importance measure found more events than *GED* with any of the measures. It is a consequence of the inclusion formula (see equation 4.2) which consists of two parts. The first one is always present, whether *GED* is run with or without user importance measure, but the second one occurs only when an importance measure is used. Therefore when calculating inclusions of two groups with an importance measure, it is almost always lower than without it. The exceptions are groups where the inclusion is equal to 100% and groups which do not share any nodes (inclusion is 0%). And here comes the question: why *GED* uses a measure of user importance, since it is obvious that it will lower the inclusion? The answer already provided in Section 4.2 is this time supported by clear evidences.

As illustrated in Figure 4.17, two communities $G_{46}$ and $G_{47}$ from time frame $T_6$ overlaps by five members and both groups have the same size – seven members. In the next time frame $T_7$ there is only one group $G_{18}$ which consists of all members from the group $G_{47}$ from the previous time frame, and one new member. Two members from the community $G_{46}$ have vanished and are missing in the following time window.

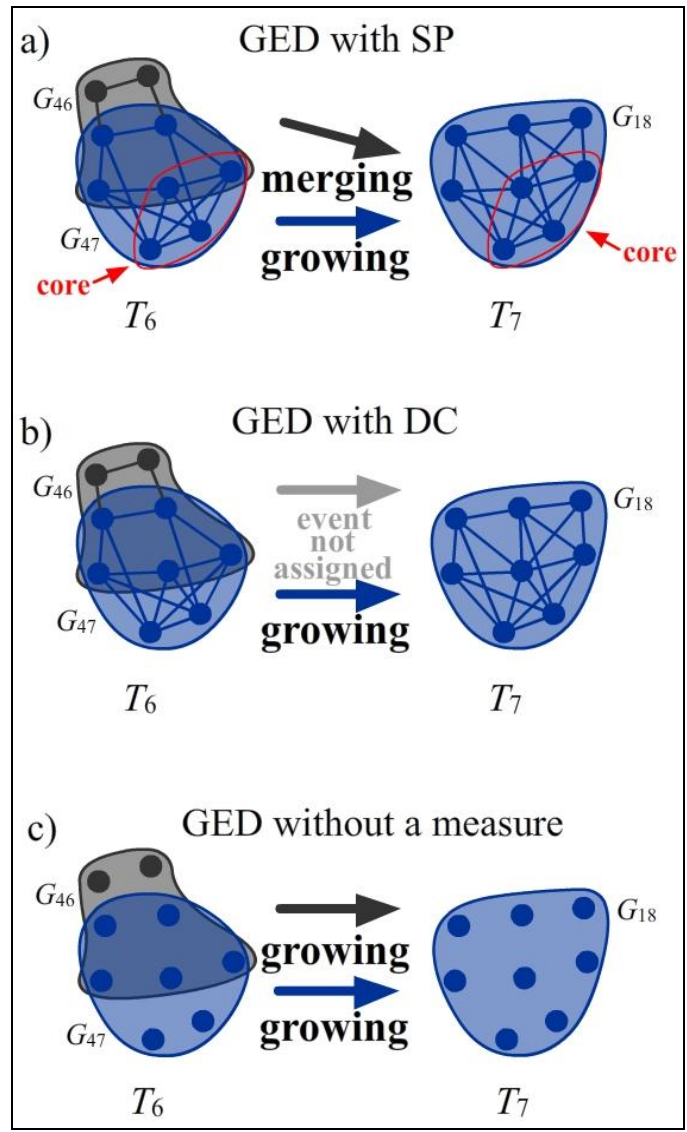

Figure 4.17. Events assigned by *GED* method with different user importance measures. a) *GED* with social position measure, red colour marks the core of the group b) *GED* with degree centrality c) *GED* without a measure.

The *GED* method with social position measure, assigned growing event to the community $G_{47}$ and merging event to the group $G_{46}$. The *GED* method with degree centrality measure also assigned growing event to the group $G_{47}$, but did not assign any event to the community $G_{46}$. Finally, *GED* without any user importance measure assigned growing events to both groups from time frame $T_6$.

To have a closer look into the first case, the social positions of members are presented in Table 4.5. It is clearly visible that the core of the blue group from time frame $T_6$ is identical to the core of the blue group from the next time window $T_7$. The situation is marked with red colour in the Figure 4.17a and with red dots in the Table 4.5. Additionally, members occurring in the all groups are marked in green. Now it is obvious that *GED* with social position measure assigned growing event to group $G_{47}$ because it is almost identical to group $G_{18}$, and "only" merging event to group $G_{46}$ because the cores of both groups have nothing in common. It has to be emphasized once again that, thanks to the user importance measure, *GED* method takes into account both the quantity and quality of the group members providing very accurate results.

| Group | Time window | Node id | SP | Rank |
|---|---|---|---|---|
| 46 | 6 | 1443 | 1,48 | 1 |
| 46 | 6 | 3145 | 1,33 | 2 |
| 46 | 6 | 7564 | 0,96 | 3 |
| 46 | 6 | 1326 | 0,86 | 4 |
| 46 | 6 | 11999 | 0,85 | 5 |
| 46 | 6 | 14151 | 0,77 | 6 |
| 46 | 6 | 621 | 0,75 | 7 |
| 47 | 6 | 2066• | 1,31 | 1 |
| 47 | 6 | 7328• | 1,30 | 2 |
| 47 | 6 | 7564• | 1,28 | 3 |
| 47 | 6 | 11999• | 1,04 | 4 |
| 47 | 6 | 1326 | 0,80 | 5 |
| 47 | 6 | 14151 | 0,67 | 6 |
| 47 | 6 | 621 | 0,60 | 7 |
| 18 | 7 | 2066• | 1,49 | 1 |
| 18 | 7 | 7328• | 1,35 | 2 |
| 18 | 7 | 7564• | 1,29 | 3 |
| 18 | 7 | 11999• | 1,24 | 4 |
| 18 | 7 | 1326 | 0,75 | 5 |
| 18 | 7 | 14151 | 0,71 | 6 |
| 18 | 7 | 621 | 0,66 | 7 |
| 18 | 7 | 4632 | 0,51 | 8 |

Table 4.5. Social position of members presented in Figure 4.17a.

The *GED* method with degree centrality measure was even more strict in the studied case, Figure 4.17b. Low degree centrality within the group $G_{46}$ causes that no event was assigned. In turn, similar structure between groups $G_{47}$ and $G_{18}$ effects in assigning the merging event. Structure of all groups and degree centrality of all members is presented in Table 4.6. Again, green colour marks members occurring in all groups.

| Group | Time window | Node | DC | Rank |
|---|---|---|---|---|
| **46** | 6 | 11999 | 3 | 1 |
| **46** | 6 | 14151 | 3 | 1 |
| **46** | 6 | 1443 | 2 | 3 |
| **46** | 6 | 3145 | 2 | 3 |
| **46** | 6 | 7564 | 2 | 3 |
| **46** | 6 | 1326 | 2 | 3 |
| **46** | 6 | 621 | 2 | 3 |
| **47** | 6 | 2066 | 5 | 1 |
| **47** | 6 | 7328 | 5 | 1 |
| **47** | 6 | 7564 | 4 | 3 |
| **47** | 6 | 11999 | 4 | 3 |
| **47** | 6 | 1326 | 4 | 3 |
| **47** | 6 | 14151 | 3 | 6 |
| **47** | 6 | 621 | 3 | 6 |
| **18** | 7 | 7564 | 7 | 1 |
| **18** | 7 | 7328 | 5 | 2 |
| **18** | 7 | 2066 | 5 | 2 |
| **18** | 7 | 11999 | 5 | 2 |
| **18** | 7 | 1326 | 5 | 2 |
| **18** | 7 | 14151 | 4 | 6 |
| **18** | 7 | 621 | 4 | 6 |
| **18** | 7 | 4632 | 3 | 8 |

Table 4.6. Degree centrality of members presented in Figure 4.17b.

Figure 4.17c presents in the best way how *GED* method without a user importance measure understands the communities. There is no core, all members are equal and relations between them are not considered at all. Such simplification causes that the events assigned to the groups are not the most adequate to situation (but only when comparing with events assigned by *GED* with user importance measure). Having information only about the members in the groups but not about their relations results in incorrect events assignment. Thus, if researchers investigating group evolution are not interested in groups structure and relations between members, a simpler and faster version of the GED Method may be successfully used. However, if there is enough time to calculate any user importance measure, it is recommended to use the *GED* method in the original version.

## 5. Prediction of Group Evolution in Social Network

In most fields of science, researchers struggle to predict the future. The future consumption of power in electric network, the future load of network grid, the future consumption of goods etc. Social networks are no different. Recently, the main focus is on *link prediction* [Liben-Nowell 07], but there are also other research directions: (i) entire future network structure modelling [Singh 07], [Juszczyszyn and **Bródka** 09], (ii) modelling social network evolution [Leskovec 08], [Michalski and **Bródka** 11], or (iii) churn prediction and its influence on the network [Wai-Ho 03], [Ruta and **Bródka** 09]. However only few researchers have considered groups in the prediction process. Some of them like Zheleva et. al. are using communities only for link prediction [Zheleva 08], the others like Kairam et. al. tries to identify and understand the factors contributing in the growth and longevity of groups within social networks [Kairam 12]. Unfortunately, there is no research directly regarding prediction of future group evolution. The main reason behind this is probably the fact that the methods for determining group history (see Section 4) have not been good enough so far. Thus in this dissertation the new approach for prediction of group evolution in the social network is presented.

### 5.1   The Concept of Using the *GED* Method for Prediction of Group Evolution.

The new approach, involves usage of the *GED* method results. It was shown that using a simple sequence which consists only of several preceding groups' sizes and events as an input for the classifier, the learnt model is able to produce very good results, even for simple classifiers. The sequences of groups sizes and events between time frames can be extracted from the *GED* results. In this dissertation 4-step sequences were used (Figure 5.1). Obviously, the event types varied depending on the individual groups, but the time frame numbers were fixed. It means that for each event four group profiles in four previous time frames together with three associated events were identified as the input for the classification model, separately for each group. A single group in a given time frame ($T_n$) was a case (instance) for classification, for which its event $T_nT_{n+1}$ was predicted (classified).

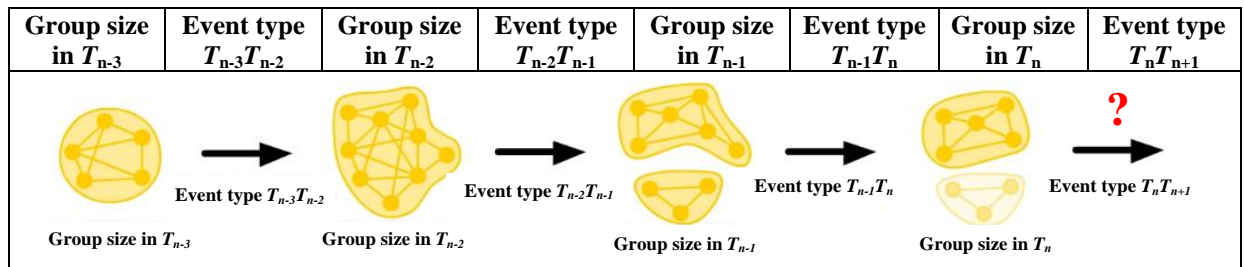

Figure 5.1 The sequence of events for a single group together with its intermediate sizes (descriptive, input variables) as well as its target class - event type in T_nT_n+1. It corresponds to one case in classification

The sequence presented in Figure 5.1 was used as an input for classification. The first part of the sequence was used as 7 input features (variables), i.e. (1) **Group size in $T_{n-3}$,** (2) **Event type $T_{n-3}T_{n-2}$,** (3) **Group size in $T_{n-2}$,** (4) **Event type $T_{n-2}T_{n-1}$,** (5) **Group size in $T_{n-1}$,** (6) **Event type $T_{n-1}T_n$,** (7) **Group size in $T_n$.** A predictive variable was the next event for a given group. Thus, the goal of classification was to predict (classify) **Event $T_nT_{n+1}$ type** – out of the six possible classes: i.e. (1) growing, (2) continuing, (3) shrinking, (4) dissolving, (5) merging and (6) splitting. Forming was excluded since it can only start the sequence.

## 5.2   Experiment Setup

Six dynamic social networks *DSN* have been extracted from four different datasets to perform and evaluate prediction of group evolution.

1. The first network was extracted from *Wroclaw University of Technology* (WrUT) email communication. The whole data set was collected within the period from February 2006 to October 2007 and consists of 5,845 nodes (distinct university employees' email addresses) and 149,344 edges (emails send from one address to another). The dynamic social network consisted of fourteen 90-day time frames extracted from this data. Timeframes have the 45-day overlap, i.e., the first time frame begins on the 1st day and ends on the 90th day, the second begins on the 46th day and ends on the 135th day and so on.

2. The second network was extracted also from *WrUT* with the difference that time frames are 45-day long and have no overlap, i.e., the first time frame begins on the 1st day and ends on the 45th day, the second begins on the 46th day and ends on the 90th day and so on.

3. The third social network was extracted from the portal *www.salon24.pl*, which is dedicated especially to political discussions, but also some other subjects from different domains may be brought up there. The network consists of 3,775 nodes and 77,932 edges. There are 12 non-overlapping time frames representing 12 months of the year 2009.

4. The forth one is the well-known *Enron* e-mail network with 150 nodes and 2,144 edges. The network was split into twelve, 90-day time frames without overlap.

5. The fifth network was extracted from the portal *www.extradom.pl*. It gathers people, who are engaged in building their own houses in Poland. It helps them to exchange best practices, experiences, evaluate various constructing projects and technologies or simply to find the answers to their questions provided by others. The data covers a period of 17 months and contains 3,690 users and 34,082 relations. 33 time frames were extracted, each of them 30-day long with 15 days overlap, similarly to the first data set.

6. The last one, sixth, network was also extracted from *extradom.pl* but consists of 16, 60-day long time frames, with 30 days overlap.

For each time frame social communities were extracted using CFinder[17] (see Section 3.3.6) and for each *DSN* the *GED* method was utilized to extract groups evolution. The *GED* method was run 36 times for each *DSN* with all combination of $\alpha$ and $\beta$ parameters from the set {50%, 60%, 70%, 80%, 90%, 100%}.

Next, the 4-step sequences where extracted from the *GED* results for all networks and every combination of $\alpha$ and $\beta$ parameters, see an example sequence in Figure 5.1.

Experiment was performed in WEKA Data Mining Software [Hall 09]. Ten different classifiers were utilized with default settings (see Table 5.1). For the method of validation 10-fold cross-validation was utilized as the most commonly used [McLachlan 04]. In WEKA, this means 100 calls of one classifier with training data, tested against the test data in order to get statistically meaningful results.

| WEKA name | Name | Default WEKA Settings[18] |
|---|---|---|
| **BayesNet** | Bayes Network classifier [Hall 09] | *estimator*: SimpleEstimator –A 0.5<br>*searchAlgorithm*: K2 –P 1 –S BAYES<br>*useADTree*: False |

---

[17] http://www.cfinder.org/
[18] **http://www.cs.waikato.ac.nz/ml/weka/**

| WEKA name | Name | Default WEKA Settings[18] |
|---|---|---|
| **NaiveBayes** | Naive Bayesian classifier [John 95] | - |
| **IBk** | k-nearest neighbor classifier [Aha 91] | *KNN*: 1<br>*Distance weighting*: No distance weigthing<br>*meanSquared*: False<br>*nearestNeighbourSearchAlgorithm*: LinearNNSearch<br>*windowSize*: 0 |
| **KStar** | Instance-Based classifier [Cleary 95] | *entropicAutoBlend*: False<br>*globalBlend*: 20<br>*missingMod*:e Averagecolumn entropy curves |
| **AdaBoost** | Adaboost M1 method [Freund 96] | J48 tree |
| **DecisionTable** | Decision table [Kohavi 95] | *crossVal*: 10<br>*evaluationMeasure*: Default: accuracy(discrete class); RMSE (numeric class)<br>*search*: BestFirst<br>*useIBK*: False |
| **JRip** | RIPPER rule classifier [Cohen 95] | *checkErrorRate*: True<br>*folds*: 3<br>*minNo*:2<br>*optimizations*:2<br>*seed*:1<br>*usePrunning*: true |
| **ZeroR** | 0-R classifier | - |
| **J48** | C4.5 decision tree [Quinlan 93] | *binarySplits*: False<br>*collapseTree*: True<br>*confidenceFactor* 0.25<br>*minNumObj*: 2<br>*numFolds*: 3<br>*reducedErrorPrunning*: False<br>*seed*: 1<br>*subtreeRaising*: True<br>*unpruned*: False<br>*useLaplace*: False<br>*useMDLcorrection*: True |
| **RandomForest** | Random forest [Breiman 01] | *maxDepth*:0<br>*numExecutionSlots*:1<br>*numFeatures*:0<br>*numTrees*: 10<br>*seed*: 1 |

Table 5.1 WEKA classifiers used during experiments and their settings. For full settings description see http://www.cs.waikato.ac.nz/ml/weka/.

All classifiers were utilized for each of 6 networks and each combination of $\alpha$ and $\beta$ parameters. The measure selected for presentation and analysis of the results is F measure which is the harmonic mean of precision and recall.

$$F = 2\frac{precision \cdot recall}{precision + recall}; \qquad precision = \frac{tp}{tp + fp}; \qquad recall = \frac{tp}{tp + fn}.$$

|  | actual class (expectation) | |
|---|---|---|
| **predicted class (observation)** | *tp* – (true positive) Correct result | *fp* – (false positive) Unexpected result |
|  | fn – (false negative) Missing result | *tn* – (true negative) Correct absence of result |

Table 5.2 The possible results of comparison of the classification results against the test data.

## 5.3 Results and Conclusions

At the beginning, the classifiers were compared for each dataset separately in order to indicate which one is the best. The results are presented in Table 5.3 and Figures 5.2-5.5.

| Dataset | Classifier | Max F measure | Min F measure | Difference |
|---|---|---|---|---|
| salon24.pl | BayesNet | 1.00 | 1.00 | 0.00 |
|  | NaiveBayes | 1.00 | 1.00 | 0.00 |
|  | IBk | 1.00 | 1.00 | 0.00 |
|  | KStar | 1.00 | 1.00 | 0.00 |
|  | AdaBoostM1 | 1.00 | 0.70 | 0.30 |
|  | DecisionTable | 1.00 | 0.90 | 0.11 |
|  | JRip | 1.00 | 0.97 | 0.03 |
|  | ZeroR | 0.82 | 0.60 | 0.23 |
|  | J48 | 1.00 | 0.99 | 0.01 |
|  | RandomForest | 1.00 | 1.00 | 0.00 |
| Enron | BayesNet | 0.83 | 0.69 | 0.15 |
|  | NaiveBayes | 0.81 | 0.72 | 0.08 |
|  | IBk | 0.79 | 0.71 | 0.08 |
|  | KStar | 0.79 | 0.72 | 0.07 |
|  | AdaBoostM1 | 0.51 | 0.32 | 0.20 |
|  | DecisionTable | 0.78 | 0.64 | 0.14 |
|  | JRip | 0.80 | 0.73 | 0.07 |
|  | ZeroR | 0.27 | 0.15 | 0.11 |
|  | J48 | 0.92 | 0.80 | 0.13 |
|  | RandomForest | 0.89 | 0.76 | 0.13 |
| extradom.pl | BayesNet | 0.87 | 0.54 | 0.32 |
|  | NaiveBayes | 0.87 | 0.50 | 0.37 |
|  | IBk | 0.88 | 0.55 | 0.33 |
|  | KStar | 0.88 | 0.52 | 0.36 |
|  | AdaBoostM1 | 0.83 | 0.50 | 0.33 |
|  | DecisionTable | 0.88 | 0.48 | 0.39 |

| Dataset | Classifier | Max F measure | Min F measure | Difference |
|---|---|---|---|---|
|  | JRip | 0.88 | 0.35 | 0.53 |
|  | ZeroR | 0.88 | 0.33 | 0.54 |
|  | J48 | 0.88 | 0.33 | 0.55 |
|  | RandomForest | 0.88 | 0.40 | 0.48 |
| **WrUT emails** | BayesNet | 0.86 | 0.76 | 0.10 |
|  | NaiveBayes | 0.86 | 0.73 | 0.13 |
|  | IBk | 0.88 | 0.79 | 0.09 |
|  | KStar | 0.88 | 0.81 | 0.08 |
|  | AdaBoostM1 | 0.68 | 0.54 | 0.14 |
|  | DecisionTable | 0.88 | 0.74 | 0.14 |
|  | JRip | 0.83 | 0.78 | 0.05 |
|  | ZeroR | 0.53 | 0.21 | 0.32 |
|  | J48 | 0.91 | 0.84 | 0.07 |
|  | RandomForest | 0.90 | 0.82 | 0.08 |

Table 5.3 The classifiers comparison for each dataset.

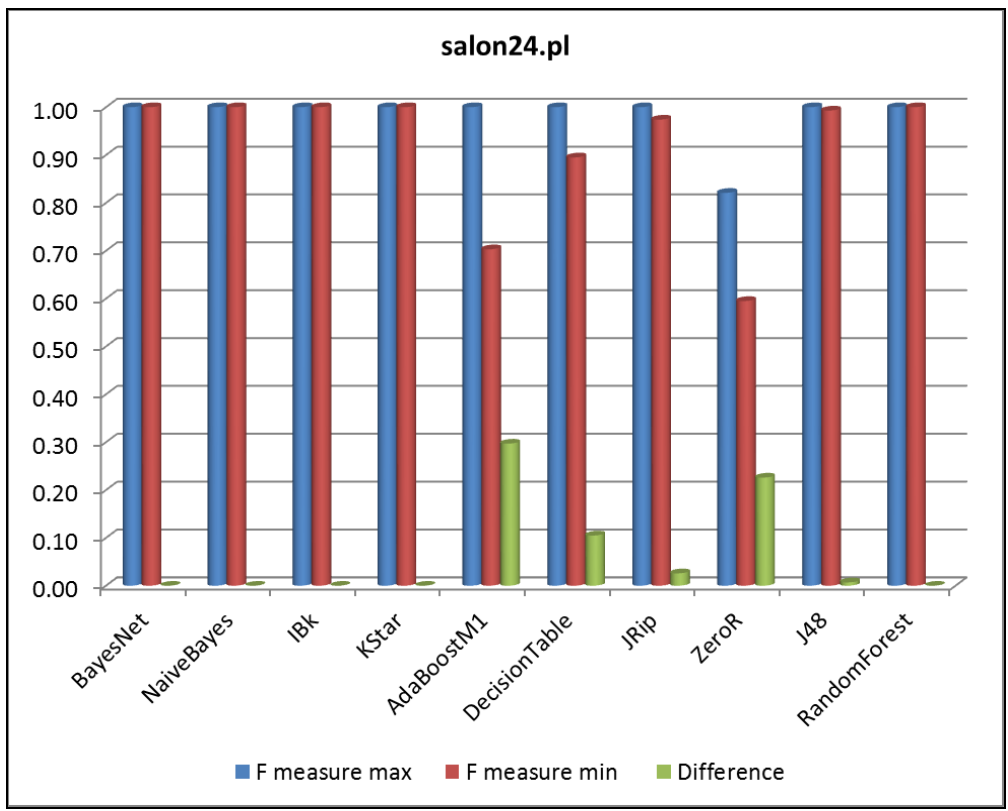

Figure 5.2 The classifiers comparison for salon24.pl.

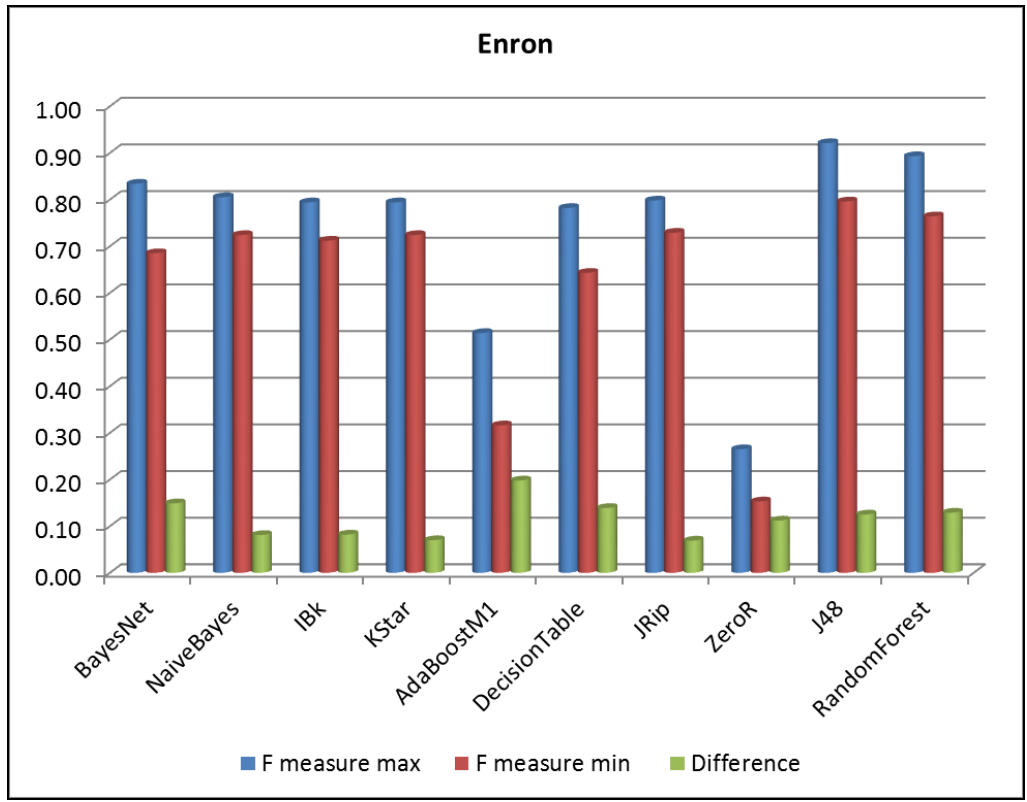

Figure 5.3 The classifiers comparison for Enron.

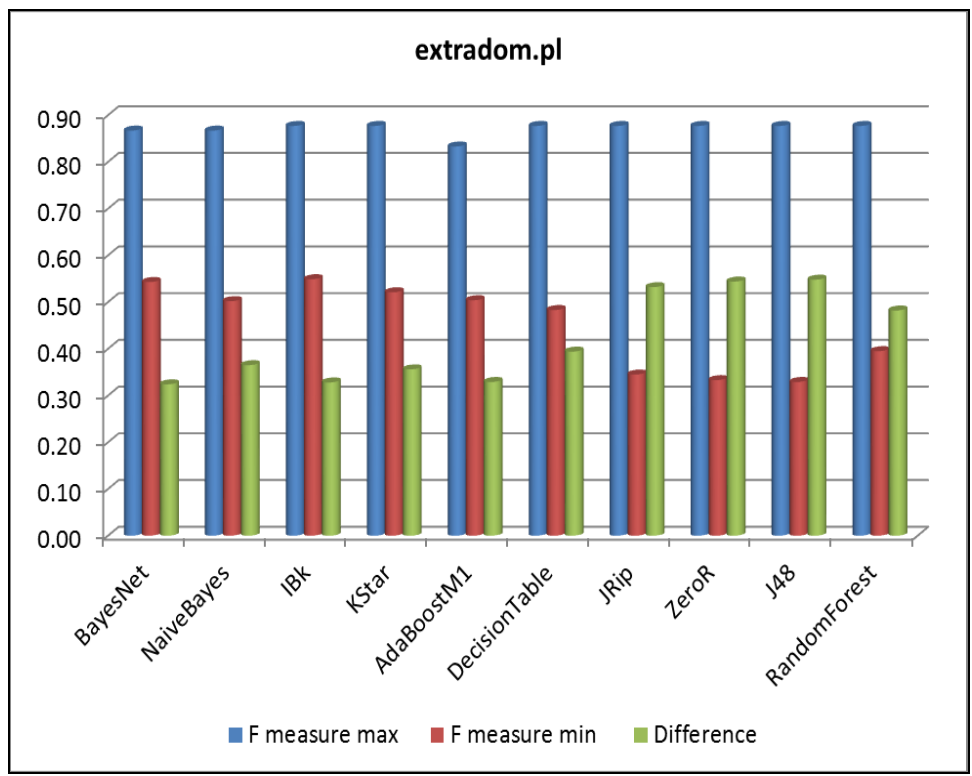

Figure 5.4 The classifiers comparison for extradom.pl.

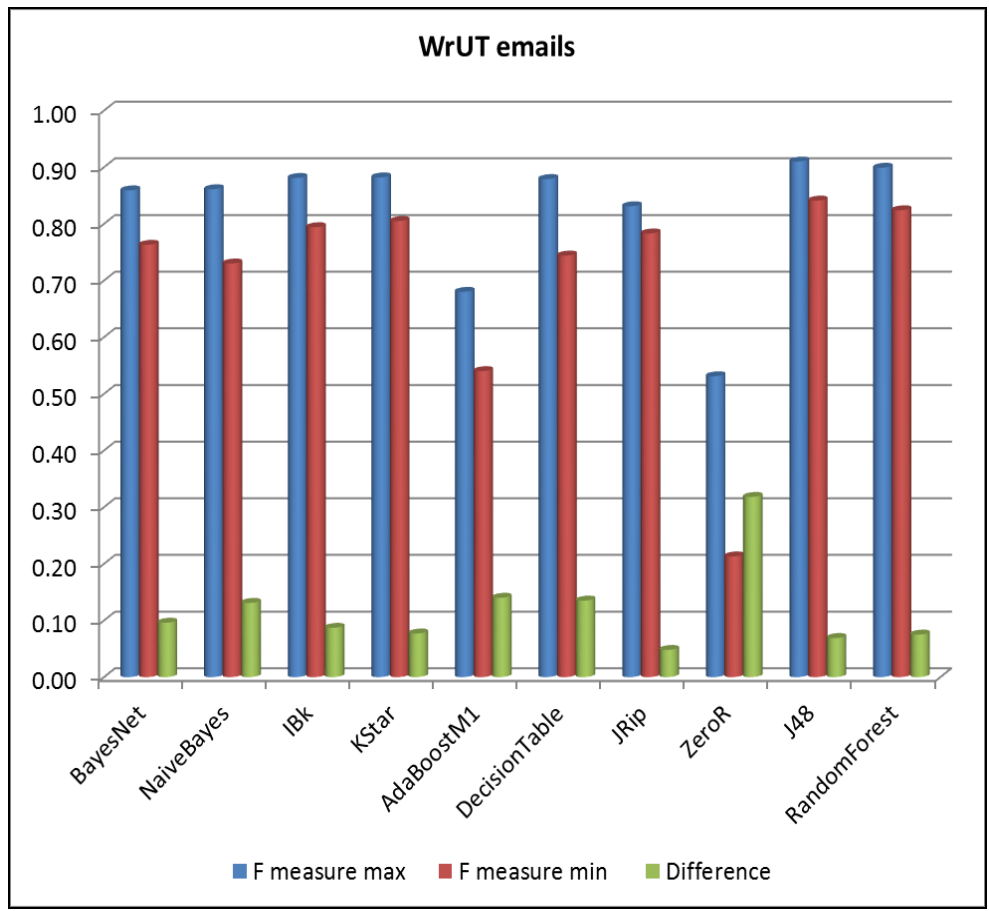

Figure 5.5 The classifiers comparison for WrUT emails.

Table 5.2 clearly indicates that for each dataset the best two classifiers are J48 (C4,5) decision trees and RandomForest ensemble of decision trees, thus, both classifiers were used for further analyses. Additionally, the results for these two classifiers are quite impressive since F measure for both of them is always around 0.8-0.9.

Next the analysis of how the $\alpha$ and $\beta$ parameters affect the classification was performed. This was done for the WrUT dataset. The first analysis was for J48 and is presented in Table 5.4 and Figures 5.6, 5.7.

| $\beta\backslash\alpha$ [%] | 50 | 60 | 70 | 80 | 90 | 100 |
|---|---|---|---|---|---|---|
| 50  | 0.881 | 0.85  | 0.887 | 0.889 | 0.884 | 0.888 |
| 60  | 0.884 | 0.879 | 0.898 | 0.885 | 0.883 | 0.91  |
| 70  | 0.886 | 0.89  | 0.897 | 0.902 | 0.897 | 0.884 |
| 80  | 0.879 | 0.885 | 0.889 | 0.91  | 0.886 | 0.882 |
| 90  | 0.87  | 0.882 | 0.871 | 0.913 | 0.892 | 0.887 |
| 100 | 0.852 | 0.869 | 0.848 | 0.907 | 0.869 | 0.841 |

Table 5.4 The weighted average of F-measure (weighted by the contribution of the class–event in the dataset) for J48 decision tree for all six possible classes.

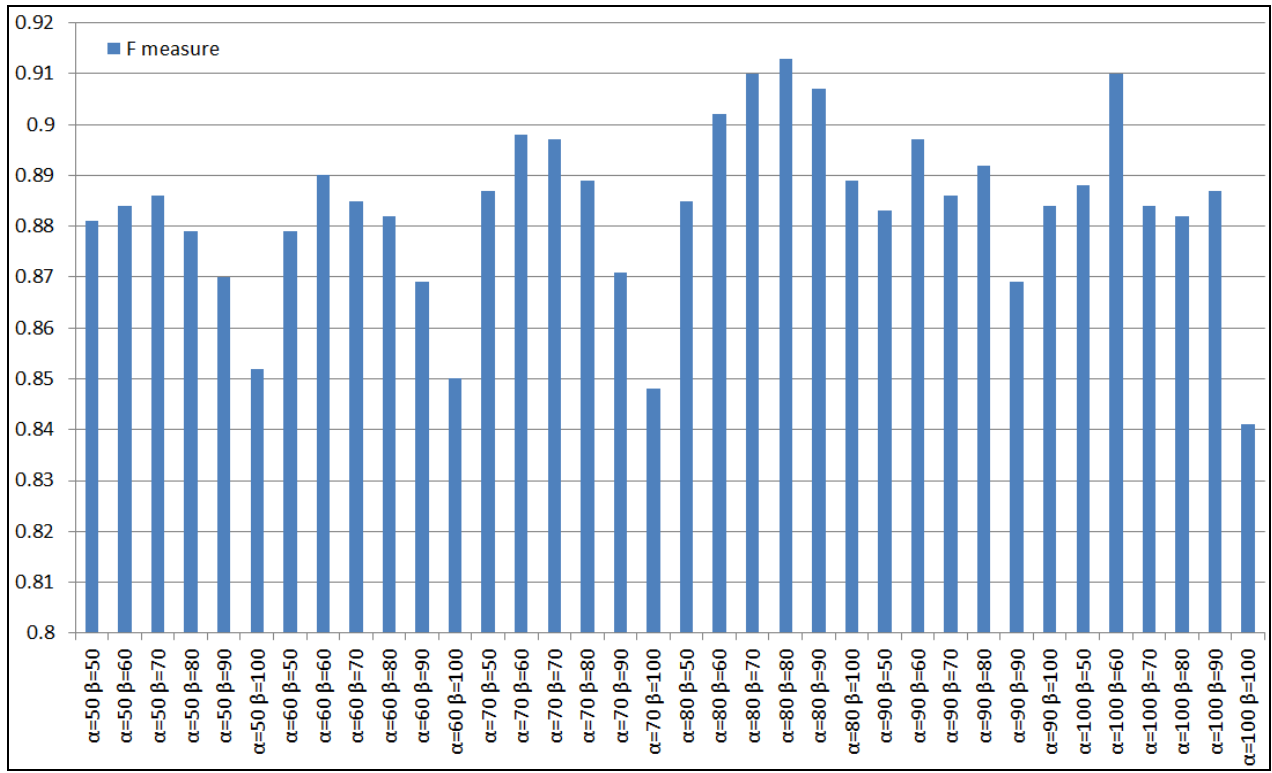

Figure 5.6 F-measure values in relation to $\beta$ and $\alpha$.

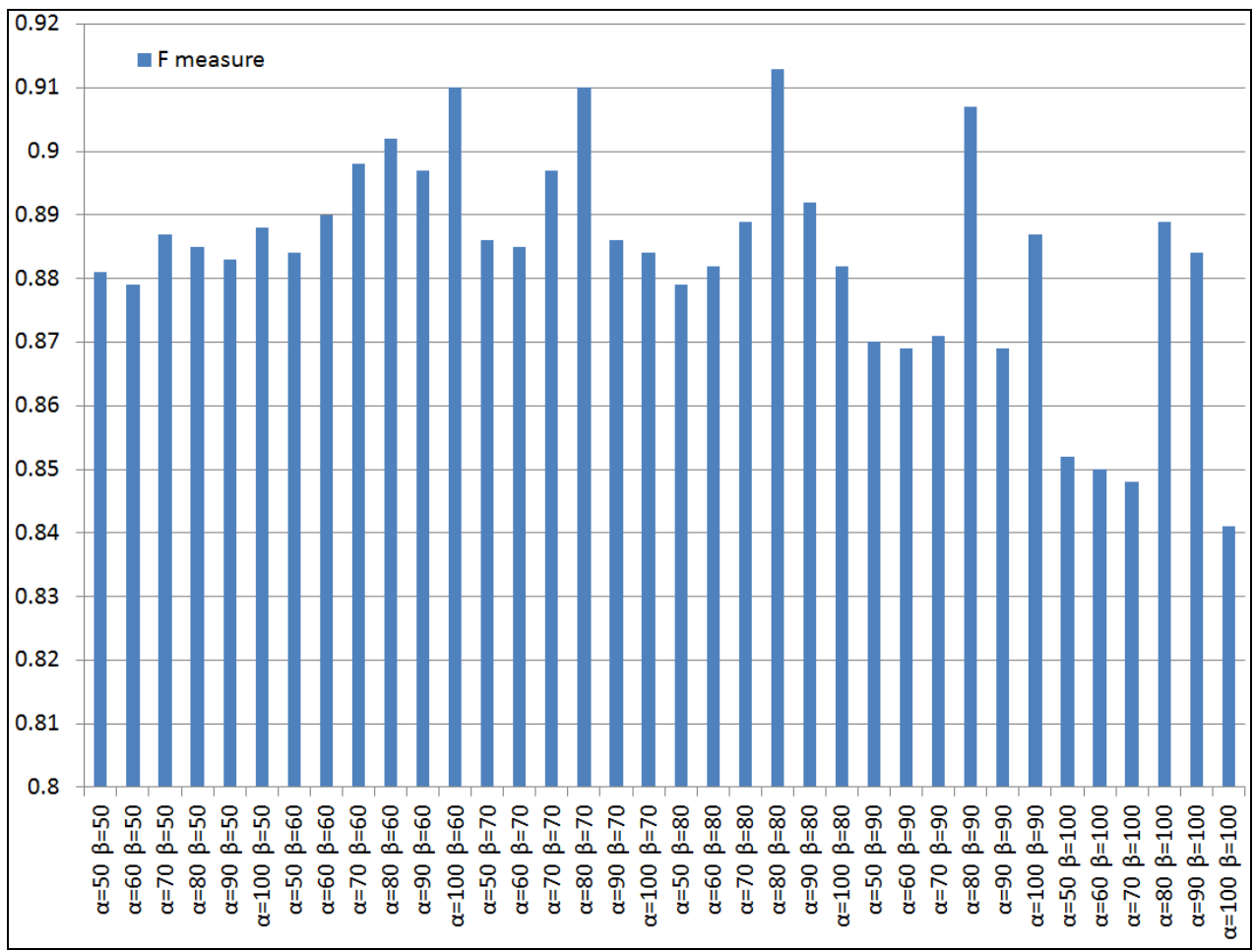

Figure 5.7 F-measure values in relation to $\alpha$ and $\beta$.

While analysing Figures 5.6 and 5.7, for the constant $\alpha$ we can observe that the best results are when $\beta$ is around 80%. However, for the constant $\beta$, it is hard to see any regular pattern. In general, the highest F-measure is for $\alpha = 80\%$. So, if the J48 decision tree is used as a classifier, it is recommended to use $\alpha = 80\%$ and $\beta$ from the set {70%, 80%, 90%} for the *GED* method parameters. The reason behind such a result can be quite simple. If we look at Figures 4.5, 4.6 and 4.7 (Section 4.3.1) we can see that the high value of $\alpha$ and $\beta$ reduce the number of split and merge events. Thus, the number of those events is similar to the number of other events. On the other hand, for the low $\alpha$ and $\beta$ the number of splits and merges overshadow the number of the other events. It means that value of about 80% appears to be the best with respect to classification quality evaluated by the F-measure.

Quite similar results were achieved by the RandomForest classifier. The parameter $\alpha$ can be from the set {80%, 90%, 100%} and $\beta$ from {60%, 70%, 80%, 90%}. Hence, the conclusion is: the *GED* method with the high $\alpha$ and $\beta$ produces better input features for classification, also if applied to the RandomForest classifier. The evaluation of $\alpha$ and $\beta$ influence on the RandomForest classifier was presented in Table 5.5, Figure 5.8 and 5.9.

| $\beta\backslash\alpha$ [%] | 50 | 60 | 70 | 80 | 90 | 100 |
|---|---|---|---|---|---|---|
| 50 | 0.846 | 0.848 | 0.857 | 0.874 | 0.868 | 0.87 |
| 60 | 0.848 | 0.852 | 0.865 | 0.881 | 0.875 | 0.899 |
| 70 | 0.846 | 0.853 | 0.872 | 0.891 | 0.879 | 0.897 |
| 80 | 0.849 | 0.854 | 0.862 | 0.893 | 0.882 | 0.867 |
| 90 | 0.843 | 0.848 | 0.849 | 0.896 | 0.872 | 0.887 |
| 100 | 0.828 | 0.824 | 0.828 | 0.869 | 0.869 | 0.849 |

Table 5.5 The weighted average of F measure for RandomForest tree for all six classes.

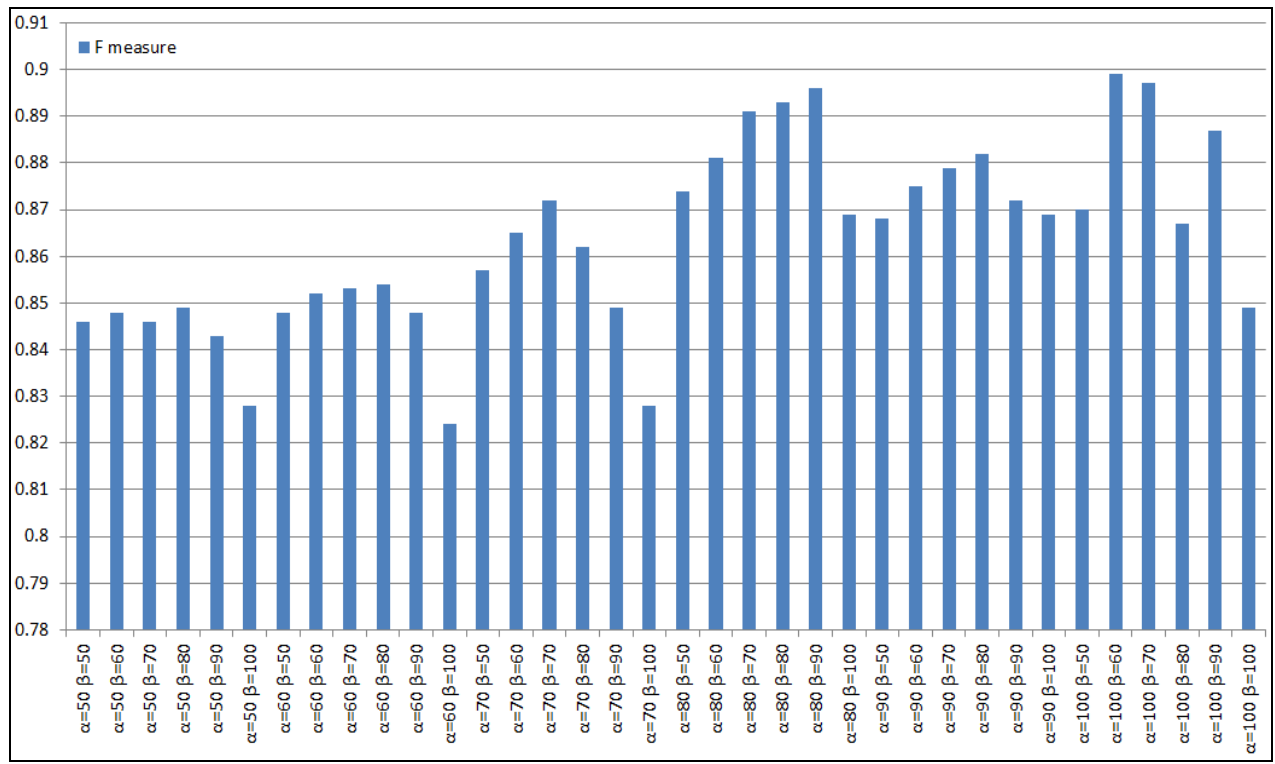

Figure 5.8 F-measure values in relation to $\beta$ and $\alpha$.

Not like for J48 tree, for RandomForest tree a specific pattern can be found for both $\alpha$ and $\beta$. For the constant $\alpha$ the best results are if $\beta$ is equal to 60, 70 or 80, see Figure 5.8, and for the constant $\beta$ the best results are when $\alpha$ is equal to 80, 90 or 100, see Figure 5.9.

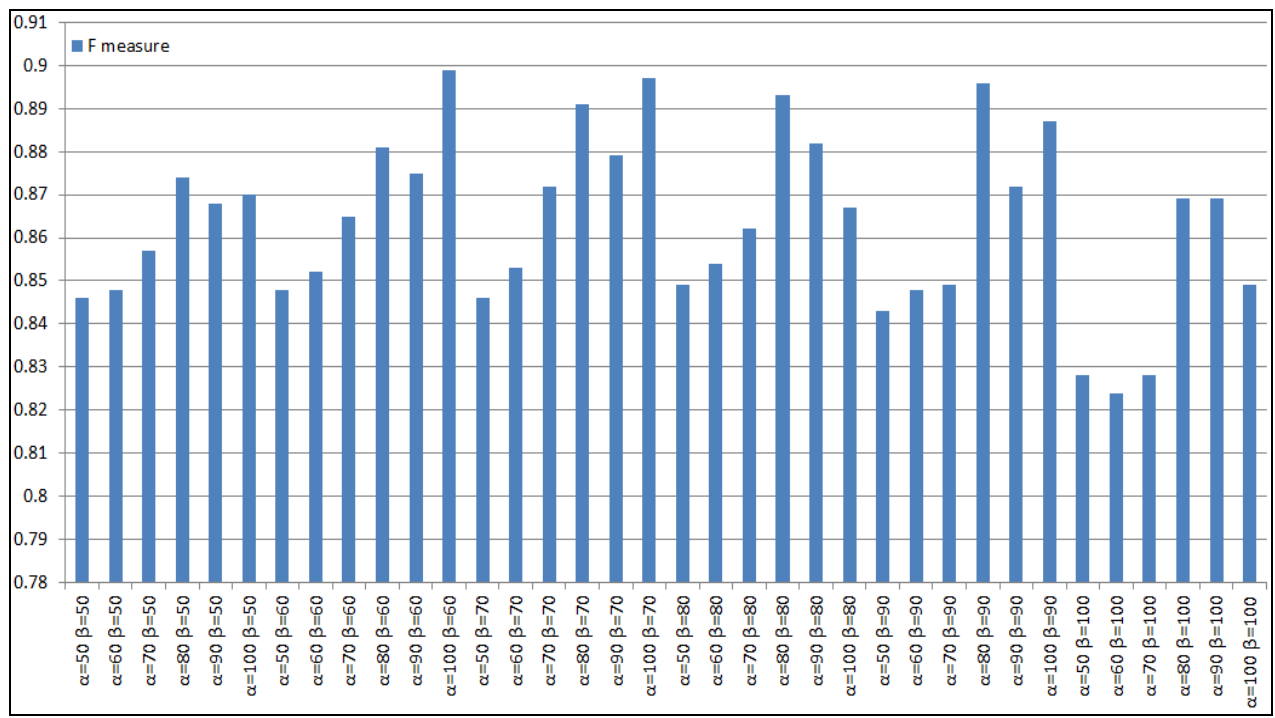

Figure 5.9 F-measure values in relation to $\alpha$ and $\beta$.

It was shown that using a simple sequence which consists only of several preceding groups' sizes and events as an input for the classifier, the learnt model is able to produce very good results, even for simple classifiers. It means that such prediction of group evolution can be very efficient in terms of prediction quality. Of course, many questions remain unsolved:

- Are similar results achievable for every network?
- What would happen, if we use different classifiers or more advanced classification concepts like competence areas (clustering of groups and application of separate classifiers to each cluster)?
- What would happen if we add more input features (measures) describing the group, like its diameter, average degree, percentage of network members which are in this group, the number of core members etc. as well as their various aggregations, e.g. average size for last 6 time frames?
- What would be the results, if we use shorter/longer sequence (more preceding events and group measures)?
- etc.

Note that above questions are suitable for the future work directions or another PhD dissertation. This thesis, however, just only aimed to show that predicting group evolution using the *GED* methods and some common classifiers is possible and effective.

# 6. Conclusions and Future Work

## 6.1 Conclusions

The main subject studied in this dissertation are multi-layered social networks (*MSN*), in which two nodes can be connected by more than one relationship (edge). Even though the multi-layered profile of networks is natural, this nature has not been studied extensively yet. In real social networks some relations may be so intertwined that it is almost impossible to analyse them separately, moreover, if considered together, they reveal additional vital information about the entire network. Especially nowadays, when it is so important to analyse information diffusion and social dependencies between people in the global network, and very often the data from only one communication channel (one type of relationships) are insufficient.

In the thesis, at first, the definition of multi-layered social network is proposed (Definition 2.2), together with the number of new measures useful to analyse such a kind of networks (Section 2.4). In particular the following new measures are proposed:

- multi-layered neighbourhood – *MN* (Section 2.4.1),
- cross layered clustering coefficient – *CLCC* (Section 2.4.2),
- four different multi-layered degree centralities – *MDCs and CDC* (Section 2.4.3),
- cross layered edge clustering coefficient – *CLECC* (Section 2.4.4),

Additionally, the guidelines on how to transform the algorithms designed for single-layered network into multi-layered social network environment are presented, together with an example of shortest path discovery algorithm (Section 2.4.5). This allows to calculate other typical measures like closeness centrality or betweenness centrality.

Based on the cross layered edge clustering coefficient measure a new algorithm called cross layered edge clustering coefficient method (*CLECC*) is proposed. Application of multi-layered neighbourhood allows the *CLECC* measure to respect all layers simultaneously. The adjustable method parameter $α$ is responsible for restrictiveness of the algorithm. This is especially suitable, if a multi-layered network consist of layers with significantly different number of edges. When layers with the high density exist along with sparse ones, the probability for two vertices to be in the multi-layered network for $α$ equal |L| is very small, thus the method output would contain only the strongest communities. However, by lowering $α$ the method limitation enforced on *CLECC* measure decreases, unveiling more of underlying community structures. Hence, the parameter $\alpha$ sets robustness of communities delivered by the *CLECC* method.

Having measures and algorithms defined, experimental studies are carried out. The first set of tests performed on reference networks (Section 3.6.1), *GN* Benchmark (Section 3.6.2) and *LFR* Benchmark (Section 3.6.3) indicates that the *CLECC* method works properly for a single-layered social network (*SSN*), and despite the fact that it is not the best method for *SSN*, it performs very well, especially for high values of mixing parameter.

The second part of tests on virtual world networks extracted from Timik.pl (Section 3.7.1) clearly shows that using the multi-layered network provides much better results than analysing each layer separately, both in algorithm quality and execution time. Thus, the statement from section 2.4 that the layers, if considered together, reveal additional vital information about the network has been confirmed.

To perform the last series of tests the extension of *LFR* Benchmark was worked out (Section 3.7.2). The new extension called m*LFR* Benchmark (multi-layered *LFR* Benchmark) provides the possibility to generate multi-layered test networks. Based on these networks, the last series of test has been carried out and presented in Section 3.7.3. The best results are almost always for the highest $\alpha$ parameter because the *CLECC* measure which for $\alpha>1$ acts as a filter and removes week connections present only on one layer. Also, the algorithm execution time is very low for $\alpha>1$. A different case is for $\alpha=1$. Then, the *CLECC* measure creates a union of all layers producing denser graph and increasing the number of connections between communities, while the number of intercommunity connection remains almost the same. This is reflected in poorer results, especially for mixing parameter $\mu > 0.6$, and very long execution time which is increasing along with mixing parameter. This is explained by network structure approaching the random graph model, thus, the layers differ from each other to a greater extent and the union network becomes denser. However, a very interesting issue happens when a mixing parameter $\mu$ exceeds the value of 0.8. The results for $\alpha=1$ no longer deteriorate and even become a little better. This might be explained by the fact that for $\mu>0.8$ there is so few intercommunity connections that the union of layers also enhance the group density.

Additionally, a very interesting property of the *CLECC* algorithm can be noticed when we compare the results for the 7-layer, 5-layer and 3-layer network. If the number of layers in the network increases, the results of the *CLECC* algorithm also become better. It means that the statement from section 2.4 that the layers, if considered together, reveal additional vital information about the network has been confirmed once again.

The next part of dissertation deals with the problem of tracking group evolution in social networks. At first the concept of dynamic social network *DSN* is presented (Section 4.1.3). *DSN* is a list of the following time frames where each frame is in fact a single-layered social network *SSN*($V,E_l,\{l\}$), or multi-layered social network *MSN*($V,E,L$).

Next, a new *GED* method for group evolution discovery is proposed (Section 4.2). It uses not only the size and equivalence of groups' members, but it also takes into account their position and importance within the group, in order to identify what has happened with the group in the successive time frames. This was mainly achieved by a new measure called *inclusion* (Equation 4.2), which respects both, the quantity and quality of the group. The quantity is reflected by the first part of the *inclusion* measure, i.e. what portion of members from one group is in another one, whereas the quality is expressed by the second part of the inclusion measure, namely what contribution of important members from one group is in another one. It provides a balance between the groups that contain many of less important members and groups with only few but key members. The *GED* method was designed to be as much flexible as possible and to be fitted to both overlapping and non-overlapping groups. Simultaneously, it preserves the low and adjustable computational complexity, because of many different user importance (centrality) indicators, which can be applied in the *inclusion* measure. Owing to its two parameters $\alpha$ and $\beta$, the full control over the method is provided.

The results of experiments and comparison with two existing methods presented in Section 4.3, lead to the conclusion that the desired effects have been achieved, and the new *GED* method may become one of the best methods for group evolution discovery.

The last problem studied in the thesis was prediction of group evolution in the social network. As a result, a new approach was presented in Section 5.1; it involves usage of the *GED* method. Having only a simple sequence which consists of several preceding groups' sizes and events as an input for the classifier, the trained model is able to produce very good results, even for simple classifiers.

The experiments performed by means of WEKA Data Mining Software with 10 different classifiers and six real world social networks, indicated that even simple classifiers like J48 (C4,5) decision trees or RandomForest ensemble of decision trees produce quite impressive results, since F-measure for both of them is always around 0.8-0.9 for all analysed datasets. Moreover, the detailed analysis of the influence of the *GED* parameters $\alpha$ and $\beta$ on the classification results pointed to conclusion that the best results are for $\alpha = 80\%$ and $\beta$ from the set $\{70\%, 80\%, 90\%\}$. This is explained by the fact that for the high values of $\alpha$ and $\beta$, the number of different events (growing, continuing, shrinking, dissolving, merging and splitting) is more or less equal.

To sum up, the main goal of this thesis, i.e. the development of a set of tools: measures, algorithms and methods which facilitate the group extraction and analysis of its evolution in multi-layered social networks has been fully achieved.

Most of the ideas and research presented in this thesis have been already published in the JCR-listed journals and proceedings of the best scientific conferences in the field. See Appendix III for the complete list of these papers.

## 6.2   Future Work

The first and main extension of ideas presented in the thesis is related to prediction of group evolution in the social network. As stated at the end of Section 5.3, many questions remain unanswered:

- Are similar prediction results achievable for every kind of network?
- What would happen, if we use different classifiers or more advanced classification concepts like competence areas (clustering of groups and application of separate classifiers to each cluster)?
- What would be the influence of adding more input features (measures) describing the group like its diameter, average degree, percentage of network members which are in this group, the number of core members etc. as well as their various aggregations, e.g. average size for last 6 time frames?
- What would be the results, if we use shorter/longer sequences (more preceding events and group measures)?
- etc.

All above will be the main direction of future research.

An additional, second possible extension involves very interesting phenomenon which occurred during the experiments on the *GED* method and was called "migration". It has been noticed that in some cases, when the leader (the member with the high position) leaves the group, in the subsequent time frames some group members with lower positions tend to follow the leader. This is quite interesting, but requires a lot of additional research to check how often it happens, why it happens and if it can be modelled using existing structural measures in the social network.

## 8. Appendix I - m*LFR* Benchmark Pseudocode

The following pseudocode description of the algorithm is a complement to the former description and pertains only to the algorithm responsible for distribution of the connections through the layers.

---

**m*LFR* Benchmark**

Create list containing all layers' identifiers except the base one.
For each layer from layer list
    Create MULTIMAP describing the same group
End for
For a randomly chosen layer *l* from layer list
    Choose last element from MULTIMAP collection for a layer *l* describing a vertex *x*
    If chosen vertex *x* belongs to the same group on layer *l* and base layer.
        Create list of neighbour propositions for vertex *x*
        For each neighbour *y* of vertex *x* on a base layer, which obeys rules:
- Vertex *x* does not already have vertex *y* as an adjacent vertex
- Number of existing connections for vertex *y* is lesser than its internal degree $k_{y,l}^{(in)}$, i.e. vertex *y* can have another neighbour
- vertex *x* is in the same group as vertex *y*
- x≠y (possible when vertices have more than one membership in common)

        Add vertex *y* to the list of neighbour propositions for vertex *x*
        For each randomly chosen neighbour *y* from list of neighbour proposition, till the list is not empty or considered vertex *x* cannot have more neighbours
        Calculate on how many layers vertex *x* and *y* are adjacent.
        If number drawn from cumulative distribution
            Add edge connecting vertex *x* and vertex *y*
            Add to vertex *y* to the list of inserted vertices.
        Remove vertex *y* from neighbour propositions
    End for
    For each vertex *z* from the list of inserted vertices.
        Remove element ($k_{z,l}^{(in)}$,*z*) from the pair list, where $k_{z,l}^{(in)}$ is the degree of the vertex *z* considering only edges yet to distribute, i.e. its internal degree downgraded by number of edges assigned to this vertex so far.
        If $k_{z,l}^{(in)} > 1$ add element ($k_{z,l}^{(in)} - 1$,*z*) to the pair list
    End for

    //SECOND PART
    Create iterator *it* pointing before the element containing information about vertex *x* (between last and next to last) on the pair list.
    Create list of elements to erase
    Calculate jumper, as a number of possible skips of elements by the iterator *it*.

---

> For $k_{x,l}^{(in)}$ minus inserted so far neighbours
>> Take the element before iterator *it* describing vertex *y*
>> While vertex *x* and *y* are neighbours or (if connecting the vertices would skew the distribution and the jump is possible).
>> Move iterator *it* and consider previous element on the list
>>
>> Add edge connecting vertex *x* and vertex *y*
>> Add vertex *y* to the pair list
>
> End for
> For each vertex *z* on to erase list
>> Remove element ($k_{z,l}^{(in)}$, *z*) from the pair list
>
> If $k_{z,l}^{(in)} > 1$ add element ($k_{z,l}^{(in)} - 1$, *z*) to the pair list
> End for
> Remove from the pair list element describing *x*
> If the pair list for layer *l* does not have any more elements, remove layer *l* from the list of layers.
End for

# 9. Appendix II – *CLECC* method Results for Football Social Network

| Name | Division | *CLECC* and *CLECC+* group id |
|---|---|---|
| **FloridaState** | Atlantic Coast | 3 |
| **NorthCarolinaState** | Atlantic Coast | 3 |
| **Virginia** | Atlantic Coast | 3 |
| **GeorgiaTech** | Atlantic Coast | 3 |
| **Duke** | Atlantic Coast | 3 |
| **NorthCarolina** | Atlantic Coast | 3 |
| **Clemson** | Atlantic Coast | 3 |
| **WakeForest** | Atlantic Coast | 3 |
| **Maryland** | Atlantic Coast | 3 |
| **MiamiFlorida** | Big East | 7 |
| **Rutgers** | Big East | 7 |
| **Temple** | Big East | 7 |
| **Syracuse** | Big East | 7 |
| **Pittsburgh** | Big East | 7 |
| **BostonCollege** | Big East | 7 |
| **WestVirginia** | Big East | 7 |
| **VirginiaTech** | Big East | 7 |
| **Michigan** | Big Ten | 4 |
| **Iowa** | Big Ten | 4 |
| **PennState** | Big Ten | 4 |
| **Northwestern** | Big Ten | 4 |
| **Wisconsin** | Big Ten | 4 |
| **Minnesota** | Big Ten | 4 |
| **Illinois** | Big Ten | 4 |
| **Purdue** | Big Ten | 4 |
| **OhioState** | Big Ten | 4 |
| **MichiganState** | Big Ten | 4 |
| **Indiana** | Big Ten | 4 |
| **OklahomaState** | Big Twelve | 4 |
| **Texas** | Big Twelve | 4 |
| **Missouri** | Big Twelve | 4 |
| **TexasA&M** | Big Twelve | 4 |
| **Oklahoma** | Big Twelve | 4 |
| **IowaState** | Big Twelve | 4 |
| **Nebraska** | Big Twelve | 4 |
| **Kansas** | Big Twelve | 4 |
| **Colorado** | Big Twelve | 4 |
| **Baylor** | Big Twelve | 4 |
| **TexasTech** | Big Twelve | 4 |
| **KansasState** | Big Twelve | 4 |
| **TexasChristian** | Conference USA | 5 |
| **AlabamaBirmingham** | Conference USA | 7 |
| **SouthernMississippi** | Conference USA | 7 |
| **Tulane** | Conference USA | 7 |
| **Army** | Conference USA | 7 |
| **Cincinnati** | Conference USA | 7 |
| **EastCarolina** | Conference USA | 7 |
| **Houston** | Conference USA | 7 |
| **Louisville** | Conference USA | 7 |
| **Memphis** | Conference USA | 7 |
| **UtahState** | Independents | 6 |
| **NotreDame** | Independents | 7 |
| **Navy** | Independents | 7 |
| **Connecticut** | Independents | 7 |
| **CentralFlorida** | Independents | 7 |
| **CentralMichigan** | Mid-American | 7 |
| **EasternMichigan** | Mid-American | 7 |
| **MiamiOhio** | Mid-American | 7 |
| **Kent** | Mid-American | 7 |
| **NorthernIllinois** | Mid-American | 7 |
| **WesternMichigan** | Mid-American | 7 |
| **Akron** | Mid-American | 7 |
| **Buffalo** | Mid-American | 7 |
| **BowlingGreenState** | Mid-American | 7 |
| **BallState** | Mid-American | 7 |
| **Toledo** | Mid-American | 7 |
| **Marshall** | Mid-American | 7 |
| **Ohio** | Mid-American | 7 |
| **NevadaLasVegas** | Mountain West | 5 |
| **AirForce** | Mountain West | 5 |
| **Utah** | Mountain West | 5 |
| **Wyoming** | Mountain West | 5 |
| **SanDiegoState** | Mountain West | 5 |
| **NewMexico** | Mountain West | 5 |
| **BrighamYoung** | Mountain West | 5 |
| **ColoradoState** | Mountain West | 5 |
| **Washington** | Pacyfic Ten | 6 |
| **Oregon** | Pacyfic Ten | 6 |
| **SouthernCalifornia** | Pacyfic Ten | 6 |
| **ArizonaState** | Pacyfic Ten | 6 |
| **UCLA** | Pacyfic Ten | 6 |
| **Arizona** | Pacyfic Ten | 6 |
| **Stanford** | Pacyfic Ten | 6 |
| **WashingtonState** | Pacyfic Ten | 6 |
| **California** | Pacyfic Ten | 6 |
| **OregonState** | Pacyfic Ten | 6 |
| **Arkansas** | Southeastern | 8 |
| **Tennessee** | Southeastern | 8 |
| **SouthCarolina** | Southeastern | 8 |
| **Mississippi** | Southeastern | 8 |
| **Georgia** | Southeastern | 8 |
| **LouisianaState** | Southeastern | 8 |
| **Alabama** | Southeastern | 8 |
| **Florida** | Southeastern | 8 |
| **Auburn** | Southeastern | 8 |
| **MississippiState** | Southeastern | 8 |
| **Kentucky** | Southeastern | 8 |
| **Vanderbilt** | Southeastern | 8 |
| **NewMexicoState** | Sun Belt | 6 |
| **Idaho** | Sun Belt | 6 |
| **NorthTexas** | Sun Belt | 6 |
| **ArkansasState** | Sun Belt | 6 |
| **MiddleTennesseeState** | Sun Belt | 7 |
| **LouisianaLafayette** | Sun Belt | 7 |
| **LouisianaMonroe** | Sun Belt | 7 |
| **SanJoseState** | Western Athletic | 5 |
| **Tulsa** | Western Athletic | 5 |
| **TexasElPaso** | Western Athletic | 5 |
| **Hawaii** | Western Athletic | 5 |
| **Nevada** | Western Athletic | 5 |
| **SouthernMethodist** | Western Athletic | 5 |
| **Rice** | Western Athletic | 5 |
| **FresnoState** | Western Athletic | 5 |
| **BoiseState** | Western Athletic | 6 |
| **LouisianaTech** | Western Athletic | 7 |

# 10. Appendix III – The List of Peer-reviewed Papers Published by the Author of This Thesis.

### Journals and book chapters

1. **Bródka P.,** Kazienko P., Musiał K., Skibicki K., *Analysis of Neighbourhoods in Multi-layered Dynamic Social Networks*. International Journal of Computational Intelligence Systems, vol. 5, no. 3, 2012 pp. 582-596. *(JCR-listed journal)*

2. **Bródka P.,** Saganowski P., Kazienko P., *GED: The Method for Group Evolution Discovery in Social Networks*, Social Network Analysis and Mining, 2012, DOI:10.1007/s13278-012-0058-8.

3. Filipowski T., **Bródka P.,** Kazienko P.: *EcoRide - the Social-based System for Car Traffic Optimization*. Chapter in Green Technologies and Business Practices: An IT Approach. IGI-Global, 2012.

4. Filipowski T., Kazienko P., **Bródka P.,** Kajdanowicz T.: *Web-Based Knowledge Exchange Through Social Links in the Workplace*. Behaviour and Information Technology, 2012, doi: 10.1080/0144929X.2011.642895. *(JCR-listed journal)*

5. Kukla G., Kazienko P., **Bródka P.,** Filipowski T.: *SocLaKE - Social Latent Knowledge Explorator*, The Computer Journal, vol. 55 no. 3, 2012, pp. 258-276, *(JCR-listed journal)*

6. Palus S., **Bródka P**., Kazienko P.: *Evaluation of Organization Structure based on Email Interactions*, International Journal of Knowledge Society Research, vol. 2, no. 1, 2011, pp. 1-13.

7. Kazienko P., Ruta D., **Bródka P.:** *The Impact of Customer Churn on Social Value Dynamics*, International Journal of Virtual Communities and Social Networking, vol. 1, no. 3, 2009, pp. 60-72.

8. Zygmunt A., **Bródka P.,** Kazienko P., Koźlak J., *Key Person Analysis in Social Communities within Blogosphere,* Journal of Universal Computer Science, vol. 18, no. 4, 2012, pp. 577-597 *(JCR-listed journal)*

### Conference proceedings

9. **Bródka P.**, Musiał K., Kazienko P., *A Performance of Centrality Calculation in Social Networks*. CASoN 2009, IEEE Computer Society, 2009, pp. 24-31.

10. **Bródka P.**, Musiał K., Kazienko P.: *Efficiency of Node Position Calculation in Social Networks*, KES 2009, LNCS, Springer, 2009, pp.455-463

11. **Bródka P.**, Musiał K., Kazienko P., *A Method for Group Extraction in Complex Social Networks*, WSKS 2010, CCIS 111, Springer, 2010, pp. 238-247.

12. **Bródka P**., Filipowski T., Kazienko P., *An Introduction to Community Detection in Multi-layered Social Network*, WSKS 2011, Springer CCIS 278, 2011, pp. 185–190

13. **Bródka P.**, Skibicki K., Kazienko P., Musiał K., *A Degree Centrality in Multi-layered Social Network*. CASoN 2011, IEEE Computer Society, 2011, pp. 237-242.

14. **Bródka P**., Stawiak P., Kazienko P., *Shortest Path Discovery in the Multi-layered Social Network*. ASONAM 2011, IEEE Computer Society, 2011, pp. 497-501.

15. **Bródka P.,** Saganowski S., Kazienko P., *Group Evolution Discovery in Social Networks*, ASONAM 2011, IEEE Computer Society, 2011, pp. 247-253.

16. **Bródka P.,** Saganowski S., Kazienko P., *Tracking Group Evolution in Social Networks*. SocInfo 2011, LNCS, Springer, 2011, pp. 316-319. ***Best Demo/Poster Paper Award, Student Travel Support Award.***

17. Juszczyszyn K., Musiał A., Musiał K., **Bródka P.:** *Molecular Dynamics Modelling of the Temporal Changes in Complex Networks*, CEC 2009, IEEE Computer Society Press, 2009 pp. 553-559.

18. Juszczyszyn K., Musiał A., Musiał K., **Bródka P.:** *Utilizing Dynamic Molecular Modelling Technique for Predicting Changes in Complex Social Networks*, IEEE/WIC/ACM 2010, IEEE Computer Society Press, 2010, pp. 1-4

19. Kazienko P., **Bródka P.,** Ruta D.: *The Influence of Customer Churn and Acquisition on Value Dynamics of Social Neighbourhoods*, WSKS 2009, LNCS, Springer, 2009, pp.491-500

20. Kazienko P., **Bródka P**., Musiał K.: *Individual Neighbourhood Exploration in Complex Multi-layered Social Network*, IEEE/WIC/ACM 2010, IEEE Computer Society Press, 2010, pp. 5-8

21. Kazienko P., **Bródka P**., Musial K., Gaworecki J.: *Multi-layered Social Network Creation Based on Bibliographic Data*, SocialCom 2010, IEEE Computer Society Press, 2010, pp. 407-412

22. Kazienko P., Musial K., Kukla E., Kajdanowicz T., **Bródka P.:** *Multidimensional Social Network: Model and Analysis*, ICCCI 2011, LNCS, Springer, 2011, pp. 378-387.

23. Kazienko P., Kukla E., Musial K., Kajdanowicz T., **Bródka P.,** Gaworecki J.: *A Generic Model for Multidimensional Temporal Social Network*, ICeND2011, CCIS, Springer, 2011, pp. 1-14